\documentclass[twocolumn,aps,pre,showpacs,epsfig,amsmath,amssymb,eqsecnum]{revtex4}

\usepackage{amsmath,amssymb,bm,epsfig,graphics,verbatim}
\usepackage[usenames,dvipsnames]{color}
\def\nopsp{{\bar{\nop}}}

\def\mass{\mathcal{M}}

\def\ncp{{\mathbf Y}}

\def\LDLa{\mathcal{A}}

\def\smear{H}
\def\V{\tilde{V}}

\def\stiff{\mathcal{L}}

\def\source{\mathfrak{H}}

\def\vertex{\Delta}
\def\nopc{Q}
\def\nop{{\rm{\mathbf\nopc}}}

\def\mycorrel{{\cal C}}

\def\correllength{\xi}

\newcommand{\be}{\begin{equation}}
\newcommand{\ee}{\end{equation}}
\newcommand{\ba}{\begin{eqnarray}}
\newcommand{\ea}{\end{eqnarray}}
\newcommand{\bw}{\begin{widetext}}
\newcommand{\ew}{\end{widetext}}

\newcommand{\rv}{{\bm{r}}}

\newcommand{\pv}{{\bm{p}}}
\newcommand{\qv}{{\bm{q}}}

\begin{document}
%-------------
%-------------

\title{Statistical physics of isotropic-genesis nematic elastomers:\\
I.~Structure and correlations at high temperatures}

\author{Bing-Sui~Lu$^1$}
\author{Fangfu Ye$^2$}
\author{Xiangjun Xing$^1$}
\author{Paul M.~Goldbart$^2$}
%\affiliation{$^1$Department of Physics and Institute for Condensed Matter Theory,
%University of Illinois at Urbana-Champaign, 1110~West Green Street, Urbana, IL~61801-3080, USA}
\affiliation{$^1$Department of Physics and Institute of Natural Sciences,
Shanghai Jiao Tong University, 800~Dongchuan Road, Minhang District, Shanghai, China}
\affiliation{$^2$School of Physics, Georgia Institute of Technology,
837~State Street, Atlanta, GA~30332-0430, USA}

%-------------
%\date{\today}
\date{\today}
%{**}

\begin{abstract}
Isotropic-genesis nematic elastomers (IGNEs) are liquid crystalline polymers (LCPs) that have been randomly, permanently cross-linked in the high--temperature state so as to form an equilibrium random solid.
Thus, instead of being free to diffuse throughout the entire volume, as they would be in the liquid state, the constituent LCPs in an IGNE are mobile only over a finite, segment specific, length--scale controlled by the density of cross-links.  %
We address the effects that such network--induced localization have on the liquid--crystalline characteristics of an IGNE, as probed via measurements made at high temperatures.
In contrast with the case of uncross-linked LCPs, for IGNEs these characteristics are determined not only by thermal fluctuations but also by the quenched disorder associated with the cross-link constraints.
To study IGNEs, we consider a microscopic model of dimer nematogens in which the dimers interact via orientation-dependent excluded volume forces.
The dimers are, furthermore, randomly, permanently cross-linked via short Hookean springs, the statistics of which we model by means of a Deam-Edwards type of distribution.
We show that at length--scales larger than the size of the nematogens this approach leads to a recently proposed, phenomenological Landau theory of IGNEs [Lu et al., {\em Phys.\ Rev.\ Lett.}~{\bf 108}, 257803 (2012)], and hence predicts a regime of short--ranged oscillatory spatial correlations in the nematic alignment, of both thermal and glassy types.
In addition, we consider two alternative microscopic models of IGNEs:
(i)~a wormlike chain model of IGNEs that are formed via the cross-linking of side-chain LCPs; and
(ii)~a jointed chain model of IGNEs that are formed via the cross-linking of main-chain LCPs.
At large length--scales, both of these models give rise to liquid--crystalline characteristics that are qualitatively in line with those predicted on the basis of the dimer-and-springs model, reflecting the fact that the three models inhabit a common universality class.
\end{abstract}

%-------------
\maketitle
%-------------

\section{Introduction}
\label{sec:intro}

\subsection{Nematic elastomeric materials}
\label{sec:materials}
Nematic elastomers are fascinating materials in which the liquid crystalline order is strongly coupled to the elasticity of the underlying elastomeric network~(see, e.g., Refs.~\cite{finkelmann,LCE:WT,urayama,Xing-Radz1,Xing-Radz2,Xing-Radz3}).  This strong nemato-elastic coupling gives rise to novel, \emph{emergent} properties in nematic elastomers that are found neither in liquid crystal nematics nor ordinary rubbery materials.  One well-known example of such properties is the soft elasticity characteristic of monodomain nematic elastomers~(see, e.g., Ref.~\cite{lubensky-xing-review}) that are formed via the K\"{u}pfer-Finkelmann procedure~\cite{KF}.

Not only are nematic elastomers fascinating but also they have proven to be a challenging subject for theoretical investigation.  Part of the challenge originates in the dependence of the physical characteristics of nematic elastomers on the conditions under which the elastomers are prepared.  For example, \emph{isotropic-genesis nematic elastomers} (or IGNEs)---nematic elastomers cross-linked in the isotropic state---exhibit the so-called supersoft version of elastic response at sufficiently low temperatures~(see, e.g., Refs.~\cite{urayamaPMtransition,biggins-short,biggins-long}), unlike their nematic-genesis nematic elastomer (or NGNE) counterparts.  Furthermore, {\it in thermal equilibrium\/} the nematic alignment in IGNEs exhibits a polydomain structure (see, e.g., Refs.~\cite{clarke,uchida-2d,uchida,selinger,feio,urayama,ye}) characterized by a length-scale determined solely via thermodynamic quantities such as the density of cross-links.

%The polydomain state is also therefore qualitatively distinct from the nonequilibrium state created via the temperature quench of a nematogenic liquid from its isotropic to its nematic state (the structure of which is observed, e.g., via the Schlieren pattern it causes), in which topological defects hinder but ultimately do not prevent relaxation to the uniform nematic state.

Another challenge to theorists comes from the fact that nematic elastomers possess a multi-level hierarchy of interdependent elements of randomness.
First, there is quenched disorder in its conventional form, associated with the permanent chemical structure that originates in the cross-linking process.
Second, as a result of sufficient cross-linking there arise the mean positions and r.m.s.~displacements of the spatially localized polymers that constitute the elastomeric network, both of these elements being random.
Third, there is the thermal disorder associated with the Brownian motion of the positional and orientational (i.e.,  nematogenic) freedoms in the state of the system {\it just prior to the instant of cross-linking\/}; in part, this thermal disorder is frozen in via the process of cross-linking.
Fourth, there is also the thermal disorder associated with the Brownian motion of the nematogens in the state of the system long after the instant of cross-linking.
It is not \textit{a priori}\/ evident how the interplay between the various types of randomness present in nematic elastomers resolve themselves, and thus determine the equilibrium structure and elastic response of such media.
As we shall try to make clear in subsequent sections of the present Paper, in order to understand nematic elastomers it is valuable to go beyond the conventional notion of quenched disorder and, instead, to consider an amalgam of the second, third and fourth types of randomness composed of a \lq\lq frozen\rq\rq\ part (due to the mean positions of the network constituents) and a \lq\lq molten\rq\rq\ part (due to the thermal fluctuations of the network constituents).

\subsection{Overview}
\label{sec:overview}
In the present Paper we investigate the static structure of nematic alignment in IGNEs that can be probed in the high-temperature regime.  This investigation is partly motivated by the experiments on polydomain structure reported in Ref.~\cite{urayamaPMtransition}.  We focus our considerations here on systems in the high-temperature regime and which are not subject to externally applied deformations.

Two types of nematic fluctuations are present in IGNEs: (i)~those that are frozen in, however imperfectly, by the network during the process of cross-linking; and (ii)~thermally driven departures away from the mean local alignment pattern that is frozen in.  In order to characterize such fluctuations, we make use of (i)~the correlator of the local nematic order that was frozen in during the cross-linking process, this correlator being appropriately averaged over realizations of the quenched disorder (and termed the glassy correlator); and (ii)~the correlator of the thermally driven departures of the nematic order from the mean frozen-in local alignment pattern, appropriately averaged over realizations of the quenched disorder (and termed the thermal correlator).

We have already mentioned that the physical properties of IGNEs depend on the conditions under which they were prepared.  To reflect this fact, we make a careful distinction between two thermodynamic ensembles in our theoretical approach: the first, in which the IGNE was {\it prepared}, which we term the \emph{preparation ensemble}; and the second, in which the system is {\it measured}, which we call the \emph{measurement ensemble}.  This structure enables our theory to capture the ability of IGNEs to \lq\lq remember\rlap,\rq\rq\ at least to some degree, the local nematic alignment pattern at the moment of preparation, and in addition enables the determination of the dependence of the strength of this memorization on (i)~the temperature at which the system was cross-linked, and (ii)~the average number of cross-links per polymer.

We note in passing that having both preparation and measurement ensembles places us in a family of disordered systems for which spontaneous replica symmetry breaking is expected to be irrelevant, in contrast with settings that do not feature a preparation ensemble; cf.~Ref.~\cite{terentjev2}.  (Technically, this expectation shows up in the need to investigate coupled replicas of the physical system in the neighborhood of not zero but one replica.)\thinspace\  Thus, we do not expect our approach to yield glassy phenomena such as hysteresis in the stress-strain behavior of IGNEs---which is not unreasonable, given the absence, to date, of experimental observations of such phenomena.

The fact that the elastomer network is thermally fluctuating means that any prospective theory of IGNEs should feature a typical localization length-scale, below which the polymer constituents of the network are effectively delocalized.  Our approach features such a length-scale, which leads to the possibility that nematic correlations undergo a novel, {\it oscillatory\/} form of decay with distance in a certain regime.  Prior theoretical approaches built on conventional random-field models do not feature the thermal fluctuations of the elastomer network (see, e.g., Refs.~\cite{uchida,uchida-2d,terentjev1,terentjev2,terentjev3}), and thus do not capture this intriguing phenomenon.  A phenomenological random-field-type model that {\it does\/} take the thermal fluctuations of the elastomer network into account was presented in Ref.~\cite{phenomenology}.  In the present Paper we derive the Landau-type free energy for IGNEs that was presented in Ref.~\cite{phenomenology}, doing so via a microscopic model that involves dimers that are randomly and permanently connected by Hookean springs.  The present Paper thus provides a microscopic justification for the ideas and results presented in Ref.~\cite{phenomenology}.

The outline of the Paper is as follows.
In Sec.~\ref{sec:ingredients} we present a microscopic dimer-spring model of IGNEs.
In Sec.~\ref{sec:replicas} we apply the replica technique and implement the Hubbard-Stratonovich scheme to decouple the interacting microscopic degrees of freedom.
In Sec.~\ref{sec:landau-wilson} we derive a Landau-Wilson type of free energy for the IGNE, which involves an order parameter field $\nop$ for the isotropic-to-nematic phase transition as well as an order parameter field $\Omega$ for the vulcanization/gelation transition.
In Sec.~\ref{sec:new} we make an expansion of the Landau-Wilson free energy for small $\nop$ and $\Omega$, which is
appropriate for exploring the physics in the vicinity of the transition to the random solid state.
In Sec.~\ref{sec:stationarity} we determine the stationary states of the Landau-Wilson free energy.
In Sec.~\ref{sec:effectivetheory} we derive an effective replica Hamiltonian describing local nematic order in IGNEs by setting $\Omega$ to its stationary value but retaining fluctuations of $\nop$ to quadratic order.
In Sec.~\ref{sec:equivalence} we then compare this effective Hamiltonian with that arising from the phenomenological Landau free energy considered in Ref.~\cite{phenomenology} and show that they are equivalent.
In Sec.~\ref{sec:diagnostics} we use the effective Hamiltonian to derive the glassy and thermal correlators.
In Sec.~\ref{sec:comparison} we describe two alternative microscopic models of the IGNE, viz.~(i)~a worm-like chain model of side-chain nematic polymer networks; and (ii)~a jointed chain model of main-chain nematic polymer networks.  As we shall see, at length-scales larger than the size of a nematogen, both of these models give rise to liquid--crystalline characteristics similar to those resulting from the dimer-and-springs model, reflecting the fact that the three models inhabit a common universality class.
In Sec.~\ref{sec:conclusion} we make some concluding remarks.

\section{Ingredients of the model}
\label{sec:ingredients}

\begin{figure}
	\centering
		\includegraphics[width=0.35\textwidth]{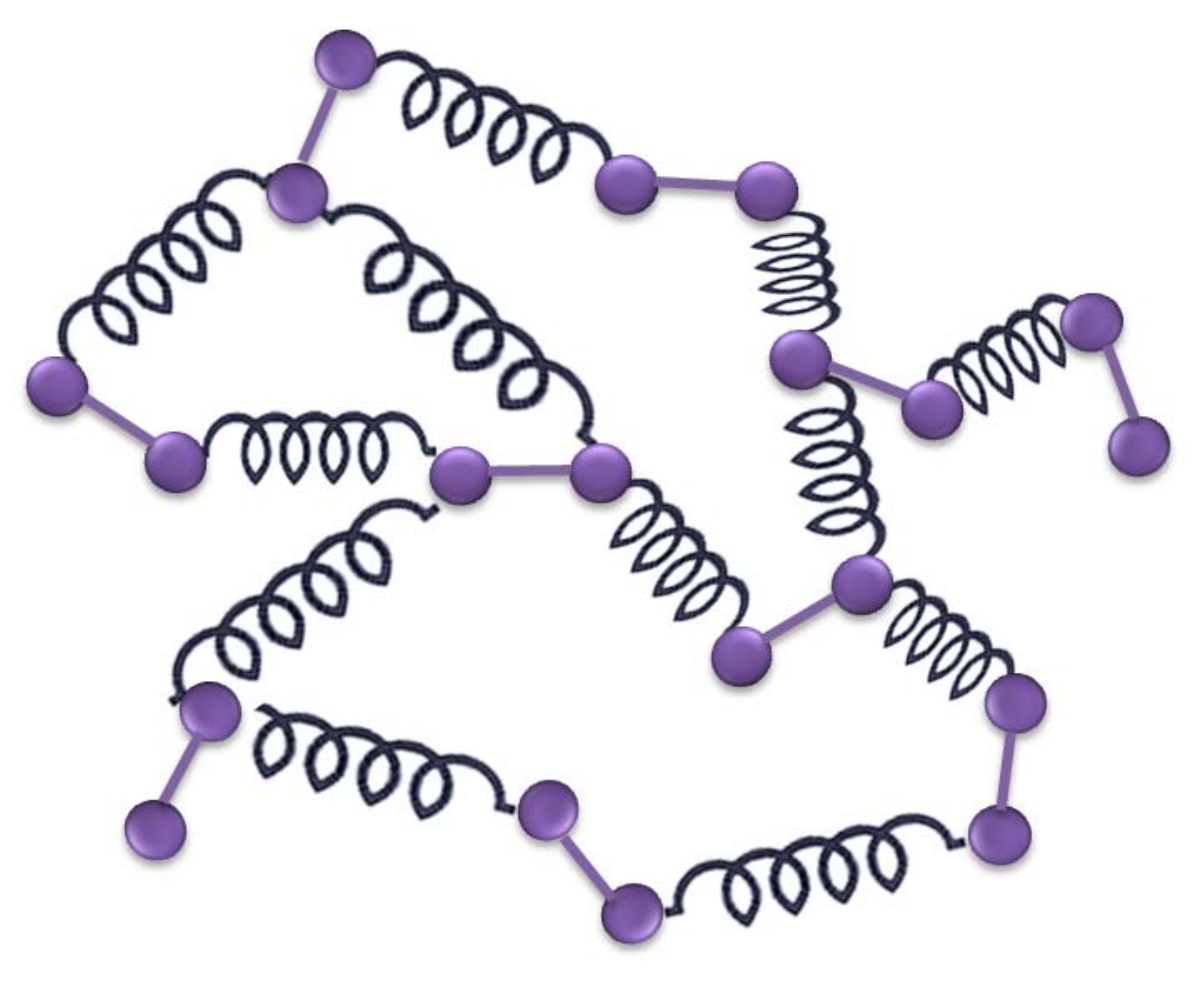}
	\caption{A model of dimers (indicated by dumbbells) cross-linked via harmonic springs.
             The dimers have length $\ell$ and the springs have r.m.s.~length $b$.}
  \label{fig:dimermodel}
\end{figure}

% [** restrict later to D=3]
We model an IGNE microscopically as a system of $N$ dimers in $D$ spatial dimensions that are randomly, permanently linked via springs (see Fig.~\ref{fig:dimermodel} and Ref.~\cite{XPMGZ-NE-VT}).  We envision the springs as mimicking the flexible constituents of liquid crystalline polymers whilst also serving as cross-links; the dimers mimic the stiff constituents of liquid crystalline polymers.  Each dimer (labeled by $j$, where $j=1,\ldots,N$) consists of two particles at position vectors $\bm{c}_{j,1}$ and $\bm{c}_{j,-1}$ separated by a fixed distance $\ell$.  The orientation of the $j$-th dimer is specified by the unit vector
\be
{\bm{n}}_j=\frac{{\bm{c}}_{j,1}-{\bm{c}}_{j,-1}}{|{\bm{c}}_{j,1}-{\bm{c}}_{j,-1}|}.
\ee
The dimers interact via three types of forces.
First, there is an {\it orientational\/} interaction between dimers that promotes parallel or antiparallel alignment.  We model this interaction via a potential of the Maier-Saupe type, viz.,
\be
\label{eq:Hnem}
H_{\rm{nem}} = -\frac{V}{2N}\sum_{i,j=1}^N J({\bm{c}}_i-{\bm{c}}_j)({\bm{n}}_i\cdot{\bm{n}}_j)^2,
\ee
where $\bm{c}_j\equiv ({\bm{c}}_{j,1}+{\bm{c}}_{j,-1})/2$ is the position of the $j$-th dimer's center of mass.
We assume that the aligning interaction is short-ranged, and model the interaction potential $J(\bm{c})$ by the form $\big( J_0 / (2\pi a^2)^{D/2} \big) \exp (-c^2/2a^2)$.  In Fourier space, the potential is given by $J_{\pv}= J_0 \exp(-p^2 a^2/2)$, where $a$ specifies the range of the interaction between dimers and $J_0$ characterizes its strength.
In addition to the orientational interaction, the dimers experience a {\it positional\/} excluded-volume interaction between particles belonging to any pair of dimers, which we model via an Edwards-type pseudo-potential~\cite{edwards1,edwards2}:
% [** really Edwards since this is not polymers but particles?]
\be
H_{\rm{ev}} = \frac{\lambda}{2}\sum_{i,j=1}^N \sum_{s,t=1,2} \delta({\bm{c}}_{i,s}-{\bm{c}}_{j,t}),
\ee
where $\lambda$ is the strength of the excluded-volume interaction~\cite{ev-footnote}.
The presence of sufficiently strong excluded-volume forces stabilizes the system against collapse to a globule, even when well cross-linked.
Third, any given two dimers that are connected by a spring are taken to interact additionally via a harmonic potential associated with the spring, which we take to have zero rest-length and native mean-square end separation $b^2$,  characteristic of Gaussian molecular chains.  These springs and, specifically, the architectural information indicating which pairs of rod ends are connected to one another by springs constitutes the quenched randomness $\chi$ of any given realization of the system.  This information takes the form
$\chi\equiv\{i{e},s_{e},j_{e},t_{e}\}_{e=1}^{M}$, where $M$ is the total number of springs and spring $e$ connects end $s_{e}$ of rod $i_{e}$ to end $t_{e}$ of rod $j_{e}$.  These springs result in the following Hookean term in the Hamiltonian:
\be
H_{\rm{xlink}} = \frac{T}{2b^2}\sum_{e=1}^M |{\bm{c}}_{i_e,s_e}-{\bm{c}}_{j_e,t_e}|^2,
\ee
where $T$ denotes the temperature and we have adopted units in which Boltzmann's constant is unity.

The total Hamiltonian for the dimers-and-springs model is then given by
\be
\label{eq:dimer_hamiltonian}
H_\chi=H_{\rm{nem}}+H_{\rm{xlink}}+H_{\rm{ev}}.
\ee
For a given realization $\chi$ of the quenched disorder, the corresponding partition function takes the form
\be
\label{eq:Zchi}
Z_{\chi}=\int
\left(\prod_{i=1}^{N}
d{\bm{c}}_{i,-1}\,
d{\bm{c}}_{i, 1}\,
\delta(|{\bm{c}}_{j,1}-{\bm{c}}_{j,-1}|-\ell)
\right)
e^{-H_\chi/T},
\ee
and the free energy $F_{\chi}$ is given by $-T\ln Z_{\chi}$.  As is well known~\cite{brout}, it is appropriate to average the free energy of the system over realizations of the quenched disorder.  Denoting this average by the square brackets $[\cdots]$, one has
\be
\label{eq:disorderaverageF}
[F_{\chi}] =
\sum_{\chi}{P_\chi}\,{F_\chi} =
- T \sum_{\chi} {P_\chi}\,\ln{Z_\chi}.
\ee

\section{Replicas and collective fields}
\label{sec:replicas}
\subsection{Statistics of quenched disorder}
What statistical distribution should one use to compute the average of the free energy over the quenched disorder?
In common with other elastomers such as isotropic rubbery systems, the quenched-disorder average for IGNEs can be performed via a variant of the Deam-Edwards distribution ${P(\chi)}$~\cite{Deam-Edwards-1976}.  Such distributions reflect situations in which systems undergo {\it instantaneous\/} cross-linking: one begins with a melt or solution at equilibrium and---so rapidly that hardly any relaxation has time to occur---one introduces permanent bonds between some random fraction of the pairs of dimers that happen, at the instant of cross-linking, to be nearby one another.
To construct the associated $P(\chi)$, one should respect the causal order of the cross-linking process (see, e.g. Ref.~\cite{xing-DE}).  Formally, this amounts to the specification
\be
\label{eq:DE}
P(\chi)=
\int
\mathcal{D}{\bm{c}^0}\,
P(\chi | \{ \bm{c}_{i,s}^0 \} )\,
P( \{ \bm{c}_{i,s}^0 \} ),
\ee
where the multiple integral $\int \mathcal{D} \bm{c}^0$ is defined via
\be
\int \mathcal{D} \bm{c}^0
\equiv
\prod_{i=1}^{N}
\int
d\bm{c}_{i,-1}^{0}\,
d\bm{c}_{i, 1}^{0},
\ee
the distribution $P( \{ \bm{c}_{i,s}^0 \} )$ describes the statistics of the liquid system at the instant of cross-linking, so that
\bw
\be
P( \{ \bm{c}_{i,s}^0 \} )
\equiv
\frac{
e^{- \left( H_{\rm{nem}}(\{ \bm{n}_{i}^0 \}) + H_{\rm{ev}}( \{\bm{c}_{i,s}^0 \}) \right)/T^0 }
\prod_{i=1}^{N} \delta(|\bm{c}_{i,1}^{0} - \bm{c}_{i,-1}^{0}|-\ell)}
{\int \mathcal{D} \bm{c}^0 \,
e^{- \left( H_{\rm{nem}}(\{\bm{n}_{i}^0\}) + H_{\rm{ev}}(\{\bm{c}_{i,s}^0\}) \right)/T^0 }
\prod_{i=1}^{N} \delta(|\bm{c}_{i,1}^{0} - \bm{c}_{i,-1}^{0}|-\ell)},
\ee
\ew
and $P(\chi | \{ \bm{c}_{i,s}^0 \} )$ is the conditional probability that cross-links are formed between pairs of dimer-ends at positions
$\{ (\bm{c}_{i_{e}, s_{e}}^{0}, \bm{c}_{j_e, t_e}^{0}) \}_{e=1}^{M}$,
given that the constituents of the liquid at the instant of cross-linking are at
$\{ \bm{c}_{i,s}^0 \}$.  It is given by~\cite{xing-DE}
\ba
\label{eq:generalized_DE}
P(\chi | \{ \bm{c}_{i,s}^0 \} ) &=& \frac{1}{M!} \left( \frac{\V\eta^2}{2N} \right)^M
e^{H_{\rm{norm}}( \{ \bm{c}_{i,s}^0 \} ) / T^0 }
\nonumber\\
&&\times \prod_{e=1}^{M} e^{-|\bm{c}_{i_e,s_e}^0 - \bm{c}_{j_e,t_e}^0|^2/2b^{2}}.
\ea
Here, $\V$ is the dimensionless volume of the system, i.e., $V/(2\pi b^2)^{D/2}$.
The term
$H_{\rm{norm}}(\bm{c}^0)$ arises from the requirement that
$P(\chi | \{ \bm{c}_{i,s}^0 \} )$
be properly normalized over $\chi$, i.e.,
$\sum_{\chi} P(\chi | \{ \bm{c}_{i,s}^0 \} ) = 1$;
it is given by
\be
\label{eq:H_norm}
H_{\rm{norm}}( \{ \bm{c}_{i,s}^0 \} )
= - T^0
\frac{\V\eta^2}{2N} \sum_{i,j=1}^{N}\sum_{s,t=-1}^{1}
e^{-|\bm{c}_{i,s}^0-\bm{c}_{j,t}^0|^2/2b^{2}}.
\ee
The probability $P(\chi | \{ \bm{c}_{i,s}^0 \} )$ features a dimensionless parameter $\eta$ that controls the likelihood that cross-links are actually formed.  In App.~\ref{app:a} we show that it is related to the average number of cross-linking springs per dimer via the formula $[M]/N \approx 2\eta^2$.

\bw
\subsection{Disorder averaging the free energy}

Now that we have constructed a suitable disorder distribution $P_{\chi}$, we use it to perform the disorder average of the logarithm of the partition function $[\ln Z_\chi]$, doing this indirectly, using the replica technique (see, e.g., Ref.~\cite{spin-glass}).  Thus, we represent the logarithm in Eq.~(\ref{eq:disorderaverageF}) as a limit to obtain
\be
\label{eq:replica_trick}
[F] = - T \lim_{n\rightarrow 0} \frac{[Z_{\chi}^n]-1}{n},
\ee
where we have interchanged the order of taking the replica limit and performing the disorder average on going from Eq.~(\ref{eq:disorderaverageF}) to Eq.~(\ref{eq:replica_trick}).  We then insert the Deam-Edwards distribution~(\ref{eq:DE}) to obtain
\ba
\label{eq:replic}
[Z_\chi^n] &=&
\sum_{M=0}^{\infty}
% \sum_{i_1,j_1=1\atop{(i_1 \neq j_1)}}^{N}
  \sum_{i_1,j_1=1}^{N}
\ldots
% \sum_{i_M,j_M=1\atop{(i_M \neq j_M)}}^{N}
  \sum_{i_M,j_M=1}^{N}
\sum_{s_1,t_1=-1}^{1}
\ldots
\sum_{s_M,t_M=-1}^{1}
\frac{1}{M!}
\left( \frac{\V \eta^2}{2N} \right)^{M}
\frac{1}{Z_{\rm{liq}}^0} e^{ H_{\rm{norm}}( \{ \bm{c}_{i,s}^0 \} )/T^0 }
\nonumber\\
&&\qquad\times
\prod_{\alpha=0}^{n} \int \mathcal{D} \bm{c}^\alpha \,
e^{ - \sum_{\alpha=0}^{n} \left( H_{\rm{nem}}(\{\bm{n}_{i}^{\alpha}\})
    + H_{\rm{ev}}(\{ \bm{c}_{i,s}^{\alpha} \}) \right)/T^0}
\prod_{e=1}^{M}
e^{-\sum_{\alpha=0}^{n} |\bm{c}_{i_e,s_e}^{\alpha} - \bm{c}_{j_e,t_e}^{\alpha}|^{2}/2b^2}
\nonumber\\
&&\qquad\quad\times
\prod_{\alpha=0}^{n} \prod_{i=1}^{N} \delta(|\bm{c}_{i,1}^{\alpha} - \bm{c}_{i,-1}^{\alpha}| - \ell).
\ea
\ew
Here, $\alpha=0,1,\ldots, n$ labels the replicas, and $Z_{\rm{liq}}^0$ is defined via
\ba
Z_{\rm{liq}}^0 &\equiv& \int \mathcal{D} \bm{c}^0 \,
e^{- \left( H_{\rm{nem}}( \{ \bm{n}_{i}^0 \} )-H_{\rm{ev}}( \{ \bm{c}_{i,s}^0 \} ) \right)/T^0}
\nonumber\\
&&\qquad\quad\times
\prod_{i=1}^{N}\delta(|\bm{c}_{i,1}^{0} - \bm{c}_{i,-1}^{0}|-\ell).
\ea
Note that in Eq.~(\ref{eq:replic}) the term $H_{\rm{norm}}( \{ \bm{c}_{i,s}^0 \} )$ limits permutation invariance to replicas $\alpha=1, \ldots, n$, with replica $\alpha=0$ being excluded.
This is consistent with our physical expectation that the preparation and measurement ensembles play distinct roles.

% ** this section needs streamlining

Next, we complete the computation of the disorder average in Eq.~(\ref{eq:replic}), which yields $[Z_\chi^n]$ in terms of the replica partition function $Z_{1+n}$, i.e.,
\ba
[Z_\chi^n]&=&
\frac{Z_{1+n}}{Z_{{\rm{liq}}}^{0}},\\
Z_{1+n}
&\equiv&
\int \left(
\prod_{\alpha=0}^n \prod_{i=1}^{N} d{\bm{c}}_{i,1}^\alpha \, d{\bm{c}}_{i,-1}^\alpha
\delta(|{\bm{c}}_{i,1}^\alpha-{\bm{c}}_{i,-1}^\alpha|-\ell)
\right)
e^{-H_{{\rm{rep}}}},
\nonumber\\
\label{eq:pre_HS}
\ea
in which $\int \, d\bm{c}\equiv \prod_{d=1}^{D} \int dc_{d}$,
where the effective replica Hamiltonian $H_{{\rm{rep}}}$ is given by
\bw
\ba
\label{eq:pre_HS1}
H_{{\rm{rep}}}
&=&
-\frac{1}{2}\sum_{\alpha=0}^n \sum_{i,j=1}^{N} \frac{J_{ij}}{T^\alpha} \big( ({\bm{n}}_i^\alpha\cdot {\bm{n}}_j^\alpha)^2 - \frac{1}{D} \big)
+\frac{1}{2}\sum_{\alpha=0}^n \sum_{i,j=1}^{N}\sum_{s,t=-1}^{1} \frac{\lambda}{T^\alpha}\delta({\bm{c}}_{i,s}^\alpha-{\bm{c}}_{j,t}^\alpha)
\nonumber\\
&&\qquad-
\frac{V\eta^2}{2N(2\pi b^2)^{D/2}}
    \sum_{i,j=1}^{N}\sum_{s,t=-1}^{1}
    \Big(
    e^{-\frac{1}{2b^2}\sum_{\alpha=0}^n |{\bm{c}}_{i,s}^\alpha-{\bm{c}}_{j,t}^\alpha|^2}
    -
    e^{-\frac{1}{2b^2}|\bm{c}_{i,s}^0 - \bm{c}_{j,t}^0|^2}
    \Big).
\ea
We note that in $H_{{\rm{rep}}}$ the replicas are coupled as a result of the disorder averaging.
See App.~\ref{app:b} for details of the derivation of Eqs.~(\ref{eq:pre_HS},\ref{eq:pre_HS1}).
\ew

\subsection{Collective fields and their physical meaning}

We note that in $H_{{\rm{rep}}}$, Eq.~(\ref{eq:pre_HS1}), the $N$ interacting (replicated) dimers are coupled with one another, which means that the trace over microscopic variables $\bm{c}_{i,s}$ in $Z_{1+n}$ [Eq.~(\ref{eq:pre_HS})] cannot be straightforwardly carried out.  We decouple these dimers by defining microscopic collective fields $\omega_{\hat{k}}$ and $\bm{q}_\pv$, and making a Hubbard-Stratonovich transformation involving fluctuating auxiliary fields conjugate to the collective fields.  Thus, we arrive at a description in terms of $N$ uncoupled copies of a single replicated dimer, for which the trace over $\bm{c}_{i,s}$ can be readily performed, order by order in an expansion in the auxiliary fields.  The collective fields are given by
\begin{subequations}
\ba
\label{eq:gamma}
\omega_{\hat{k}}&\equiv&
\frac{1}{2N}\sum_{i=1}^{N}\sum_{s=-1,1}
\prod_{\alpha=0}^n
e^{-i\hat{k}\cdot \hat{c}_{i,s}},\\
\label{eq:q}
q_{d_1 d_2}(\pv)
&\equiv&
\frac{1}{N}\sum_{i=1}^N
e^{i\pv\cdot\bm{c}_i}
(n_{i\,d_1} n_{i\,d_2} - D^{-1}\delta_{d_1 d_2}).
\ea
\end{subequations}
Here, we have used the symbol $\hat{k}$ to denote the $n+1$-fold replicated wave-vector
$(\bm{k}^{0}, \bm{k}^{1},\ldots,\bm{k}^{n})$,
and have restricted the value of $\hat{k}$ in $\omega_{\hat{k}}$ to the \lq\lq higher-replica sector\rq\rq\ (HRS), viz., the set of replicated vectors each having at least two non-zero vector entries.  Details of the Hubbard-Stratonovich transformation are given in App.~\ref{app:b}.  The result is that the replica partition function $Z_{1+n}$ becomes a functional integral over the auxiliary fields $\Omega_{\hat{k}}$ and $\nop_{\pv}$, whose expectation values are related to the expectation values of $\omega_{\hat{k}}$ and $\bm{q}_{\pv}$ via
\begin{subequations}
\label{eq:HS}
\ba
\langle \Omega_{\hat{k}}\rangle_f
&=&
\left[\big\langle
\omega_{\hat{k}}
\big\rangle_\chi\right];
\label{eq:Omega}
\\
\langle \nop_\pv^\alpha \rangle_f
&=&
\left[\big\langle
\bm{q}_\pv^\alpha
\big\rangle_\chi\right],
\ea
\end{subequations}
as we demonstrate in App.~\ref{app:c}.  Note that
$\langle \cdots \rangle_f$
indicate an expectation value taken with respect to the Landau-Wilson free energy $f_{1+n}$ discussed in Sec.~\ref{sec:landau-wilson} and defined in Eq.~(\ref{eq:landauwilsonf}), and
$\langle \cdots \rangle_\chi$
denotes an expectation value taken with respect to the Hamiltonian $H_\chi$, Eq.~(\ref{eq:dimer_hamiltonian}).

\bw
To gain some intuition for the physical significance of
$\left[\langle \Omega \rangle_f\right]$ and
$\left[\langle \nop^\alpha \rangle_f\right]$,
we transform Eqs.~(\ref{eq:HS}) to replicated real space:
\begin{subequations}
\ba
\left[\langle \omega(\hat{r})\rangle_\chi \right]
&=&
\left[
\left\langle
\frac{1}{2N}\sum_{i=1}^{N}\sum_{s=-1}^{1}
\prod_{\alpha=0}^n
\delta\big(
\rv^{\alpha} - \bm{c}_{i,s}^{\alpha}
\big)
\right\rangle
\right]
- \frac{1}{V^{1+n}}
=\left[\langle \Omega(\hat{r})\rangle_f \right];
\\
\label{eq:q}
\left[ \langle q_{d_1 d_2}(\rv)\rangle_\chi \right]
&=&
\left[
\left\langle
\frac{1}{N}\sum_{i=1}^N
(n_{i\,d_1} n_{i\,d_2} - D^{-1}\delta_{d_1 d_2})
\thinspace
\delta(\bm{r} - \bm{c}_i)
\right\rangle
\right]
=
\left[ \langle \nopc_{d_1 d_2}(\rv)\rangle_f \right].
\ea
\end{subequations}
\ew
Thus, except for a trivial constant, we can interpret $\langle \Omega(\hat{r}) \rangle_f$ as the joint probability  that a given dimer end is found at position $\rv^0$ at the instant of cross-linking, and that the same dimer end would be found at $n$ subsequent widely-separated time instants at the positions $\rv^1, \ldots, \rv^n$~\cite{XPMGZ-NE-VT}.
$\langle \Omega(\hat{r}) \rangle_f$ vanishes if all dimers are delocalized, and has a nonzero value if a fraction of them are localized.  Thus,
$\langle \Omega(\hat{r}) \rangle_f$
serves as the order parameter that detects the phase transition from the liquid state to the random solid state.  Similarly,
$\langle \nop^0 (\rv) \rangle_f$
is the nematic order parameter for the preparation state, whilst
$\langle \nop^\alpha (\rv) \rangle_f$ (for $\alpha=1,\ldots, N$)
is the nematic order parameter for the measurement state.

We note that when defining $\Omega$ we have chosen to exclude the components associated with the {\it one-replica sector\/} (denoted \lq\lq 1RS\rq\rq ), i.e., the set of replicated wave-vectors that each have only one non-zero vector entry [i.e., $\hat{k}=(\bm{0},\ldots,\bm{k}^\alpha,\ldots,\bm{0})$].  This is because this sector of the field corresponds to fluctuations in the macroscopic density of dimers, and is strongly stabilized as a result of the excluded volume interactions, regardless of the extent of the cross-linking.  The physical content of the decomposition of fields into higher and lower replica- sectors is that if condensation occurs in the higher sector only, this implies the random localization of particles.  On the other hand, if condensation occurs in the 1RS as well, this indicates the formation of a state having some kind of spatially modulated density structure.  We are focusing on highly incompressible systems, for which density fluctuations are negligibly small.  Incompressibility is enforced by taking $\lambda$ to have a large value, so that fluctuations of 1RS counterpart to $\Omega$ are strongly suppressed.
% [** If the following needs to be said there must be a better place for it? A footnote?
We also define the {\it zero-replica sector\/} (denoted \lq\lq 0RS\rq\rq ) to be the set whose only member is the replicated wave-vector that has zero for every entry.
The {\it lower-replica sector\/} (denoted \lq\lq LRS\rq\rq ) would then refer to the union of the one-replica sector and the zero-replica sector.

\subsection{Intermezzo on replicas}

Compared with the more familiar example of spin glasses (see, e.g., Ref.~\cite{spin-glass}), the effective replica Hamiltonian~(\ref{eq:pre_HS1}) in our theory contains an extra replica.  What meaning can we ascribe to this \lq\lq zeroth\rq\rq\ replica as well as to the other $n$ replicas?  Physically, the zeroth replica, which originates in the Deam-Edwards distribution, corresponds to the state of the system at the instant of preparation, whilst the other $n$ replicas correspond to the state of the system when it is measured.  The dependence of the measured properties on the state of the system at preparation corresponds, operationally, to the coupling in the effective replica Hamiltonian (\ref{eq:pre_HS1}) between the freedoms belonging to the zeroth replica and those belonging to the other $n$ replicas.  On the other hand, such a coupling does not imply that the measured properties influence the preparation state, as one can show via a careful consideration of the $n\rightarrow 0$ limit~\cite{xing-DE}.

Because the replica framework accounts for both the preparation and measurement ensembles vis-{\`a}-vis the zeroth and other $n$ replicas, it is well adapted to investigations of nematic elastomers, whose measured properties are known to depend on their preparation histories; see, e.g., Ref.~\cite{urayamaPMtransition}.  Such a dependence was already understood by Deam and Edwards and by de~Gennes more than thirty years ago~(see, e.g., Refs.~\cite{Deam-Edwards-1976,polymer:deGennes}).

\section{Landau-Wilson free energy}
\label{sec:landau-wilson}
As we demonstrate in App.~\ref{app:b}, the local incompressibility of IGNEs allows us to express the effective replica theory in terms of the auxiliary fields $\Omega$ and $\nop^\alpha$ as
\be
Z_{1+n} \propto \int \mathcal{D}\Omega_{\hat{k}}\prod_{\alpha=0}^n \mathcal{D}Q_{\bm{p}}^\alpha
         \exp\big(-Nf_{1+n} (\Omega, \nop^\alpha) \big),
\ee
where the Landau-Wilson free energy per dimer $f_{1+n}$ is given by
\ba
\label{eq:landauwilsonf}
f_{1+n}(\Omega, \nop)
&=& \frac{{\tilde{\eta}^2}}{2 V^n}\overline{\sum_{\hat{k}}}
    \vertex_{\hat{k}}
    |\Omega_{\hat{k}}|^2
  + \frac{1}{2}\sum_{\alpha=0}^n \sum_{\bm{p}}
    \frac{J_\pv}{T^\alpha}
    \{ \nop_{\pv}^\alpha\, \nop_{-\pv}^\alpha \}
  \nonumber\\
  &&\quad - \ln\langle \exp \big( G_1(\Omega) + G_2(\nopc) \big) \rangle_{1,1+n}.
\ea
We have introduced the notation
\ba
G_1(\Omega) &\equiv&
\frac{{\tilde{\eta}}^2}{2 V^n} {\overline{\sum_{\hat{k}}}}
\vertex_{\hat{k}}\,\Omega_{\hat{k}}
\sum_{s=-1}^{1} e^{-i\hat{k}\cdot {\hat{c}}_{s}},
\\
G_2(\nopc) &\equiv&
\sum_{\alpha=0}^n \sum_{\bm{p}} \frac{J_\pv}{T^\alpha}\,
Q_{d_1 d_2}^\alpha(\bm{p})\,
e^{-i\bm{p}\cdot {\bm{c}}^\alpha}
\nonumber\\
&&\quad\times\left( n_{d_1}^\alpha n_{d_2}^\alpha - \frac{1}{3}\delta_{d_1 d_2} \right),
\ea
and have specialized to three spatial dimensions (i.e., $D=3$).  The curly braces $\{{\mathbf S}\,{{\mathbf S}^{\prime}}\}$ in Eq.~(\ref{eq:landauwilsonf}) denote the trace of the product of the tensors ${\mathbf S}$ and ${{\mathbf S}^{\prime}}$, i.e., $\sum_{d_1,d_2=1}^{D} S_{d_1 d_2} S_{d_2 d_1}^{\prime}$.
The symbols ${\tilde{\eta}}^2$, $\vertex_{k}$ and $\langle \cdots \rangle_{1,1+n}$ are defined via
\begin{subequations}
\ba
{\tilde{\eta}}^2 &\equiv& 4\eta^2,
\\
\vertex_{k} &\equiv& \exp(-b^2 k^2/2),
\\
\langle \cdots \rangle_{1,1+n}
&\equiv&\!\!\!
\prod_{\alpha=0}^{n} \int \! \frac{d\bm{c}_{1}^{\alpha}\, d\bm{c}_{-1}^{\alpha}}{4\pi V \ell^2} \,\delta(|\bm{c}_{1}^{\alpha} - \bm{c}_{-1}^{\alpha}| - \ell).
\ea
\end{subequations}
Moreover, ${\overline{\sum}}_{\hat{k}}$ denotes the sum over replicated wave-vectors that are restricted to the higher-replica sector.

\begin{widetext}
\subsection{Expanding the Landau-Wilson free energy}
\label{sec:new}
To develop the expansion of the Landau-Wilson free energy we expand the log-trace term in Eq.~(\ref{eq:landauwilsonf}) in powers of $\nop$ and $\Omega$ to obtain (see App.~\ref{app:d} for details):
\be
\label{eq:landauwilsone}
f_{1+n} (\Omega, \nop)
\approx
f_{\Omega}(\Omega) + f_{Q}(\nop) + f_{C}(\Omega, \nop).
\ee
%%%%%%%%%%%%%%%%%%%%%%%%%%
The first term on the right hand side, $f_{\Omega}(\Omega)$, describes the vulcanization/random solidification transition for an isotropic  elastomer~\cite{Vulcan_Goldbart,peng,xiaoming} and is given by
\ba
f_{\Omega}(\Omega)
&=&
\frac{{\tilde{\eta}}^2}{2} \overline{\sum\limits_{\hat{k}}}
\Big( \big( 1 - {\tilde{\eta}}^2 \big) + \frac{1}{2}\big( b^2 + \frac{\ell^2}{6} \big) |\hat{k}|^2 \Big)
|\Omega_{\hat{k}}|^2
%\nonumber \\ &&
-\frac{{\tilde{\eta}}^6}{6} {\overline{\sum}}_{\hat{k}_1,\hat{k}_2,\hat{k}_3}\delta_{\hat{k}_1+\hat{k}_2+\hat{k}_3,\hat{0}}
\,
\Omega_{\hat{k}_1}\Omega_{\hat{k}_2}\Omega_{\hat{k}_3}.
\label{eq:vulc}
\ea
It exhibits a linear instability at the critical value ${\tilde{\eta}}_c^2=1$,~\cite{instability-footnote} which reflects the destabilization of the liquid state with respect to a gel/random solid state when the average number of cross-links per dimer is increased beyond a certain critical value.  As IGNEs are random solids, they must have undergone a vulcanization transition (controlled by the density of cross-links) from the liquid state to the random solid state.  Thus, we keep the $\Omega$-only terms to cubic order; this is in line with the approach adopted in Ref.~\cite{xiaoming}.
%%%%%%%%%%%%%%%%%%
The second term, $f_{Q}$, is given by
\ba
\label{eq:f_Q_Full}
f_{Q}(\nop)
&=&\sum_{\alpha=0}^n \sum_{\bm{p}}
\frac{1}{2T^\alpha}
\mass(T^{\alpha}, \pv)
\{ \nop_{\pv}^\alpha \, \nop_{-\pv}^\alpha \},
\ea
These terms describe the free-energy cost of inducing nematic alignment from the unaligned state.  As the focus of the present Paper is on the liquid-crystalline properties of IGNEs at high preparation and measurement temperatures, we retain the $\nop$-only terms to quadratic order.  This collection of terms is an $(1+n)$-fold replicated version of the (quadratic part of the) Landau-de Gennes free energy for a liquid of nematogens at high temperatures~\cite{prost}.  In Eq.~(\ref{eq:f_Q_Full}),
$T^{0}$ is the {\it preparation temperature\/}
(i.e., the temperature at which the system was cross-linked), and
$T^{\alpha}$ (for $\alpha= 1,\ldots,n$) is the {\it measurement temperature\/} (i.e.,  the temperature at which the system is measured, long after cross-linking).  As replicas $\alpha=1,\ldots,n$ are actually copies of a single system measured at one temperature, we have $T^\alpha = T$ for $\alpha=1,\ldots,n$.  The kernel $\mass(T^{\alpha}, \pv)$ is given by
\be
\mass(T^{\alpha},\pv)
\equiv
J_0 e^{-p^2 a^2/2}
\left( 1 - \frac{2J_0}{15T^\alpha}e^{-p^2 a^2/2} \right),
\ee
which, expanded in powers of wave-vectors, yields~\cite{ref:massfunction}
\be
\label{eq:massfunctionexpansion}
\mass(T^{\alpha},\pv)\approx \LDLa^{\alpha} t^{\alpha} + \stiff^{\alpha} p^2,
\ee
where $t^0$ is the reduced preparation temperature $(T^0-T^{\ast})/T^{\ast}$
(where $T^{\ast}$ has the value $4 J_0/15$), and
$t^\alpha$ (for $\alpha=1,\ldots,n$) is reduced measurement temperature, $(T-T^{\ast})/T^{\ast}$.
For simplicity, we write $t^\alpha = t$ for $\alpha=1, \ldots, n$.
%%%%%%%%%%%%%%%%%%%%%%%%%%%%%
The parameter $\LDLa^{\alpha}$ ($\equiv J_0 T^{\ast}/T^{\alpha}$) characterizes the aligning tendencies of nematogens, and $\stiff^{\alpha}$ [$\equiv (2T^{\ast}-T^{\alpha})a^2\,J_0/2T^{\alpha}$] is the generalized nematic stiffness in the (Landau-de~Gennes equivalent of the) one-Frank-constant approximation~\cite{prost}.
Hence, the nematic free energy $f_Q$ becomes
\ba
f_{Q}(\nop)
\approx
\frac{1}{2T^0}\sum_{\pv}\mass(T^0, \pv)\{ \nop_{\pv}^{0} \, \nop_{-\pv}^{0}\}+
\sum_{\alpha=1}^n \sum_{\bm{p}}
\frac{1}{2T}
( \LDLa t + \stiff p^2 )
\{ \nop_{\pv}^\alpha \, \nop_{-\pv}^\alpha \},
\ea
where $\LDLa$ and $\stiff$ are, respectively, $\LDLa^\alpha$ and $\stiff^\alpha$ for $\alpha=1,\ldots,n$.
%%%%%%%%%%%%%%%
The third term of the right hand side of Eq.~(\ref{eq:landauwilsone}), $f_{C}$, describes the coupling between the nematic freedoms and the elastomer and is given by
\ba
f_C(\Omega, \nop) &=&
\sum_{\alpha\neq\beta}^n \sum_{\bm{p},\bm{q}}
\Phi_{d_1 d_2 d_3 d_4}^{\alpha\beta}(\pv,\qv)
\,
 \Omega_{-{\bm{p}}{\hat{\epsilon}}^\alpha-{\bm{q}}{\hat{\epsilon}}^\beta}\,
 Q_{d_1 d_2}^\alpha(\bm{p}) \, Q_{d_3 d_4}^\beta(\bm{q})
 \nonumber\\
&&+
\sum_{\alpha=0}^n \sum_{\bm{p}}\overline{\sum_{\hat{k}}}
\Psi_{d_1 d_2}^{\alpha}(\pv,\qv,\hat{k}) \,
 \Omega_{\hat{k}} \, \Omega_{-\hat{k}-{\bm{p}}{\hat{\epsilon}}^\alpha}
\, Q_{d_1 d_2}^\alpha(\bm{p})
\nonumber\\
&&+
\sum_{\alpha,\beta=0\atop{(\alpha\neq\beta)}}^{n} \sum_{\bm{p},\bm{q}}\overline{\sum_{\hat{k}}}
\Gamma_{d_1 d_2 d_3 d_4}^{\alpha \beta}(\pv,\qv,\hat{k}) \,
 \Omega_{\hat{k}}\Omega_{-\hat{k}-{\bm{p}}{\hat{\epsilon}}^\alpha-{\bm{q}}{\hat{\epsilon}}^\beta}
\, Q_{d_1 d_2}^\alpha(\bm{p}) \, Q_{d_3 d_4}^\beta(\bm{q})
  \nonumber\\
&&+\sum_{\alpha=0}^{n}
    \sum_{\bm{p},\bm{q}}
    \overline{\sum\limits_{\hat{k}}}
    %\sum_{\hat{k}\in HRS}
     \left(
     \Upsilon_{d_1 d_2 d_3 d_4}^{\alpha}(\pv,\hat{k})\,\delta_{\pv+\qv,\bm{0}}
    +\Theta_{d_1 d_2 d_3 d_4}^{\alpha}(\pv,\qv,\hat{k})
     \right)
     \nonumber\\
     &&\quad\times \,
    \Omega_{\hat{k}} \, \Omega_{-\hat{k}-(\bm{p}+\bm{q}){\hat{\epsilon}}^\alpha}
    \, Q_{d_1 d_2}^\alpha(\bm{p}) \, Q_{d_3 d_4}^\alpha(\bm{q}),
\ea
where the wave-vector-dependent coefficients $\Phi$, $\Psi$, $\Theta$, $\Upsilon$ and $\Gamma$ are defined as
\begin{subequations}
\ba
\Phi_{d_1 d_2 d_3 d_4}^{\alpha\beta}(\pv,\qv)
&\equiv&
 - \vertex_{-{\bm{p}}{\hat{\epsilon}}^\alpha-{\bm{q}}{\hat{\epsilon}}^\beta}
 \frac{{\tilde{\eta}}^{2} J_\pv\, J_\qv}{1600 T^\alpha\, T^\beta}
 p_{d_1} p_{d_2} q_{d_3} q_{d_4};
\\
\Psi_{d_1 d_2}^{\alpha}(\pv,\qv,\hat{k}) &\equiv& -\vertex_{\hat{k}} \vertex_{-\hat{k}-{\bm{p}}{\hat{\epsilon}}^\alpha}
 \frac{{\tilde{\eta}}^4 \ell^2\, J_\pv}{10 T^\alpha}
% \nonumber\\
% &&\quad\times
 \left( p_{d_1} p_{d_2} + \left(k^\alpha + \frac{1}{2}p \right)_{d_1} \left(k^\alpha + \frac{1}{2}p \right)_{d_2} \right)
 \nonumber\\
\\
 \Upsilon_{d_1 d_2 d_3 d_4}^{\alpha}(\pv,\hat{k})
 &\equiv&
 -\vertex_{\hat{k}}^2 \frac{2 {\tilde{\eta}}^4}{15}
 \left( \frac{J_\pv}{T^\alpha} \right)^2
 \,\delta_{d_1 d_3}\delta_{d_2 d_4};
\\
\Gamma_{d_1 d_2 d_3 d_4}^{\alpha \beta}(\pv,\qv,\hat{k})
&\equiv&
 \frac{{\tilde{\eta}}^4 \ell^4\,J_\pv\, J_\qv}{450 T^\alpha\, T^\beta}
 \vertex_{\hat{k}}\vertex_{-\hat{k}-{\bm{p}}{\hat{\epsilon}}^\alpha-{\bm{q}}{\hat{\epsilon}}^\beta}
 \nonumber\\
 &&\times
\left(
k_{d_1}^\alpha k_{d_2}^\alpha k_{d_3}^\beta k_{d_4}^\beta
+ \frac{1}{4} \left(p_{d_1} k_{d_2}^\alpha + k_{d_1}^\alpha p_{d_2}\right)
  \left(q_{d_3} k_{d_4}^\beta + k_{d_3}^\beta q_{d_4}\right)\right.
\nonumber\\
&&\quad+\left.
\frac{1}{8} p_{d_1} p_{d_2} q_{d_3} q_{d_4}
\right);
\ea
% \nonumber\\
\ba
 \Theta_{d_1 d_2 d_3 d_4}^{\alpha}(\pv,\qv,\hat{k})
 &\equiv&
 -\frac{{\tilde{\eta}}^4\,J_\pv\, J_\qv}{30(T^\alpha)^2}
 \vertex_{\hat{k}}\vertex_{-\hat{k}-(\bm{p}+\bm{q}){\hat{\epsilon}}^\alpha}
 \nonumber\\
&&\times
\left( \left( 2 - \frac{\ell^2 |\pv + \qv|^2}{56} + \frac{\ell^4 |\pv + \qv|^4}{8064}
- \frac{\ell^2 |\pv + \qv + 4\bm{k}^\alpha|^2}{56}\right.\right.
\nonumber\\
&&\quad
+
\left.\frac{\ell^4 |\pv + \qv + 4\bm{k}^\alpha|^4}{8064} \right)
\delta_{d_1 d_3}\delta_{d_2 d_4}
\nonumber\\
&&
+\left(- \frac{1}{14} + \frac{\ell^2 |\pv+\qv|^2}{1008} \right)
\ell^2 (p+q)_{d_1} (p+q)_{d_3} \delta_{d_2 d_4}
\nonumber\\
&&
+\left(- \frac{1}{14} + \frac{\ell^2 |\pv+\qv + 4\bm{k}^\alpha |^2}{1008} \right)
\nonumber\\
&&\quad\times
\ell^2 (p+q+4k^\alpha)_{d_1} (p+q+4k^\alpha)_{d_3} \delta_{d_2 d_4}
\nonumber\\
&&+\frac{\ell^4}{2016}(p+q)_{d_1} (p+q)_{d_2}(p+q)_{d_3} (p+q)_{d_4}
\nonumber\\
&&
+
\frac{\ell^4}{2016}(p+q+4k^\alpha)_{d_1} (p+q+4k^\alpha)_{d_2}
\nonumber\\
&&\quad
\left.
\times
(p+q+4k^\alpha)_{d_3} (p+q+4k^\alpha)_{d_4}
\right).
\ea
\end{subequations}
Here, we have retained the lowest-order terms that couple $\Omega$ and $\nop$, i.e., those proportional to
$\nop\, \nop\, \Omega$ and $\nop\, \Omega\, \Omega$.  We have, moreover, kept only the lowest-order wave-vector dependencies of their coefficients.  We have, in addition, retained the term proportional to $\nop\, \nop\, \Omega\, \Omega$ because we have found that it is the leading one responsible for the qualitatively new possibility of thermal and glassy correlators that not only decay but also oscillate with distance.  The physical significance of $f_{C}$ is that it encodes into the theory the fact that the nematic degrees of freedom inhabit an environment that is at the microscopic level anisotropic, inhomogeneous and thermally fluctuating, but that no remnants of this anisotropy or inhomogeneity survive to the macroscopic level.
$f_{C}$ thus also describes the mutual effects of random localization and nematic alignment.  In particular, it enables us to study the impact of the memorization of the nematic fluctuations in the preparation state on the nematic fluctuations in the measurement state~\cite{phenomenology}.

\end{widetext}

\section{Saddle-point approximation}
\label{sec:stationarity}

Next, we consider the saddle-point equations for $\nop$ and $\Omega$, which follow from the stationarity of Eq.~(\ref{eq:landauwilsone}):
\begin{subequations}
\label{eq:stationarityy}
\ba
&&\left.\frac{\delta f_{1+n}}{\delta\Omega}\right\vert_{{\bar{\Omega}},{\bar{Q}}}=0;
\label{eq:stationarityOmegaA}
\\
&&\left.\left(
\frac{\delta f_{1+n}}{\delta\nopc_{d_1 d_2}^\alpha}
- \frac{1}{3}\delta_{d_1 d_2}{\mathrm{Tr}}\frac{\delta f_{1+n}}{\delta\nop^\alpha}
\right)
\right\vert_{{\bar{\Omega}},{\bar{Q}}}=0.
\label{eq:stationarityQA}
\ea
\end{subequations}
%where in Eq.~(\ref{eq:stationarityQA}), only the traceless, symmetric part is to be considered.
As we are concerned with macroscopically isotropic states, we require that the saddle-point value of $\nop$  vanishes: $\nopsp = \mathbf{0}$.  As a result, the saddle-point equation~(\ref{eq:stationarityOmegaA}) for ${\bar{\Omega}}$ is the same as it would be for ordinary (i.e., non-nematogenic) elastomers, and reads
\ba
\label{eq:Omega_sp}
\Big(
1-{\tilde{\eta}}^2
+ \frac{1}{2}
\ell_r^{\prime 2}
|\hat{k}|^2 \Big)
{\Omega}_{\hat{k}}
-
\frac{{\tilde{\eta}}^{4}}{2}
\overline{\sum\limits_{\hat{k}^{\prime}}}
{\Omega}_{\hat{k}^{\prime}}
{\Omega}_{\hat{k}-\hat{k}^{\prime}}
=0,
\ea
in which we have defined the length-scale $\ell_r^\prime$ via $\ell_r^{\prime 2}\equiv b^2 + (\ell^2/6)$.
Next, by defining the quantities $\epsilon\equiv 3({\tilde\eta}^2-1)/{\tilde\eta}^4$ and $\ell_r\equiv \sqrt{3}\ell_r^{\prime}/{\tilde\eta}^2$,
the saddle-point equation takes the form analyzed in Ref~\cite{peng}:
\ba
\label{eq:Omega_sp}
\Big(
-\epsilon
+ \frac{1}{2}
\ell_r^2
|\hat{k}|^2 \Big)
{\Omega}_{\hat{k}}
-
\frac{3}{2}
\overline{\sum\limits_{\hat{k}^{\prime}}}
{\Omega}_{\hat{k}^{\prime}}
{\Omega}_{\hat{k}-\hat{k}^{\prime}}
=0,
\ea
% [** check this def]
and is therefore solved by mean of the the Ansatz
\be
\label{eq:peng_ansatz}
{\bar{\Omega}}_{\hat{k}} = G \int\frac{d\bm{z}}{V} \, \int d\tau\, P(\tau)\, e^{i\sum_{\alpha} \bm{k}^{\alpha}\cdot \bm{z}
- |\hat{k}|^2 / 2\tau }
\ee
(with $\hat{k}\in\mathrm{HRS}$), provided that the gel fraction obeys
\be
\label{eq:gelfraction}
G = 2\epsilon / 3,
\ee
and  $P(\tau)$, which has the meaning of the distribution of inverse square localization lengths $\tau$, obeys
\be
\frac{\tau^2}{2}\frac{dP(\tau)}{d\tau} =
\left(\frac{\epsilon}{2} - \tau\right)P(\tau)
- \frac{\epsilon}{2}\int_0^\tau d\tau_1 \, P(\tau_1)\, P(\tau-\tau_1).
\ee
The typical value $\xi_{L}$ of the localization length is $\epsilon^{-1/2}\ell_r$, which diverges as the vulcanization transition is approached from the solid side.
The saddle-point equation for $\nop^\alpha$, viz., Eq.~(\ref{eq:stationarityQA}), yields
\ba
&&\overline{\sum_{\hat{k}}}{\bar\Omega}_{\hat{k}}{\bar\Omega}_{-\hat{k}-\pv\epsilon^\alpha}
\Big( \big( p_{d_1}\, p_{d_2} + (k^\alpha + p/2)_{d_1}(k^\alpha + p/2)_{d_2} \big)
\nonumber\\
&&\qquad- \frac{1}{D}\big( |\pv|^2 + |{\bm{k}}^{\alpha} + \pv/2| \big)\delta_{d_1 d_2} \Big)
=0.
\ea
This equation is automatically satisfied, which we can see as follows.  The product ${\bar\Omega}_{\hat{k}}{\bar\Omega}_{-\hat{k}-\pv\epsilon^\alpha}$ involves a product of two Kronecker deltas $\delta_{\sum_{\alpha=0}^{n}\bm{k}^\alpha,\bm{0}}\delta_{-\pv-\sum_{\alpha=0}^{n}\bm{k}^\alpha,\bm{0}}$, which reflects the macroscopic translational invariance of the random solid, and implies that $\pv=\bm{0}$.  Next, we note that the restricted wave-vector sum ${\overline{\sum}}_{\hat{k}}$ can be replaced by the full wave-vector sum $\sum_{\hat{k}}$, as the 1RS and 0RS contributions to the restricted sum vanish. Lastly, ${\bar\Omega}_{\hat{k}}{\bar\Omega}_{-\hat{k}}$ is a scalar, implying that
\be
\sum_{\hat{k}}{\bar\Omega}_{\hat{k}}{\bar\Omega}_{-\hat{k}}
\big( k_{d_1}^\alpha\, k_{d_2}^\alpha
- \frac{1}{3} |{\bm{k}}^\alpha|^2 \delta_{d_1 d_2}
\big)
=0.
\ee

For simplicity, instead of working with the full distribution $P(\tau)$, we replace it by one that is sharply peaked at $1/\xi_{L}^{2}$.  This approximation is valid as long as we are concerned with the broad implications for the nematogens of the very {\it presence\/} of localized network constituents.  We expect that the spread of localization lengths would at most result in the quantitative but not qualitative modification of our results.  If the localization is sharp, the Ansatz for the order parameter becomes
\be
\label{eq:barredOPmom}
{{\bar{\Omega}}}_{\hat{k}} \approx G\int\frac{d\bm{z}}{V} e^{i\sum_{\alpha} \bm{k}^{\alpha}\cdot \bm{z}
-\frac{1}{2}|\hat{k}|^2 \xi_{L}^{2}}.
\ee

%%%%%%%%%%%%%%%%%%%%%%%%%%%%%%%%%%%%%%%%%%%%%%%%%%%%%%%%%%%%%%%%%%%
\section{Effective theory of structure and correlations in IGNEs}
\label{sec:effectivetheory}

In App.~\ref{app:eff_theory}, we derive an effective replica Hamiltonian for the liquid crystalline behavior of IGNEs, using an approximation in which $\Omega$ is set to its saddle-point value $\bar{\Omega}$ in the Landau-Wilson free energy (\ref{eq:landauwilsone}), but allowing $\nop^\alpha$ to undergo fluctuations.  By making this approximation we are neglecting the impact of fluctuations in $\Omega$ on the nematic freedoms.  We adopt this level of description because it is the least complicated one that is capable of revealing the impact of the localized network on the liquid crystallinity characteristic of IGNEs.  As it remains constant, we do not need to consider the $f_{\Omega}$ contribution to the free energy.  Furthermore, at the present level of approximation, the term in $f_{C}$ proportional to $\bar{\Omega}\bar{\Omega}\nop^\alpha$ vanishes, as we show in App.~\ref{app:eff_theory}, as a result of the macroscopic translational and translational invariance of the random solid state.
Moreover, as we show in the same appendix, at wavelengths long compared with $\xi_{L}$ the dominant contribution to $f_{C}$ comes from a term in ${\bar\Omega}\,{\bar\Omega}\,\nop^\alpha\,\nop^\beta$.
Thus, in forming an the effective replica Hamiltonian for the high-temperature liquid crystallinity of IGNEs, which we denote by $H_{1+n}[\{\nop^{\alpha}(\cdot)\}_{\alpha=0}^{n}]$, we need only consider the $f_{Q}$ term together with a term in ${\bar\Omega}\,{\bar\Omega}\,\nop^\alpha\,\nop^\beta$ term:
\ba
\label{eq:H_eff0}
\!\!\!\!\!\!H_{1+n}[\{ \nop^\alpha \}_{\alpha=0}^{n}]
&\approx&
 \sum_{\bm{p}}
 \frac{1}{2T^0}(\LDLa^0 t^0 + \stiff^0 p^2)
 \{ \nop_{\pv}^{0} \, \nop_{-\pv}^{0} \}
\nonumber\\
&&\!\!\!\!\!\!\quad
+\sum_{\alpha=1}^n \sum_{\bm{p}}
 \frac{1}{2T}(\LDLa t + \stiff p^2)
 \{ \nop_{\pv}^{\alpha} \, \nop_{-\pv}^{\alpha} \}
\nonumber\\
&&\!\!\!\!\!\!\quad
- \frac{1}{T^0}\sum_{\alpha=1}^{n}\sum_{\bm{p}}
  \smear_{\pv} \{ \nop_{\pv}^{0} \, \nop_{-\pv}^{\alpha} \}
\nonumber\\
&&\!\!\!\!\!\!\quad
- \frac{1}{2T}\sum_{\alpha,\beta=1\atop{(\alpha\neq\beta)}}^{n}\sum_{\bm{p}}
  \smear_{\pv} \{ \nop_{\pv}^{\alpha} \, \nop_{-\pv}^{\beta} \}.
\ea
%In App.~\ref{app:h}, we show that the corrections generated by terms in $f_C$ that are sub-dominant to $\bm{k}^\alpha\bm{k}^\alpha\bm{k}^\beta\bm{k}^\beta\,{\bar\Omega}\,{\bar\Omega}\,\nop^\alpha\,\nop^\beta$ do not change the central qualitative prediction, viz., that thermal and glassy correlations undergo both spatial oscillations and decay in a certain regime.
Let us draw attention to the kernel $\smear_{\pv}$, which is defined via
\be
\label{eq:smear_definition}
\smear_\pv \equiv \smear_{\bf 0} e^{-\frac{1}{2}p^2\xi_L^2}.
\ee
Here, $\smear_{\bm{0}}$, which we call the \emph{disorder strength}, characterizes the strength with which the network influences the liquid crystallinity, and has the value $G^2 J_0^2 {\tilde{\eta}}^4 \ell^4 /(900 T \xi_L^4)$.
The kernel $\smear_\pv$ is the manifestation of the presence of the thermally fluctuating random elastomeric network and, in particular, encodes the central physical characteristic of the network, viz., that at long length-scales the localization appears perfect but at length-scales shorter than $\xi_{L}$ thermal  fluctuations render the network effectively \lq\lq molten\rlap.\rq\rq\thinspace\
Note that the characteristic length-scale beyond which $\smear(\rv)$ is suppressed is $\correllength_L$.
%
%This can be interpreted as arising from an unconventional, \lq\lq\,soft\,\rq\rq type of random field which goes to zero at short length-scales but assumes a non-vanishing value at long length-scales.
%Such an unconventional, soft random field is to be contrasted with the conventional, \lq\lq hard\rq\rq\ random field that is present at all length-scales.  The softness of the former reflects the fact that the elastomeric background that gives rise to quenched disorder is itself thermally fluctuating.  This is discussed in detail in Ref.~\cite{phenomenology}.

To determine the effect that the preparation history has on the equilibrium liquid crystalline properties post-cross-linking, we integrate out the zeroth-replica element, $\nop^0$, thus obtaining the effective Hamiltonian
%the reason being that the fluctuations of $\nop^0$ in the high-temperature isotropic \emph{liquid} state (such being the preparation ensemble) cannot be controlled:
%this is equivalent to averaging over prior unknowns
\ba
\label{eq:H_eff}
&&\!\!\!\!\!\!\!\!\!\!\!\!H_{{\rm eff}}\left[ \{ \nop^\alpha \}_{\alpha=1}^{n} \right]
\nonumber\\
&&\!\!\!\!\!\!\!\!\!\!\!\!
\equiv
\sum_{\alpha=1}^{n}\sum_{\bm{p}}
\left( \frac{\LDLa t + \stiff p^2}{2T}
- \frac{|\smear_{\pv}|^2}{2T^0(\LDLa^0 t^0 + \stiff^0 p^2)} \right)
\big\{\nop_{\pv}^{\alpha}\nop_{-\pv}^{\alpha}\big\}
 \nonumber\\
%&&\!\!\!\!\!\!\quad -\frac{\ldgthree}{3}\sum_{\alpha=1}^n\sum_{123}
% \nopc_{ab}^\alpha({\bm{p}}_1) \nopc_{bc}^\alpha({\bm{p}}_2) \nopc_{ca}^\alpha({\bm{p}}_3)
%\nonumber\\
%&&\!\!\!\!\!\!\quad
%+ \frac{\ldgfour}{4}
%  \sum_{\alpha=1}^n
%  \sum_{1234}
%  \nopc_{ab}^\alpha({\bm{p}}_1) \nopc_{ba}^\alpha({\bm{p}}_2) \nopc_{cd}^\alpha({\bm{p}}_3) \nopc_{dc}^\alpha({\bm{p}}_4)
%  \nonumber\\
&&\!\!\!\!\!\!\!\!\!\!\!\!\quad
- \sum_{\alpha,\beta=1\atop{(\alpha\neq\beta)}}^{n}\sum_{\bm{p}}
  \left( \frac{\smear_{\pv}}{2T}
  + \frac{|\smear_{\pv}|^2}{2T^0(\LDLa t^0 + \stiff^0 p^2)} \right)
  \big\{\nop_{\pv}^{\alpha}\nop_{-\pv}^{\beta}\big\}.
\ea
A noteworthy feature of $H_{\rm eff}$ is that the replica-diagonal contribution to the quadratic term in $\nop^\alpha$ is structurally distinct from the replica off-diagonal contribution.
%It is important to note that a conventional replica treatment of a phenomenological theory in which the nematic order is coupled linearly to a quenched Gaussian random field typically leads to a situation in which both the replica off-diagonal and replica-diagonal contributions to the quadratic term have the same kernel.
This structural feature, as well as the short length-scale liquidity encoded in $\smear_\pv$, enables us to capture a richer range of physical behavior (such as oscillatory-decaying nematic correlations) than can be predicted via conventional random-field approaches, for which the replica-diagonal and off-diagonal terms contain identical coefficients.

\section{Phenomenological content of the microscopic replica theory}
\label{sec:equivalence}
We now show that the effective replica Hamiltonian~(\ref{eq:H_eff}) that we have derived on the basis of the microscopic dimers-and-springs model can be interpreted as having arisen from a more phenomenological, continuum description of a liquid crystalline systems subject to a novel pair of interrelated random fields.  This phenomenological description and its implications were explore in Ref.~\cite{phenomenology}.  Our purpose here is not to revisit these issues in detail but rather to reveal the microscopic underpinnings of the phenomenological theory.  In that latter theory, the free energy of an IGNE having a given realization $\chi^\prime$ of the quenched disorder is given by
\begin{eqnarray}
\label{eq:landau}
H_{\chi^\prime}[\nop]&=&
\frac{1}{2}\sum_{\pv}
\Big(
\big(
\LDLa t+\stiff p^{2}+{\smear}_{\pv}
\big)
\big\{ \nop_{{\pv}}\nop_{-{\pv}}\big\}
\nonumber\\
&&
\qquad-
2\big\{
\big(\ncp_{{\pv}}+({T}/{T^0})\,{\smear}_{{\pv}}\,\nop_{{\pv}}^{0}\big)
\nop^{\phantom{0}}_{-{\pv}}
\big\}
\Big),
\nonumber\\
\end{eqnarray}
where $\chi^\prime = \{ \nop^0, \ncp \}$, and $\ncp$ and $\nop^{0}$ are independent, Gaussian-distributed random fields, with zero means and non-zero variances, the latter being given by
\begin{subequations}
\label{eq:disorderstats}
\begin{eqnarray}
\big[\big\{\nop^{0}_{{\pv}}\,\nop^{0}_{{\pv}^{\prime}}\big\}\big]
&=&
%T^0\mu_D\,\frac{\delta_{{\pv}+{\pv}^{\prime},{\bf 0}}}{\LDLa^{0}t_{p}+\stiff^0 k^{2}},
5T^0\,\frac{\delta_{{\pv}+{\pv}^{\prime},{\bf 0}}}{\LDLa^{0}t_{p}+\stiff^0 k^{2}},
\\
\big[\big\{\ncp_{{\pv}}\,\ncp_{{\pv}^{\prime}}\big\}\big]
&=&
T\,{\smear}_{\pv}\,
\delta_{{\pv}+{\pv}^{\prime},{\bf 0}}\,.
\end{eqnarray}
\end{subequations}
Here, $\nop^0$ describes the impact of the configuration of the nematogens that is present at the instant of cross-linking on the post-cross-linking nematic alignment pattern, and $\ncp$ accounts for the impact of the local anisotropic environment created by the localized network constituents post-cross-linking on this nematic alignment pattern. For a given realization of $\chi^\prime$, the partition function is given by
\be
Z_{\chi^\prime} \propto \int \mathcal{D}\nop \exp(-H_{\chi^\prime}[\nop]/T),
\ee
and by suitably averaging over $\chi^\prime$ using the replica technique we obtain
\ba
\label{eq:Z_chi_prime}
\left[ Z_{\chi^\prime}^n \right]
&\propto& \left[ \int \prod_{\alpha=1}^{n} \mathcal{D}\nop^\alpha \exp\left(-\sum_{\alpha=1}^{n} H_{\chi^\prime}[\nop^\alpha]/T\right) \right]
\nonumber\\
&&= \int \prod_{\alpha=1}^{n} \mathcal{D}\nop^\alpha \exp(-H^\prime[\nop^\alpha]),
\ea
where the effective replica Hamiltonian $H_{\mathrm{phen}}[\nop^\alpha]$ is given by
\ba
\label{eq:H_prime}
&&H_{\mathrm{phen}}[\nop^\alpha]
\nonumber\\
&&=
\frac{1}{2T}\sum_{\alpha=1}^{n}\sum_{\bm{p}}
\left( \LDLa t+\stiff p^2 + \smear_{\pv} \right) \{\nop_{\pv}^\alpha \, \nop_{-\pv}^\alpha\}
\nonumber\\
&&\quad
- \frac{1}{2T^2}\sum_{\alpha,\beta = 1}^{n}\sum_{\pv}[\{\ncp_{\pv} \, \ncp_{-\pv}\}]
\{\nop_{\pv}^\alpha \, \nop_{-\pv}^\beta\}
 \nonumber\\
&&\quad
- \frac{1}{2(T^0)^2}\sum_{\alpha,\beta=1}^{n}\sum_{\bm{p}}
  |\smear_{\pv}|^{2}\,[\{\nop^0 \, \nop^0\}]
  \{\nop_{\pv}^\alpha \, \nop_{-\pv}^\beta\}.
  \nonumber\\
\ea
Applying the disorder statistics specified in Ref.~(\ref{eq:disorderstats}), we arrive at the result that
$H_{\mathrm{phen}}[\nop^\alpha]=H_{{\rm eff}}[\nop^\alpha]$, i.e., the phenomenological continuum description
originally reported in Ref.~\cite{phenomenology} contains the same physics as the microscopic dimers-and-springs model.

\section{Structure and correlations in IGNEs}
\label{sec:diagnostics}

\subsection{Diagnostic quantities}
\label{sec:diagnostic_quantities}
%(i)~the average, taken over realizations of the cross-linking (i.e., over conventional quenched disorder, denoted $[\cdots]$), of the equilibrium expectation value of the local nematic alignment tensor at position ${\bm r}$, i.e., $[\langle \bm{q}({\rv})\rangle]$.

To describe the essential features of the pattern nematic ordering characteristic of IGNEs in the high-temperature regime we focus on the following pair of thermally- and disorder-averaged correlators:
\hfil\break\noindent
(i)~the average, taken over realizations of the cross-linking, of the product of the local nematic order at two points, i.e.,
\begin{subequations}
\label{eq:coredefs}
\begin{equation}
\label{eq:glasscorrdef}
\mycorrel^{G}({\bm r},{\bm r}^{\prime})
\equiv
[\{\langle \bm{q}({\bm r})\rangle\,
  \langle \bm{q}({\bm r}^{\prime})\}\rangle],
\end{equation}
which we term the \lq\lq glassy correlator\rq\rq; and
\hfil\break\noindent
(ii)~the disorder average of the familiar correlator of the fluctuations in the nematic order, i.e.,
\begin{equation}
\label{eq:thermcorrdef}
\mycorrel^{T}({\bm r},{\bm r}^{\prime})
%-\mycorrel^{G}_{d_{1}d_{2}d_{1}^{\prime}d_{2}^{\prime}}({\bm r},{\bm r}^{\prime})]
\equiv
\big[\big\langle
\big\{
\big( \bm{q}({\bm r})-\langle \bm{q}({\bm r})\rangle \big)\,
\big( \bm{q}({\bm r}^{\prime})
-\langle \bm{q}({\bm r}^{\prime})\rangle \big)
\big\}
\big\rangle\big],
\end{equation}
\end{subequations}
which we term the \lq\lq thermal correlator.\rq\rq

\begin{figure}
	\centering
		\includegraphics[width=0.4\textwidth]{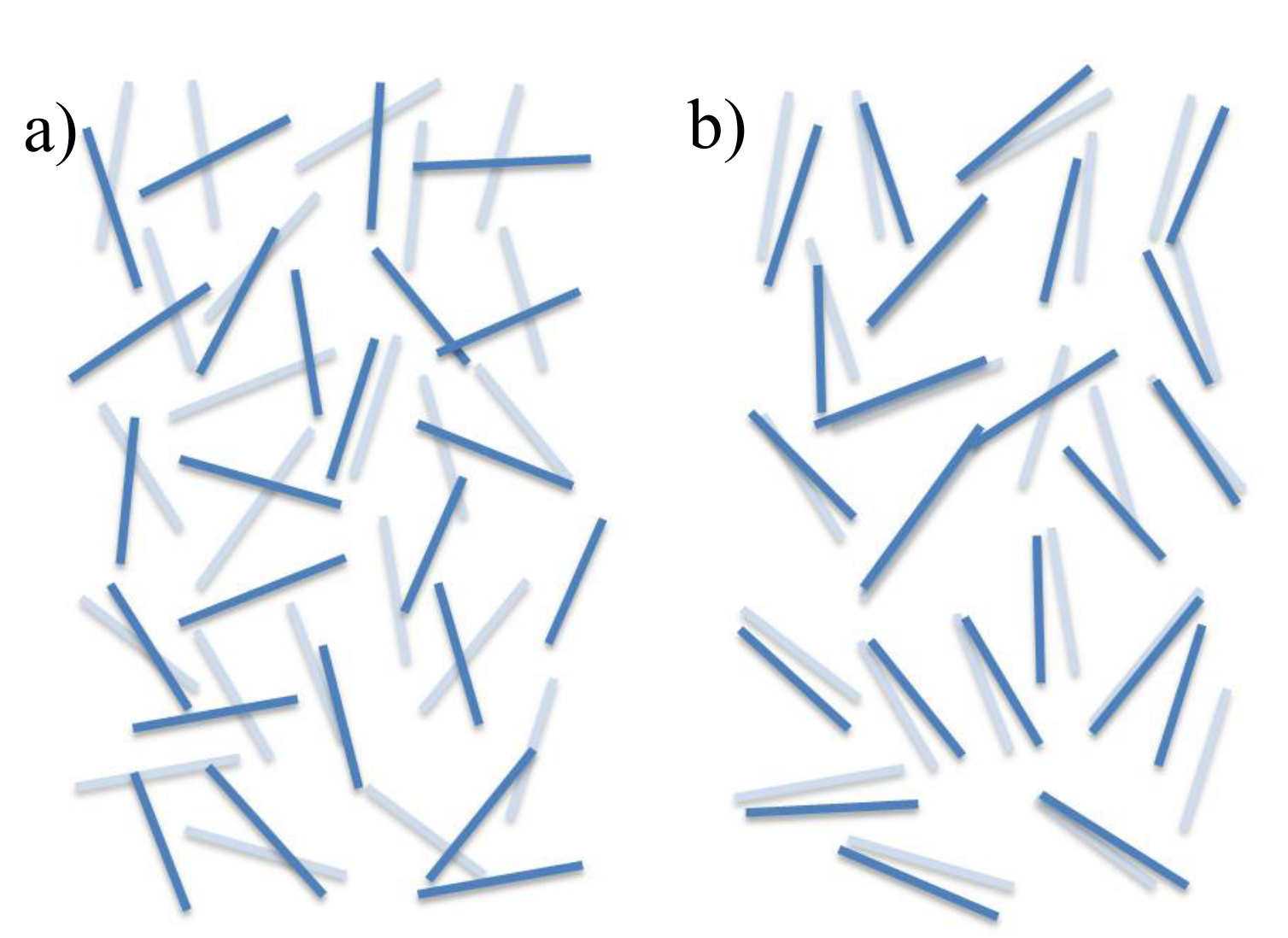}
	\caption{%[** Warning: choose for blue and shaded renderings that will appear distinct in black and white]
Schematic depictions of snapshots of nematogen locations and orientations at a particular instant (blue, full),
and at a much earlier instant (gray, shaded).
(a)~A conventional liquid crystal in the isotropic state.
Such systems do not memorize the local pattern of nematogen alignment indefinitely.
There is no correlation between the orientations of blue and shaded nematogens that are depicted near one another.  Nor is there any preference for blue and shaded nematogens that are depicted near one another to be the same nematogen.
(b)~A liquid cystalline elastomer in the macroscopically isotropic state.
Such systems do memorize the local pattern of nematogen alignment indefinitely.
The orientations of blue and gray nematogens depicted near one another are likely to be correlated.
For systems in which the nematogens are chemically bonded to an elastomer network, blue and shaded nematogens depicted near one another are likely to be the same nematogen.
However, for systems in which the nematogens are not chemically bonded to a network there is no preference for blue and shaded nematogens depicted near one another to be the same nematogen.}
  \label{fig:trappingin}
\end{figure}
The correlator $\mycorrel^{T}$ characterizes the strength of the thermal fluctuations of the nematic alignment away from the local mean value as well as the spatial range over which these fluctuations are correlated.  Inter alia, through its range, $\mycorrel^{T}$ is capable of signaling the occurrence of a continuous phase transition. The correlator $\mycorrel^{G}$ is a diagnostic of particular value for nematic elastomers, as it detects the occurrence of randomly frozen (i.e., time-persistent) local nematic order.  For the case where ${\bm r}$ and ${\bm r}^{\prime}$ are co-located, it is the nematic analog of the Edwards-Anderson order parameter, introduced long ago for spin glasses~\cite{Edwards-Anderson-SpinGlass-1975}, in the sense that it measures the magnitude of local nematic ordering, regardless of the orientation of that ordering.  Moreover, how $\mycorrel^{G}$ varies with the separation of ${\bm r}$ and ${\bm r}^{\prime}$ determines the spatial extent of regions that share a roughly common nematic alignment.  Two mechanisms are responsible for the existence of these aligned regions.  First, the formation of a random network causes a local breaking of rotational invariance, which has the effect of creating randomly anisotropic environments that tend to align the nematogens locally.  Second, although the equilibrium state of the system at the instant prior to cross-linking is, on average, isotropic, a \lq\lq snapshot\rq\rq\ of its microscopic configuration at that instant would reveal local nematic order of the type that we normally call thermal fluctuations.  The cross-linking process can trap these fluctuations in, either partially or fully, the extent depending on the strength of the cross-linking and the temperature at the moment of cross-linking.

The correlators $\mycorrel^{G}$ and $\mycorrel^{T}$ have been computed in Ref.~\cite{phenomenology} via Eq.~(\ref{eq:landau}).  In the present section, we re-derive the results via the replica approach to the microscopic dimers-and-springs model that we have developed in the foregoing sections.  To proceed, we make use of the effective replica Hamiltonian~(\ref{eq:H_eff}) and the following identities (which we prove in App.~\ref{app:c}):
\begin{subequations}
\label{eq:correlrel}
\ba
&&
[\{\langle \bm{q}(\rv)         \rangle
   \langle \bm{q}(\rv^{\prime})\rangle\}]
=
\lim_{n\rightarrow 0} \langle\!\langle
\nop^\alpha(\rv)
\nop^\beta (\rv^{\prime})
\rangle\!\rangle
\nonumber\\
&&
\hspace{4cm} \qquad (\alpha\neq\beta),
\\
&&
[\{\langle \bm{q}(\rv)\,
           \bm{q}(\rv^{\prime})
\rangle\}]
=
\lim_{n\rightarrow 0}
\langle\!\langle
\nop^\alpha(\rv)
\nop^\alpha(\rv^{\prime})
\rangle\!\rangle,
\ea
\end{subequations}
where $\langle\!\langle\ldots\rangle\!\rangle$
denotes an average performed with respect to the effective Hamiltonian $H_{{\rm eff}}$, Eq.~(\ref{eq:H_eff}).
To compute the quantities on the right hand side of Eqs.~(\ref{eq:correlrel}), we invoke the quadratic form of Eq.~(\ref{eq:H_eff}), invert the corresponding kernel, and use the replica diagonal and off-diagonal parts to obtain
\begin{subequations}
\label{eq:correl}
\begin{eqnarray}
&&\!\!\!\!\!\!\!\!\!\!\!\!\!\!\!
\mycorrel^{T}_{\pv}=
%T\mu_D\,\frac{1}{\LDLa t + \stiff p^{2}+{\smear}_{\pv}},
5T\,\frac{1}{\LDLa t + \stiff p^{2}+{\smear}_{\pv}},
\label{eq:correlT}
\\
&&\!\!\!\!\!\!\!\!\!\!\!\!\!\!\!
\mycorrel^{G}_{\pv}=
%T\mu_D\,
5T\,
\frac{
\frac{T}{T^0} (\LDLa^{0}t^0 + \stiff^0 p^{2})^{-1}\vert\smear_{\pv}\vert^{2}
+\smear_{\pv}
}
{\left(\LDLa t\!+\!\stiff p^{2}\!+\!{\smear}_{\pv}\right)^{2}}.
\label{eq:correlG}
\end{eqnarray}
\end{subequations}
%where, $\mu_D\equiv (D-1)(D+2)/2$ counts the number of degrees of freedom of $\nop$ and has value $5$ for $D=3$.
Note the presence of the scale-dependent kernel $\smear_{\pv}$ in the denominators of the correlators, which plays an essential role in determining their behavior.  This may appear surprising when we compare with the result one would obtain via a conventional random-field approach (for details, see App.~\ref{app:rf_approach}).  At length-scales larger than $\xi_L$, the presence of $\smear_\pv$ in the denominator of $\mycorrel_\pv^T$ leads to a {\it downward\/} renormalization of the bare critical temperature $T^*$ by an amount proportional to the disorder strength (and hence grows with the cross-linking density).  This indicates that the nematogens are more strongly inhibited from aligning with one another if the density of cross-links is higher.  Interestingly enough, a similar result has been obtained using the molecular level neo-classical elasticity theory of nematic elastomers \cite{WarTer96}.
This disordering effect of the random polymer network (on the nematic alignment) is not contained in the conventional random-field type of models~\footnote{This distinction arises because our effective replica Hamiltonian (\ref{eq:H_eff}) does not contain replica-diagonal contributions to the term proportional to $\smear_\pv \{ \nop_\pv^\alpha \, \nop_{-\pv}^\beta \}$, whereas the effective replica Hamiltonian that corresponds to the conventional random-field approach contains both replica-diagonal and replica off-diagonal contributions, all having the same coefficients. }.
In addition, a larger value of $\smear_{\bm{0}}$ leads to a larger amplitude of $\mycorrel_\pv^G$, indicating that a stronger localization of the network results in a more strongly trapped-in nematic pattern.  Also note  that the correlators $\mycorrel_\pv^T$ and $\mycorrel_\pv^G$ meet the physically sensible requirement that they revert to the forms appropriate to a nematic liquid in the absence of a network at length-scales very short compared with $\xi_L$, for which $\smear_{\pv}$ becomes very small, indicative of the molten character of the network at such scales.

\begin{figure}
	\centering
        \includegraphics[width=0.5\textwidth]{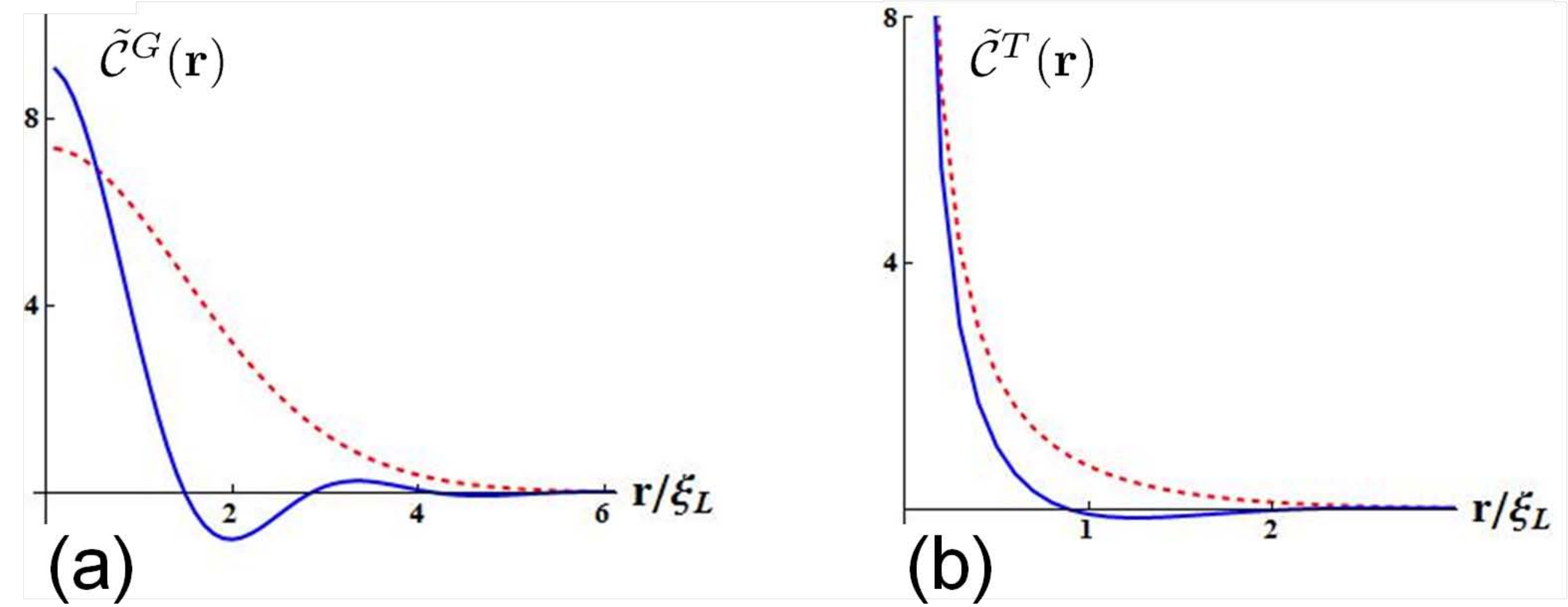}
	\caption[Real-space decay behavior of the glassy correlator and the thermal correlator]{Real-space decay behavior of
(a)~the glassy correlator (rescaled)
${\widetilde{\mycorrel}}^{G}(r)\equiv
%(60\pi^2 \stiff /T\mu_D) \, \mycorrel^{G}(r)$,
(12\pi^2 \stiff /T) \, \mycorrel^{G}(r)$,
for $t^0 \gg T\smear_{\bf 0}/T^0\LDLa^0$; $t=0.1 \stiff/\LDLa \xi_L^2$,
at (i)~$\smear_{\bf 0}/\smear^{(c)}=0.5$ (weak disorder; red, dashed)
and (ii)~$\smear_{\bf 0}/\smear^{(c)}=40$ (strong disorder; blue, solid).
(b)~the thermal correlator (rescaled)
${\widetilde{\mycorrel}}^{T}(r)\equiv
%(2\pi^2 \stiff /T\mu_D) \, \mycorrel^{T}(r)$,
(2\pi^2 \stiff /5T) \, \mycorrel^{T}(r)$,
for the same parameters.
On going from weak to strong disorder, both correlators cross over from simple exponential decay
to oscillatory decay of wavelength of order $\xi_L$.}
	\label{fig:twocorrelators}
\end{figure}
\begin{figure}
	\centering
        \includegraphics[width=0.4\textwidth]{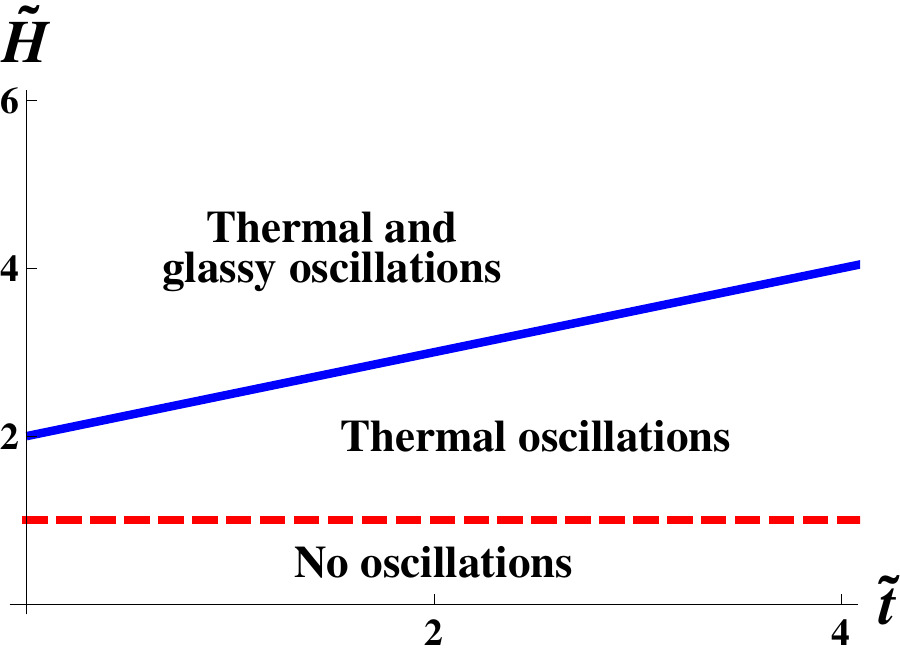}
	\caption[Crossover diagram for the glassy and thermal correlators, indicating the three
    qualitatively distinct regimes of behavior for a system that is cross-linked at a very high temperature]
    {Crossover diagram for the glassy and thermal correlators, indicating the three
    qualitatively distinct regimes of behavior for a system that is cross-linked at a very high temperature.
    Here, ${\tilde{H}} \equiv \smear_{\bf 0}/\smear^{(c)}$ is a measure of the disorder strength, and ${\tilde{t}}$ is the rescaled reduced temperature, with the value $(\LDLa\xi_L^2/\stiff)t$.
    Above the blue solid line, both correlators oscillate and decay as a function of separation.
    Between the blue solid and red dashed lines, both correlators decay but only the thermal one
    also oscillates.
    Below the red dashed line, both correlators decay but neither oscillates.}
	\label{fig:phaseMFT}
\end{figure}
\subsection{Oscillatory-decaying correlations}
\label{sec:oscillatory_decay}
The scale-dependent kernel $\smear_\pv$ in the denominator of each of the correlators in Eq.~(\ref{eq:correl}) also gives rise to the possibility that the correlators undergo both oscillation and decay with distance.
As discussed in Ref.~\cite{phenomenology}, $\mycorrel^T(\rv)$ undergoes oscillatory decay with distance for $H_0 > H^{(c)} ( =  2\stiff/\xi_L^2)$, and the length-scale of the oscillation is given by $\xi_{T,o}=\xi_L/\sqrt{2\ln (H_0/H^{(c)})}$, which is independent of $T$.
%The cross-over boundary between the oscillatory and non-oscillatory regimes for $\mycorrel^T$, which is the threshold at which $1/\xi_{T,o}$ increases from zero to a non-value, is given by
%\be
%\smear_{\bf 0}=\smear^{(c)}.
%\ee
On the other hand, $\mycorrel^G(\rv)$ undergoes oscillatory decay for sufficiently large values of $H_0$ and sufficiently low values of $T$ (see Fig.~\ref{fig:twocorrelators}), with its oscillation scale $\xi_{G,o}$ given implicitly by the following equation:
\begin{equation}
\label{eq:nonlinear}
1 + (\xi_N/\xi_{G,o})^2 + 4(\xi_N/\xi_L)^2 - (\smear_{\bf 0}/\LDLa t) e^{-\xi_L^2/2\xi_{G,o}^2}=0.
\nonumber
\end{equation}
The cross-over boundary between the oscillatory and non-oscillatory regimes for $\mycorrel^G$, which is the threshold at which the inverse oscillation wavelength $1/\xi_{G,o}$, increases from zero to a non-zero value, is given by
\be
\smear_{\bf 0}=\LDLa t + 2\smear^{(c)}.
\ee
The different regimes of behavior of $\mycorrel^T(\rv)$ and $\mycorrel^G(\rv)$ are shown in Fig.~\ref{fig:phaseMFT}.

\section{Alternative microscopic models of IGNEs}
\label{sec:comparison}
Instead of the dimers-and-springs model of IGNEs, one could have started with alternative microscopic models that may at first sight appear to be more faithful representation genuine IGNEs.  For example, one could have started with either of the following two microscopic models:
(i)~Model~A: a system comprising worm-like chains---pairs of which are permanently, randomly bonded by point-like cross-links---as well as stiff rods that dangle fromn each chain at regular arc-length intervals; and
(ii)~Model~B: a system of chains, each constructed from stiff rods that are connected in series and then permanently bonded by point-like cross-links between randomly chosen rod ends.
Models~A and B are, respectively, caricatures of side- and main-chain nematic polymer networks;
starting from either Model~A or Model~B, we can apply the Hubbard-Stratonovich decoupling procedure described in Secs.~II to IV, and thus derive a corresponding Landau-Wilson free energy that is structurally equivalent (i.e., having the same symmetries and types of couplings) to that derived for the dimers-and-springs model; cf.~Eq.~(\ref{eq:landauwilsonf}).  The purpose of this section is to establish this structural equivalence.  We can then, in principle, compute the coefficients of the terms of the expansion of the Landau-Wilson free energy separately for Models~A and B, apply a similar saddle-point analysis, and derive an effective replica Hamiltonian that would enable us to explore nematic fluctuations.  However, the coefficients of the terms of such expansions are technically more difficult to compute than those for the dimers-and-springs model.  Moreover, even if one were to succeed in computing such coefficients, the expansion would still lead, at sufficiently large length-scales, to predictions that are identical, up to an overall length-scale, to those made on the basis of the dimers-and-springs model, Eq.~(\ref{eq:landauwilsone}).  This is why we chose to work with the dimers-and-springs model in deriving an effective Hamiltonian for the liquid crystallinity of IGNEs.

In what follows, we consider Models~A and B separately.  We define the microscopic Hamiltonian, introduce the corresponding collective fields, perform the Hubbard-Stratonovich decoupling procedure and, lastly, carry out the log-trace expansion.  We then show that the auxiliary fields and the terms in the Landau-Wilson free energy corresponding to either model have the same symmetries and structure as those obtained from the dimers-and-springs model, and thus, give rise to effective Hamiltonians that are structurally equivalent to $H_{1+n}$; cf.~Eq.~(\ref{eq:H_eff0}).  To streamline the presentation we display only the essential equations, as the formal procedure employed in this section is identical to the one employed in Secs.~II and III.

\subsection{Model A: side-chain nematic polymer network}

\begin{figure}
	\centering
		\includegraphics[width=0.35\textwidth]{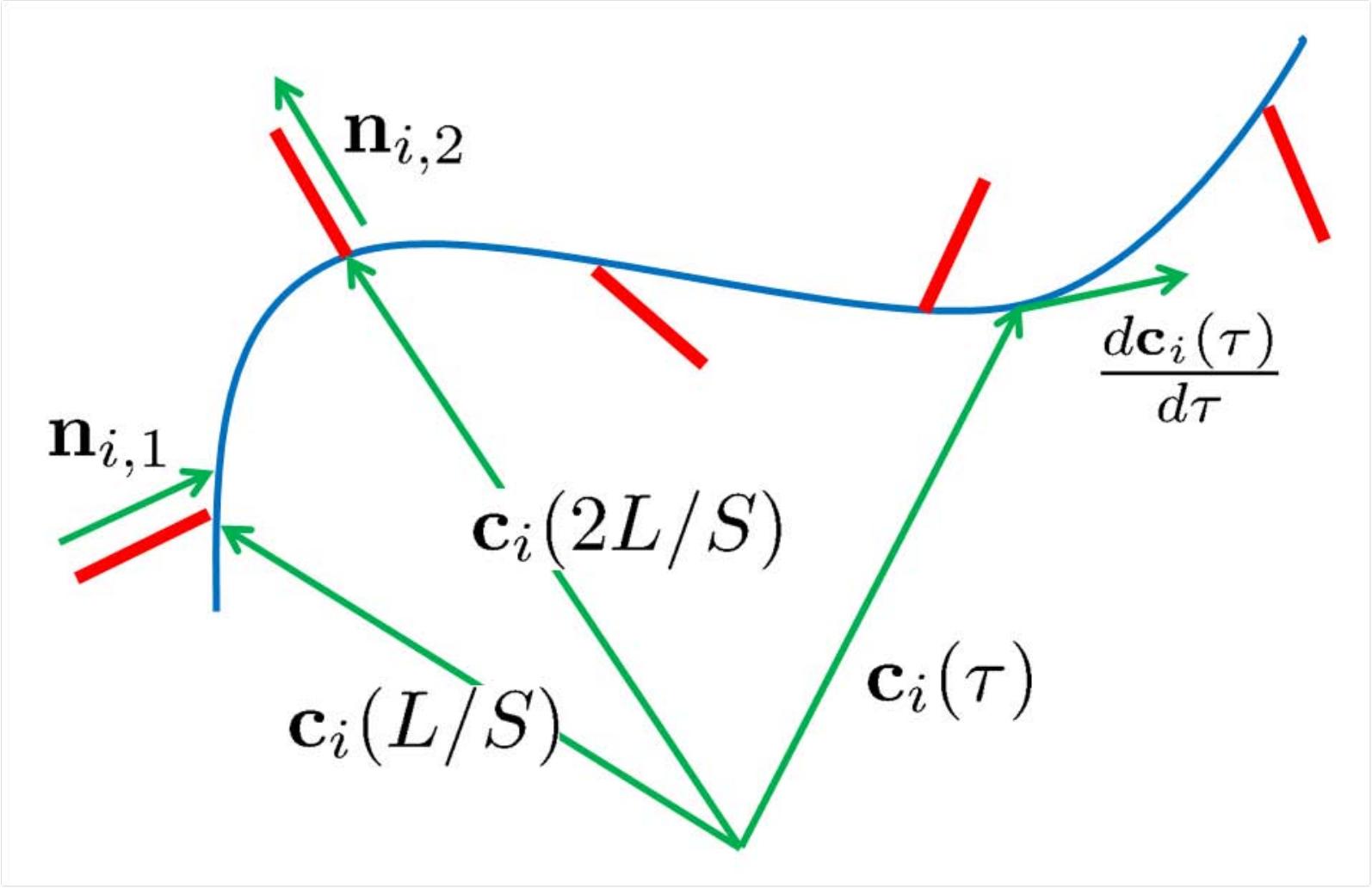}
	\caption{%
Model of a side-chain nematic polymer consisting of nematogens (red online) and worm-like chain backbone (blue online).
The chains are labeled by $i$ ($=1,\ldots,N$), is of length $L$, and has $S$ nematogens attached.
The segment at arc-length $\tau$ of chain $i$ has position $\bm{c}_i(\tau)$.
The nematogens are attached to the backbone at equally spaced arc-length intervals.}
\label{fig:wlc}
\end{figure}

We represent a side-chain nematic polymer as a worm-like chain (see, e.g., Ref.~\cite{huthmann}) of length $L$ with nematogens attached at equal intervals of arc-length $L/S$ along the chain.  Each nematogen is represented by a unit vector $\bm{n}_{i,s}$, where $i=1,\ldots, N$ labels the chains, and $s=1,\ldots, S$ labels the nematogens.
We denote the arc-length measured from one end of the chain by $\tau$ (with $0 < \tau < L$).  A segment at arc-length $\tau$ along chain $i$ has a position vector $\bm{c}_i(\tau)$; the $s$-th nematogen resides at arc-length $sL/S$, and its base, which is attached to chain $i$, has a position vector $\bm{c}_{i}(sL/S)$; see Fig.~\ref{fig:wlc}.  We require that nematogens and tangent vectors to the chains interact via orientational forces that favor parallel or anti-parallel alignment.

The side-chain nematic polymer network is formed by randomly, permanently cross-linking $M$ pairs of such chains via point-like cross-links.  The pairs of cross-linked chains and the number $M$ are fixed for a given realization of quenched disorder, but vary across such realizations of it.  In the absence of cross-links the system is specified by the following Hamiltonian:
\bw
\ba
\label{eq:H_SC}
H_{{\rm SC}}
&=&
\frac{T}{2}\sum_{i=1}^{N}\int_{0}^{L} \frac{d\tau}{L}\left\vert \frac{d^2 \bm{c}_{i}(\tau)}{d\tau^2} \right\vert^2
                 + \frac{\lambda}{2}\sum_{i,j=1}^{N} \int \frac{d\tau \, d\tau^\prime}{L^2} \delta(\bm{c}_i(\tau)-\bm{c}_j(\tau^\prime))
                 \nonumber\\
                 &&
                 -\frac{V}{2N}\sum_{i,j=1}^{N} \int \frac{d\tau \, d\tau^\prime}{L^2}
                 J(\bm{c}_i(\tau)-\bm{c}_j(\tau^\prime))
                 \left( \frac{d\bm{c}_i(\tau)}{d\tau}\cdot \frac{d\bm{c}_j(\tau^\prime)}{d\tau} \right)^2
                \nonumber\\
                &&-\frac{V}{2N}\sum_{i,j=1}^{N} \sum_{s,t=1}^{S}
                  J(\bm{c}_i(sL/S)-\bm{c}_j(tL/S))
                  \left( \bm{n}_{i,s}\cdot \bm{n}_{j,t} \right)^2
                  \nonumber\\
                &&
                -\frac{V}{N}\sum_{i,j=1}^{N} \sum_{s=1}^{S}\int \frac{d\tau}{L}
                  J(\bm{c}_i(\tau)-\bm{c}_j(sL/S))
                \left( \frac{d\bm{c}_i(\tau)}{d\tau}\cdot \bm{n}_{j,s} \right)^2.
\ea
\ew
The partition function corresponding to this system, subject to $M$ cross-linking constraints, is given by
\ba
Z_{\rm{SC}} &\propto& \prod_{i=1}^{N} \int \mathcal{D}\bm{c}_{i}
    \prod_{s=1}^{S} \int d\bm{n}_{i,s}
         e^{-H_{{\rm SC}}/T}
         \nonumber\\
         &&\times
       \prod_{e=1}^{M} \delta \big( \bm{c}_{i_e}(\tau_e)-\bm{c}_{j_e}(\tau_e^\prime) \big).
\ea
Here,
$\int \mathcal{D}\bm{c} \equiv \prod_{0 \leq \tau \leq L}
 \int d\bm{c}(\tau)$, and
 $\int d\bm{n}$
denotes integration over the unit sphere of nematogen orientations.  To proceed further, we can follow the strategy described in Secs.~II and III, including making a Hubbard-Stratonovich transformation to introduce the auxiliary fields $\{\Omega_{\hat{k}},\Omega_{\pv}^{\alpha},\nop_{\pv}^{\alpha}\}$, to arrive at the corresponding Landau-Wilson free energy $f_{\rm{SC}}$:
\bw
\ba
\label{eq:f_SC}
f_{\rm{SC}}
%\nonumber\\
&=& \frac{\eta^2}{\tilde{V}^{n}}{\overline{\sum}_{\hat{k}}} |\Omega_{\hat{k}}|^2
+ \frac{N}{2 \tilde{V}} \sum_{\alpha=0}^{n} \sum_{\pv}^{\prime} \lambda_1^\alpha |\Omega_{\pv}^{\alpha}|^2
+ \frac{1}{2}\sum_{\alpha=0}^{n} \sum_{\pv} \frac{J_{\pv}}{T^{\alpha}} {\rm{Tr}} |\nop_{\pv}^{\alpha}|^2
\nonumber\\
&-& \ln \Big\langle
\exp \Big(
\frac{2\eta^2}{\tilde{V}^n} {\overline{\sum}_{\hat{k}}} \Omega_{\hat{k}} \int_{0}^{L} \frac{d\tau}{L} e^{i\hat{k}\cdot \hat{c}(\tau)}
%\nonumber\\
+ \frac{i N}{\tilde{V}} \sum_{\alpha=0}^{n} \sum_{\pv}^{\prime} \lambda_1^\alpha \Omega_{\pv}^{\alpha}
    \int_{0}^{L} \frac{d\tau}{L} e^{i\pv\cdot \bm{c}^{\alpha}(\tau)}
    \nonumber\\
&+& \sum_{\alpha=0}^{n} \sum_{\pv} \frac{J_{\pv}}{T^{\alpha}} \nop_{d d^\prime}^{\alpha}(\pv)
    \Big(  \int \frac{d\tau}{L} \left( \frac{d\bm{c}_{d}^{\alpha}(\tau)}{d\tau} \frac{d\bm{c}_{d^\prime}^{\alpha}(\tau)}{d\tau} - \frac{1}{3}\delta_{d d^\prime} \right) e^{i\pv\cdot \bm{c}^{\alpha}(\tau)}
    \nonumber\\
    &&
+ \frac{1}{S} \sum_{s=1}^{S} (n_{s,d}^{\alpha} n_{s,d^\prime}^{\alpha} - \frac{1}{3}\delta_{d d^\prime})
    e^{i\pv\cdot \bm{c}^{\alpha}(sL/S)}
    \Big)
\Big)
\Big\rangle_{\rm{SC}}.
\ea
\end{widetext}
Here, $\lambda_1^\alpha \equiv \lambda - \eta^2 (\delta^{\alpha,0} - 1) /(N\tilde{V}^{n-1})$, reflecting the fact that the excluded-volume parameter $\lambda$ is not renormalized by the effects of cross-linking in the preparation state.  Note that the Boltzmann average $\langle\ldots\rangle_{\rm{SC}}$ is defined via
\bw
\be
\langle\cdots\rangle_{\rm{SC}}
\equiv
\frac{\prod_{\alpha=0}^{n} \int\mathcal{D}\bm{c}^{\alpha}
\prod_{s=1}^{S} \int d\bm{n}_{s}^{\alpha}
\cdots \exp\left( - \frac{1}{2}\int_{0}^{L} \frac{d\tau}{L}
\left\vert \frac{d^2\bm{c}(\tau)}{d\tau^2}\right\vert^{2} \right)}
{\prod_{\alpha=0}^{n} \int\mathcal{D}\bm{c}^{\alpha}
\prod_{s=1}^{S} \int d\bm{n}_{s}^{\alpha}
\exp\left( - \frac{1}{2}\int_{0}^{L} \frac{d\tau}{L}
\left\vert \frac{d^2\bm{c}(\tau)}{d\tau^2}\right\vert^{2} \right)}.
\ee
\ew
We observe that $f_{\rm{SC}}$ is invariant under transformations of the auxiliary fields that correspond to independent translations and rotations of the replicas, as in the case of the Landau-Wilson free energy for the dimers-and-springs model~(\ref{eq:landauwilsonf}).
In addition,
$\Omega_{\hat{k}}$ is a HRS field,
$\Omega_{\pv}^{\alpha}$ is a 1RS field, and
$\nop_{\pv}^{\alpha}$ is a traceless and symmetric second-rank tensor field, all as with the corresponding auxiliary fields for the dimers-and-springs model.
Furthermore, terms developed by expanding the log-trace part in Eq.~(\ref{eq:f_SC}) are structurally equivalent to those arising from the corresponding expansion of the Landau-Wilson free energy~(\ref{eq:landauwilsonf}) of the dimers-and-springs model.  In other words, as fields theories, Model~A and the dimers-and-springs model are identical, up to elementary rescalings of their coefficients. We thus conclude that at length-scales larger than the microscopic scales at which they are defined, these models yield predictions that are structurally identical.

\subsection{Model B: main-chain nematic polymer network}

We apply the procedure used in Sec.~\ref{sec:replicas} and the previous subsection to to derive the Landau-Wilson free energy for Model~B, a model for main-chain nematic polymer networks.  This model consists of jointed chains (see, e.g., Ref.~\cite{huthmann}) comprising $S-1$ rods each of length $\ell$.  In addition, adjacent pairs of rods interact via a bending energy that promotes their parallel alignment, and  arbitrary pairs of rods also interact via orientational forces that favor parallel or anti-parallel alignment.  The network is formed via the random, instantaneous cross-linking of pairs of jointed chains via permanent point-like cross-links, located at the ends of the rods.  The chains are labeled by $i$ ($=1,\ldots, N$) and the rod-rod junctions are labeled by $s$ ($=1,\ldots,S$); the position of rod end $s$ on chain $i$ is $\bm{c}_{i,s}$, as shown in Fig.~\ref{fig:fjc}.  The system is then specified by the following Hamiltonian:
\bw
\ba
H_{{\rm MC}}
&=&
-\frac{T}{2\ell^2}\sum_{i=1}^{N}\sum_{s=1}^{S-1}
   \big( \bm{c}_{i,s+1} - \bm{c}_{i,s} \big) \cdot \big( \bm{c}_{i,s} - \bm{c}_{i,s-1} \big)
%   \nonumber\\
%&&
+ \frac{\lambda}{2}\sum_{i,j=1}^{N}\sum_{s,t=1}^{S}
    \delta(\bm{c}_{i,s}-\bm{c}_{j,t})
    \nonumber\\
&-& \frac{1}{2 \ell^4}\sum_{i,j=1}^{N}\sum_{s,t=1}^{S-1}
    J\big( (\bm{c}_{i,s+1}+\bm{c}_{i,s})/2 - (\bm{c}_{j,t+1}+\bm{c}_{j,t})/2 \big)
%    \nonumber\\
%&&\qquad\times
\left( (\bm{c}_{i,s+1} - \bm{c}_{i,s})\cdot (\bm{c}_{j,t+1} - \bm{c}_{j,t}) \right)^2.
\ea
\ew
The partition function corresponding to this system, subject to $M$ cross-linking constraints, is given by
\ba
Z_{\rm{MC}} &\propto& \prod_{i=1}^{N}\prod_{s=1}^{S} \int d\bm{c}_{i,s} e^{-H_{{\rm MC}}/T}
\delta(|\bm{c}_{i,s+1}-\bm{c}_{i,s}|-\ell)
\nonumber\\
&&\times \prod_{e=1}^{M}\delta(\bm{c}_{i_e,s_e}-\bm{c}_{i_e^\prime, s_e^\prime}).
\ea
Using the replica method and a Deam-Edwards type of distribution for the quenched randomness, together with a Hubbard-Stratonovich decoupling, we arrive at the following Landau-Wilson free energy in terms of the auxiliary fields $\{\Omega_{\hat{k}},\Omega_{\pv}^{\alpha},\nop_{\pv}^{\alpha}\}$:
%%%%%%%%%%%%%%%%%%%%%%%%%%%%%%%%%%%%%%%
% \begin{subequations}
% \ba
% \omega_{\hat{k}}
% &=& \frac{1}{NS}\sum_{i=1}^{N} \sum_{s=1}^{S}
% e^{-i\hat{k}\cdot \hat{c}_{i,s}};
% \\
% \omega_{\pv}^{\alpha}
% &=& \frac{1}{N}\sum_{i=1}^{N} \sum_{s=1}^{S}
% e^{-i\pv\cdot \bm{c}_{i,s}^{\alpha}};
% \\
% q_{d d^{\prime}}^{\alpha}(\pv)
% &=&
% \frac{1}{NS}
% \sum_{i=1}^{N}
% \sum_{s=1}^{S-1}
% \Big( \frac{1}{\ell^{2}}(c_{s+1,d}^{\alpha} - c_{s,d}^{\alpha})
% (c_{s+1,d^\prime}^{\alpha} - c_{s,d^\prime}^{\alpha})
% - \frac{1}{D}\delta_{d d^\prime} \Big)
%     e^{-i\pv\cdot \bm{c}^{\alpha}(sL/S)}.
% \ea
% \end{subequations
%%%%%%%%%%%%%%%%%%%%%%%%%%%%%%%%%%%%%%%
\begin{widetext}
\ba
\label{eq:f_MC}
f_{\rm{MC}}
&=& \frac{S^2 \eta^2}{\tilde{V}^{n}}{\overline{\sum}_{\hat{k}}} |\Omega_{\hat{k}}|^2
+ \frac{N S^2}{2 \tilde{V}} \sum_{\alpha=0}^{n} \sum_{\pv}^{\prime} \lambda_2^\alpha |\Omega_{\pv}^{\alpha}|^2
+ \frac{S}{2}\sum_{\alpha=0}^{n} \sum_{\pv} \frac{J_{\pv}}{T^{\alpha}} {\rm{Tr}} |\nop_{\pv}^{\alpha}|^2
\nonumber\\
&-& \ln \Big\langle
\exp \Big(
\frac{2 S \eta^2}{\tilde{V}^n} {\overline{\sum}_{\hat{k}}} \Omega_{\hat{k}} \sum_{s=1}^{S} e^{i\hat{k}\cdot \hat{c}_{s}}
%\nonumber\\
+ \frac{i N S}{\tilde{V}} \sum_{\alpha=0}^{n} \sum_{\pv}^{\prime}
     \lambda_2^\alpha \Omega_{\pv}^{\alpha}
     \sum_{s=1}^{S} e^{i\pv\cdot \bm{c}_{s}^{\alpha}}
    \nonumber\\
&+& \sum_{\alpha=0}^{n} \sum_{\pv} \frac{J_{\pv}}{T^{\alpha}} \nop_{d d^\prime}^{\alpha}(\pv)
    \sum_{s=1}^{S-1}
    \big(
    \frac{1}{\ell^{2}}
    (c_{s+1,d}^{\alpha} - c_{s,d}^{\alpha})(c_{s+1,d^\prime}^{\alpha} - c_{s,d^\prime}^{\alpha})
    - \frac{1}{3}\delta_{d d^\prime}
    \big)
    e^{i\pv\cdot (\bm{c}_{s}^{\alpha}+\bm{c}_{s+1}^{\alpha})/2}
\Big)
\Big\rangle_{\rm{MC}},
\ea
where $\lambda_2^\alpha \equiv \lambda - \eta^2 (\delta^{\alpha,0} - 1) /(NS\tilde{V}^{n-1})$, and the Boltzmann average $\langle\ldots\rangle_{\rm{MC}}$ is defined to be
\ba
\langle\cdots\rangle_{\rm{MC}}
\equiv
\frac{\prod_{\alpha=0}^{n} \prod_{s=1}^{S} \int d\bm{c}_{s}^{\alpha}
\cdots
\exp\left( \frac{1}{2\ell^{2}}\sum_{\alpha=0}^{n} \sum_{s=2}^{S-1}
(\bm{c}_{s+1}^{\alpha} - \bm{c}_{s}^{\alpha})\cdot (\bm{c}_{s}^{\alpha} - \bm{c}_{s-1}^{\alpha}) \right)
\prod_{\alpha=0}^{n} \prod_{s=1}^{S-1} \delta(\left\vert \bm{c}_{s+1}^{\alpha} - \bm{c}_{s}^{\alpha} \right\vert - \ell)}
{\prod_{\alpha=0}^{n} \prod_{s=1}^{S} \int d\bm{c}_{s}^{\alpha}
\exp\left( \frac{1}{2\ell^{2}}\sum_{\alpha=0}^{n} \sum_{s=2}^{S-1}
(\bm{c}_{s+1}^{\alpha} - \bm{c}_{s}^{\alpha})\cdot (\bm{c}_{s}^{\alpha} - \bm{c}_{s-1}^{\alpha}) \right)
\prod_{\alpha=0}^{n} \prod_{s=1}^{S-1} \delta(\left\vert \bm{c}_{s+1}^{\alpha} - \bm{c}_{s}^{\alpha} \right\vert - \ell)}.
\ea
\ew
%%%%%%%%%%%%%%%%%%%%%%%%%%%%%%%%%%%
% The auxiliary fields $\Omega_{\hat{k}}$, $\Omega_{\pv}^{\alpha}$ and $\nop_{\pv}^{\alpha}$ are related to the collective fields $\omega_{\hat{k}}$, $\omega_{\pv}^{\alpha}$ and $\bm{q}_{\pv}^{\alpha}$ via
% \begin{subequations}
% \label{eq:MC_relations}
% \ba
% \langle \Omega_{\hat{k}}\rangle_f
% &=&
% \langle \omega_{\hat{k}}\rangle_{\rm{MC}};
% \\
% \langle \Omega_{\pv}^{\alpha}\rangle_f
% &=&
% \langle \omega_{\pv}^{\alpha}\rangle_{\rm{MC}};
% \\
% \langle Q_{d d^{\prime}}^{\alpha}(\pv) \rangle_f
% &=&
% \langle Q_{d d^{\prime}}^{\alpha}(\pv) \rangle_{\rm{MC}},
% \ea
% \end{subequations}
% where the subscript $f$ in the left-hand side of Eq.~(\ref{eq:MC_relations}) means that the expectation value is evaluated with respect to $f_{{\rm{MC}}}$ [Eq.~(\ref{eq:f_MC})].
%%%%%%%%%%%%%%%%%%%%%%%%%%%%%%%%%%%%%%%%

\begin{figure}
	\centering
		\includegraphics[width=0.35\textwidth]{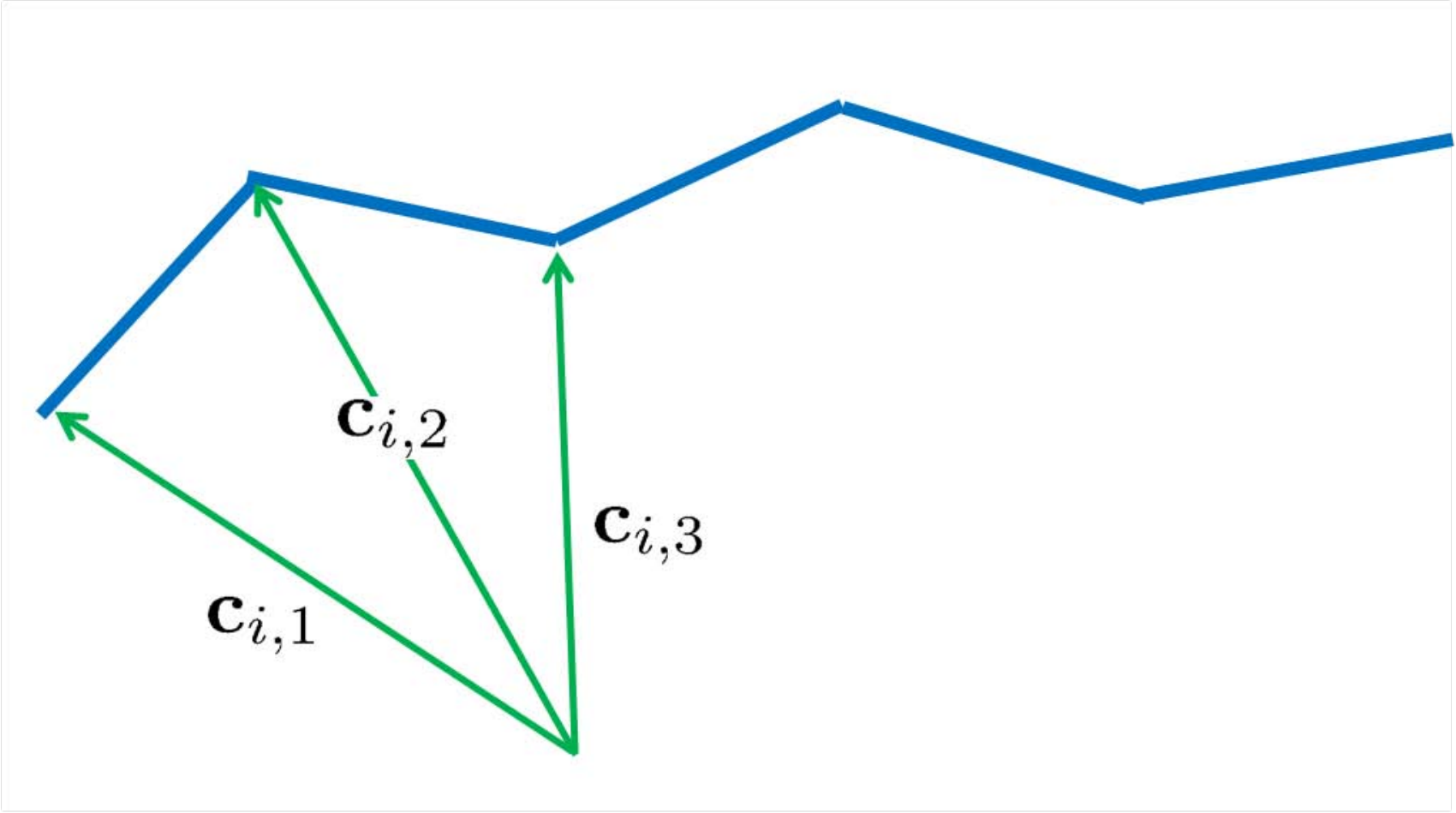}
	\caption{Model of a main-chain nematic polymer consisting of $S-1$ rods of fixed length $\ell$.
The chain is labeled $i$ ($= 1, \ldots, N$) and the ends of the rods are labeled by $s$ (where $s=1,\ldots, S$). Rod-end $s$ on chain $i$ is located at $\bm{c}_{i,s}$.}
  \label{fig:fjc}
\end{figure}
The comments made at the end of the previous subsection concerning symmetries of the Landau-Wilson free energy under transformations of the auxiliary fields hold for Model~B, too.

\section{Concluding remarks}
\label{sec:conclusion}
The objective of this Paper has been to develop a microscopic approach to the liquid crystalline properties of isotropic-genesis nematic elastomers (IGNEs), in which local nematic order---both in the preparation and the measurement ensembles---and random localization---induced by the presence of an elastomer network---are naturally incorporated.  This development, which takes as its starting point a system of dimers that are permanently connected at random by Hookean springs, serves as the underpinning to the phenomenological approach presented in Ref.~\cite{phenomenology} by providing a systematic derivation of the formulas on which the phenomenological description is based.

Specifically, by deriving an effective Hamiltonian of liquid crystallinity in IGNEs, we have shown that this microscopic approach leads to the phenomenological Landau theory of IGNEs proposed in Ref.~\cite{phenomenology} which, {\it inter alia\/}, predicts that at sufficiently large disorder strengths, both the thermal and glassy spatial correlations of nematic alignments can undergo oscillation with decay.  The development has as a core feature an ensemble---the preparation ensemble---that is distinct from the usual (measurement) ensemble of Gibbs statistical mechanics.  This feature enables us to determine in detail the influence of the preparation history of IGNEs on their subsequent equilibrium behavior.  The consequences of these two ensembles were analyzed in detail in Ref.~\cite{phenomenology}.  However, the appearance of the two ensembles took the form of a hypothesis in Ref.~\cite{phenomenology}, whereas in the present Paper they come into play naturally.  Lastly, we have argued that at sufficiently large length-scales, predictions made on the basis of a simple dimers-and-springs model are qualitatively identical to those made on the basis of two more realistic (but more complicated) microscopic models of IGNEs: one for side-chain liquid crystalline elastomers and one for main-chain liquid crystalline elastomers.

Apart from its relevance to the specific subject of liquid crystalline elastomers, the present work brings to light a more general issue, viz., that the concept of a quenched random field should be broadened to incorporate not only the conventional, \lq\lq frozen\rq\rq\ type, which does not fluctuate thermally, but also the type necessary for understanding media such as liquid crystalline elastomers, in which the frozen nature of the random field is present only at longer length-scales, fading out as the length-scale progresses through a characteristic localization length, owing to the thermal position fluctuations of the network's constituents.  The framework elucidated in the present work can be extended, with suitable modifications, to explore the statistical physics of other randomly cross-linked systems, such as smectic elastomers and various biological materials.

\begin{acknowledgments}
We thank
Tom Lubensky,
Leo Radzihovsky
Kenji Urayama,
and Mark Warner
for informative discussions.
This work was supported by
the U.S.~National Science Foundation via grants DMR~09 06780 and DMR~12 07026,
the Institute for Complex Adaptive Matter,
Shanghai Jiao Tong University, and
the National Science Foundation of China via Grants 11174196 and 91130012.
\end{acknowledgments}

\begin{widetext}

% \newpage

\appendix

\section{  Calculation of the average linking number per dimer}
\label{app:a}
In this series of Appendices we show various components of the calculations that are necessary for deriving the results presented in the main body of the Paper.
In the present Appendix we show that the average number of springs connected to a dimer, $[M]/N$, is approximately given by $2\eta^2$.
By definition,
\ba
\label{eq:M}
[M] &=&
\sum_{\chi} P(\chi) M
\nonumber\\
&=&
\sum_{\chi} \int \mathcal{D} \bm{c}^0 P(\chi | \{ \bm{c}^0 \} ) P(\{ \bm{c}^0 \}) M
\nonumber\\
&=&
\frac{1}{Z_{\rm{liq}}^0}\sum_{M=0}^{\infty}\sum_{i_1,j_1=1}^{N}\sum_{s_1,t_1=-1,1}
\ldots\sum_{i_M,j_M=1}^{N}\sum_{s_M,t_M=-1,1}
\frac{M}{M!}\Big( \frac{\eta^2 V}{2N(2\pi b^2)^{D/2}} \Big)^{M}
\nonumber\\
&&\times
\int \mathcal{D} \bm{c}^0 \, e^{H_{\rm{norm}} - H_{\rm{nem}} - H_{\rm{ev}}}
\prod_{e=1}^{M} e^{-\frac{1}{2b^2}|\bm{c}_{i_e,s_e}^0 - \bm{c}_{j_e,t_e}^0|^2}
\prod_{i=1}^{N}\delta(|\bm{c}_{i,1}-\bm{c}_{i,-1}|-\ell)
\nonumber\\
&=&
\frac{\eta^2 V}{2N(2\pi b^2)^{D/2} Z_{\rm{liq}}^0}
\int \mathcal{D} \bm{c}^0
\Big(\sum_{i, j=1}^{N} \sum_{s,t=-1,1} e^{-\frac{1}{2b^2}|\bm{c}_{i,s}^0 - \bm{c}_{j,t}^0|^2}\Big)
\nonumber\\
&&\times
e^{H_{\rm{norm}} - H_{\rm{nem}} - H_{\rm{ev}}}
\sum_{M=1}^{\infty}
\frac{1}{(M-1)!}
\Big(\frac{\eta^2 V}{2N(2\pi b^2)^{D/2}}
\sum_{i, j=1}^{N} \sum_{s,t=-1,1} e^{-\frac{1}{2b^2}|\bm{c}_{i,s}^0 - \bm{c}_{j,t}^0|^2}\Big)^{M-1}
\prod_{i=1}^{N}\delta(|\bm{c}_{i,1}^0 - \bm{c}_{i,-1}^0|-\ell)
\nonumber\\
&=&
\frac{\eta^2 V}{2N(2\pi b^2)^{D/2} Z_{\rm{liq}}^0}
\int \mathcal{D} \bm{c}^0
\Big(\sum_{i, j=1}^{N} \sum_{s,t=-1,1} e^{-\frac{1}{2b^2}|\bm{c}_{i,s}^0-\bm{c}_{j,t}^0|^2}\Big)
\nonumber\\
&&\times
e^{H_{\rm{norm}} - H_{\rm{nem}} - H_{\rm{ev}}}
\exp\Big( \frac{\eta^2 V}{2N(2\pi b^2)^{D/2}} \sum_{i, j=1}^{N} \sum_{s,t=-1,1} e^{-\frac{1}{2b^2}|\bm{c}_{i,s}^0-\bm{c}_{j,t}^0|^2} \Big)
\prod_{i=1}^{N}\delta(|\bm{c}_{i,1}^0-\bm{c}_{i,-1}^0|-\ell)
\nonumber\\
&=&
\frac{\eta^2 V}{2N(2\pi b^2)^{D/2} Z_{\rm{liq}}^0}
\int \mathcal{D} \bm{c}^0
\sum_{i, j=1}^{N}\sum_{s,t=-1,1}
e^{-\frac{1}{2b^2}|\bm{c}_{i,s}^0-\bm{c}_{j,t}^0|^2 - H_{\rm{nem}} -H_{\rm{ev}}}
\prod_{i=1}^{N}\delta(|\bm{c}_{i,1}^0-\bm{c}_{i,-1}^0| - \ell).
\ea
To find a rough estimate for $[M]/N$, we shall approximate the system as a dilute gas of dimers, which involves approximating the quantity $\exp(-|\bm{c}_{i}-\bm{c}_{j}|^2/(2b^2))$ by $(2\pi b^2)^{D/2}/V$.  This leads to
\be
\sum_{i, j=1}^{N} \sum_{s,t=-1,1}
\int \mathcal{D} \bm{c}^0
e^{-\frac{1}{2b^2}|\bm{c}_{i,s}^0-\bm{c}_{j,t}^0|^2 - H_{\rm{nem}} -H_{\rm{ev}}}
\prod_{i=1}^{N}\delta(|\bm{c}_{i,1}^0-\bm{c}_{i,-1}^0| - \ell)
\approx
4N^2 \frac{(2\pi b^2)^{D/2}}{V} \, Z_{\rm{liq}}^0.
\ee
Substituting this into Eq.~(\ref{eq:M}), we obtain
\be
[M]/N \approx 2\eta^2,
\ee
which shows that $\eta^2$ provides a quantitative measure of the average number of cross-linking springs per dimer.

\section{Hubbard-Stratonovich scheme}
\label{app:b}
In this Appendix, we present the details of the Hubbard-Stratonovich scheme that leads to Eq.~(\ref{eq:landauwilsonf}).  We first consider the replica partition function $Z_{1+n}$ in Eq.~(\ref{eq:pre_HS}), which can equivalently be expressed as follows:
\ba
\label{eq:Z_1+n}
Z_{1+n}
&=& \sum_{M=0}^\infty\sum_{i_1,j_1=1}^{N}\dots \sum_{i_M,j_M=1}^{N} \sum_{s_1,t_1=1,-1}\dots\sum_{s_M,t_M=1,-1}
\frac{1}{M!}\left(\frac{V\eta^2}{2N(2\pi b^2)^{D/2}}\right)^M
\nonumber\\
&&\times
\prod_{\alpha=0}^n \int\prod_{i=1}^{N} d{\bm{c}}_{i,1}^\alpha \, d{\bm{c}}_{i,-1}^\alpha
e^{\frac{1}{2}\sum_{\alpha=0}^n \sum_{i,j=1}^{N} \frac{J_{ij}}{T^\alpha} \big( ({\bm{n}}_i^\alpha\cdot {\bm{n}}_j^\alpha)^2 - \frac{1}{D} \big)-\frac{1}{2b^2}\sum_{\alpha=0}^n \sum_{e=1}^M |{\bm{c}}_{i_e,s_e}^\alpha - {\bm{c}}_{j_e,t_e}^\alpha|^2}
\nonumber\\
&&\times
\exp
\Big(
-\frac{V\eta^2}{2N(2\pi b^2)^{D/2}} \sum_{i,j=1}^{N}\sum_{s,t=1,-1} e^{-\frac{1}{2b^2}|\bm{c}_{i,s}^0 - \bm{c}_{j,t}^0|^2}
-\frac{1}{2}\sum_{\alpha=0}^n \sum_{i, j=1}^{N} \sum_{s,t=-1,1} \frac{\lambda}{T^\alpha}\delta({\bm{c}}_{i,s}^\alpha - {\bm{c}}_{j,t}^\alpha)
\Big)
\nonumber\\
&&\times
\prod_{\alpha=0}^{n} \prod_{j=1}^N\delta(|{\bm{c}}_{j,1}^\alpha-{\bm{c}}_{j,-1}^\alpha|-\ell)
\nonumber\\
&=& \prod_{\alpha=0}^n \int\prod_{i=1}^{N} d{\bm{c}}_{i,1}^\alpha \, d{\bm{c}}_{i,-1}^\alpha
e^{\frac{1}{2}\sum_{\alpha=0}^n \sum_{i, j=1}^{N} {J_{ij}}{T^\alpha} \big( ({\bm{n}}_i^\alpha\cdot {\bm{n}}_j^\alpha)^2 - \frac{1}{D} \big)
-\frac{1}{2}\sum_{\alpha=0}^n \sum_{i, j=1}^{N} \sum_{s,t=-1,1} \frac{\lambda}{T^\alpha}\delta({\bm{c}}_{i,s}^\alpha-{\bm{c}}_{j,t}^\alpha)}
\nonumber\\
&&\times
\exp
\bigg( \frac{V\eta^2}{2N(2\pi b^2)^{D/2}} \sum_{i,j=1}^{N} \sum_{s,t=1,-1}
   \Big(
    e^{-\frac{1}{2b^2}\sum_{\alpha=0}^n |{\bm{c}}_{i,s}^\alpha-{\bm{c}}_{j,t}^\alpha|^2}
    -
    e^{-\frac{1}{2b^2}|\bm{c}_{i,s}^0 - \bm{c}_{j,t}^0|^2}
   \Big)
\bigg)
\nonumber\\
&&\times
\prod_{\alpha=0}^{n} \prod_{j=1}^{N} \delta(|{\bm{c}}_{j,1}^\alpha-{\bm{c}}_{j,-1}^\alpha|-\ell).
\ea
Note that the two terms in the first exponential factor in the first equality are replicated versions of the Maier-Saupe interaction $H_{\rm{nem}}$ and Hookean spring potential $H_{\rm{ev}}$.
Also note that the first term in the last exponential factor in the first equality, $-\left( V\eta^2/2N(2\pi b^2)^{D/2}\right) \sum_{i,j}\sum_{s,t} \exp\left( -|\bm{c}_{i,s}^{0} - \bm{c}_{j,t}^{0}|^2/2b^2 \right)$, is the term $H_{\rm{norm}}(\{ \bm{c}_{i,s}^{0} \})/T^0$ in Eq.~(\ref{eq:H_norm}), whilst the second term in the same exponential factor is a replicated version of the excluded-volume interaction, $H_{\rm{ev}}$.
In the last step we have summed over all realizations of the quenched disorder.  This results in an exponentiation of the Hookean energy term, and leads to the following expression:
%The expression for $Z_{1+n}$ involves a constraint that fixes the dimer rod length at $\ell$.
%We can eliminate the constraint by directly working with the centre-of-mass coordinate $\bm{c}\equiv ({\bm{c}}_1+{\bm{c}}_2)/2$ and dimer orientation $\bm{n}\equiv ({\bm{c}}_2-{\bm{c}}_1)/\ell$.  In terms of these new coordinates, Eq.~(\ref{eq:Z_1+n}) becomes
\ba
\label{eq:Z_1+nNew}
Z_{1+n}
&=&
\prod_{\alpha=0}^n
\int\prod_{i=1}^{N}
d{\bm{c}}_{i,1}^\alpha d{\bm{c}}_{i,-1}^\alpha
e^{\frac{V}{2N}\sum_{\alpha=0}^{n} \sum_{i,j=1}^{N}
 \frac{J({\bm{c}}_i^\alpha-{\bm{c}}_j^\alpha)}{T^\alpha}
\big( n_{i,d_1}^\alpha n_{i,d_2}^\alpha - \frac{1}{D}\delta_{d_1 d_2} \big)
\big( n_{j,d_1}^\alpha n_{j,d_2}^\alpha - \frac{1}{D}\delta_{d_1 d_2} \big)}
\nonumber\\
&&\times
\exp
\Big(
-\sum_{\alpha=0}^n\frac{\lambda}{2T^\alpha}\int d{\bm{r}}^\alpha
\sum_{i=1}^{N} \sum_{s=-1,1}
\delta\big( {\bm{c}}_{i,s}^\alpha - {\bm{r}}^\alpha \big)
\sum_{j=1}^{N} \sum_{t=-1,1}
\delta\big( {\bm{c}}_{j,t}^\alpha - {\bm{r}}^\alpha \big)
\Big)
\nonumber\\
&&\times
\exp
\bigg(
\frac{2N V\eta^2}{(2\pi b^2)^{D/2}}
\Big(
\int d\hat{x}\,d\hat{y}
e^{-\frac{|\hat{x}-\hat{y}|^2}{2b^2}}
\frac{1}{2N}\sum_{i=1}^{N} \sum_{s=-1,1}
\delta\big( \hat{x} - {\hat{c}}_{i,s} \big)
\frac{1}{2N}\sum_{j=1}^{N} \sum_{t=-1,1}
\delta\big( \hat{y} - {\hat{c}}_{j,t} \big)
\nonumber\\
&&\quad
-
\int d\bm{x}\,d\bm{y}
e^{-\frac{|\bm{x}-\bm{y}|^2}{2b^2}}
\frac{1}{2N}\sum_{i=1}^{N} \sum_{s=-1,1}
\delta\big( \bm{x} - {\bm{c}}_{i,s}^{0} \big)
\frac{1}{2N}\sum_{j=1}^{N} \sum_{t=-1,1}
\delta\big( \bm{y} - {\bm{c}}_{j,t}^{0} \big)
\Big)
\bigg)
\prod_{\alpha=0}^{n} \prod_{j=1}^N\delta(|{\bm{c}}_{j,1}^\alpha-{\bm{c}}_{j,-1}^\alpha|-\ell).
\nonumber\\
\ea
Here, we have replaced $J_{i,j}$ by its continuum limit $J({\bm{c}}_i^\alpha-{\bm{c}}_j^\alpha)$, viz., $J_{i,j}\approx (V/N)J({\bm{c}}_i^\alpha-{\bm{c}}_j^\alpha)$, where we approximate $J(\bm{c})$ by a potential of Gaussian form, i.e., $J(\bm{c})\approx \big( J_0 / (2\pi a^2)^{d/2} \big) \exp(-c^2/2a^2)$, suitable for describing short-range interactions.
Next, we define the following collective fields in Fourier space:
\begin{subequations}
\label{eq:collective_fields}
\ba
\label{eq:collective_fields_q}
q_{d_1 d_2}^\alpha(\bm{p})
&\equiv& \frac{1}{N}\sum_{i=1}^N e^{i\bm{p}\cdot{\bm{c}}_i^\alpha}(n_{i,d_1}^\alpha n_{i,d_2}^\alpha - D^{-1}\delta_{d_1 d_2}),
\\
\label{eq:collective_fields_density}
\omega_{\bm{p}}^\alpha
&\equiv& \frac{1}{2N}\sum_{i=1}^{N}\sum_{s=-1,1} e^{-i\bm{p}\cdot {\bm{c}}_{i,s}^\alpha },
\\
\label{eq:collective_fields_omega}
\omega_{\hat{k}}
&\equiv& \frac{1}{2N}
\sum_{i=1}^{N}\sum_{s=-1,1} e^{-i\hat{k}\cdot {\hat{c}}_{i,s}}.
\ea
\end{subequations}
In Eq.~(\ref{eq:collective_fields_omega}), the argument of $\omega_{\hat{k}}$ can take any value in replicated Fourier space. However, in Eq.~(\ref{eq:decomp}) below, we shall decompose the field $\omega_{\hat{k}}$ into the part whose argument takes values from the HRS, the part whose argument takes values from the 1RS, and the part whose argument takes values from the 0RS.  We shall then redefine $\omega_{\hat{k}}$ to be only those parts whose arguments belong to the HRS.
Note that $\omega_{\pv{\hat{\epsilon}}^\alpha}=\omega_{\pv}^{\alpha}$ and that the 0RS part of $\omega_{\hat{k}}$ is a constant.
To simplify the notation, we define the following normalized averages:
\ba
\langle \cdots \rangle_{N, 1+n}
&\equiv&
\prod_{\alpha=0}^n \prod_{i=1}^N
\int \frac{d{\bm{c}}_{i,1}^\alpha d{\bm{c}}_{i,-1}^\alpha}{4\pi V \ell^2} \delta(|\bm{c}_{i,1}^{\alpha} - \bm{c}_{i,-1}^{\alpha}| - \ell)\cdots;
\nonumber\\
\langle \cdots \rangle_{1, 1+n}
&\equiv&
\prod_{\alpha=0}^n
\int \frac{d{\bm{c}}_{1}^\alpha d{\bm{c}}_{-1}^\alpha}{4\pi V \ell^2} \delta(|\bm{c}_{1}^{\alpha} - \bm{c}_{-1}^{\alpha}| - \ell)\cdots.
\ea
We substitute the collective fields into Eq.~(\ref{eq:Z_1+nNew}), taking care to separate out the 0RS, 1RS and HRS parts of the terms involving $\omega_{\hat{k}}$.  The replica partition function then becomes
\ba
\label{eq:decomp}
&&Z_{1+n}
=
(4\pi V \ell^2)^{N}
\bigg\langle
\exp\left(
  \frac{N}{2}\sum_{\alpha=0}^n \sum_{\bm{p}}
  \frac{J_\pv }{T^\alpha}|q_{d_1 d_2}^\alpha (\bm{p})|^2
 -\frac{2N^2}{V}\sum_{\alpha=0}^n
 \frac{\lambda}{T^\alpha}
 |\omega_{\bm{0}}^{\alpha}|^2
 - \frac{2N^2}{V}\sum_{\alpha=0}^n {\sum_{\bm{p}}}^\prime
 \lambda^\alpha |\omega_{\bm{p}}^\alpha|^2
 \right)
 \nonumber\\
 &&\times
\exp\bigg(
    \frac{2N\eta^2}{V^n}
    |\omega_{\hat{0}}|^2
    + \frac{2N\eta^2}{V^n} \sum_{\alpha=1}^n {\sum_{\bm{p}}}^\prime
    \vertex_{\bm{p}}
    |\omega_{\pv\epsilon^\alpha}|^2
    + \frac{2N\eta^2}{V^n}
    \overline{\sum_{\hat{k}}}
    \vertex_{\hat{k}} |\omega_{\hat{k}}|^2
    \bigg)
\bigg\rangle_{N,1+n}
\nonumber\\
&&\propto
\bigg\langle
\exp\left(
  \frac{N}{2}\sum_{\alpha=0}^n \sum_{\pv} \frac{J_\pv }{T^\alpha}|q_{d_1 d_2}^\alpha (\pv)|^2
 -\frac{N}{2}\sum_{\alpha=0}^n{\sum_{\pv}}^\prime \frac{{\tilde{\lambda}}^\alpha_{\pv}}{T^\alpha}
   |\omega_{\pv}^\alpha|^2
+ \frac{N{\tilde{\eta}}^2 }{2 V^n} \overline{\sum_{\hat{k}}} \vertex_{\hat{k}}
   |\omega_{\hat{k}}|^2 \right)
\bigg\rangle_{N,1+n},
\nonumber\\
\ea
where ${\sum_{\bm{p}}}^\prime$ denotes a sum over all wave-vectors $\bm{p}$ excluding $\bm{p}=\bm{0}$, we have made the following definitions:
$\vertex_{k}\equiv \exp(-\frac{1}{2} b^2 k^2)$,
${\tilde{\eta}}^2\equiv 4\eta^2$,
and ${\tilde{\lambda}}^\alpha_\pv/T^\alpha \equiv (4N/\V)\lambda^\alpha/T^\alpha - (4\eta^2/\V^{n})(1-\delta^{\alpha,0})\vertex_{\pv}$, and
${\tilde{\lambda}}^\alpha_\pv$ is the effective excluded--volume interaction between dimers.
For $\alpha \neq 0$, the excluded--volume interaction is renormalized downwards; owing to the presence of attractive interactions induced by cross-linking, whereas for $\alpha = 0$, the excluded--volume interaction has no such renormalization.  In the case of replica $\alpha = 0$, the downward renormalization due to cross-linking is canceled exactly by the correction from the term $H_{\rm{norm}}$; physically, this is expected as the dimers cannot feel the cross-link induced attractive forces at the instant just prior to cross-linking.

We now implement the Hubbard-Stratonovich transformation, which is based on the following set of equalities for complex variables $q$ and $\omega$:
\begin{subequations}
\label{eq:hubbard}
\ba
&&e^{-J|q|^2}=\frac{J}{\pi}\int d({\rm{Re}}\omega)d({\rm{Im}}\omega) e^{-J|\omega|^2 + 2i J {\rm{Re}}q\omega^*};
\\
&&e^{+J|q|^2}=\frac{J}{\pi}\int d({\rm{Re}}\omega)d({\rm{Im}}\omega) e^{-J|\omega|^2 + 2 J {\rm{Re}}q\omega^*}.
\ea
\end{subequations}
Here, $q$ may be thought of as a complex-variable analogue of a collective field and $\omega$ is the complex-variable analogue of its conjugate auxiliary field.
The Hubbard-Stratonovich procedure allows us to express the replica partition function $Z_{1+n}$ as a functional integral in terms of the auxiliary fields
$\Omega$
[whose argument we restrict to the HRS via the constraint $\Omega_{\pv\epsilon^{\alpha}}$, in accordance with the replica-sector division laid out in Eq.~(\ref{eq:decomp})],
$\Omega^\alpha$
(which is formally $\Omega$ but with its argument taking values in the 1RS), and
$\nop^\alpha$
[conjugate to (resp.) $\omega$, $\omega^\alpha$, and $\bm{q}^\alpha$].
In terms of these auxiliary fields, the replica partition function $Z_{1+n}$ is given by
\be
Z_{1+n} \propto \int \mathcal{D}\Omega\prod_{\alpha=0}^n \mathcal{D}\Omega^\alpha\mathcal{D}\nop^\alpha
         \exp \big( -Nf_{1+n}[ \Omega, \Omega^\alpha, \nop^\alpha] \big).
\ee
Here, the Landau-Wilson free energy per dimer $f_{1+n}$ (scaled in units of $T$) is given by
\ba
\label{eq:landauwilsonff}
f_{1+n}(\Omega, \Omega^\alpha, \nop^\alpha)
&=& \frac{{\tilde{\eta}^2}}{2 V^n}\overline{\sum_{\hat{k}}}
    \vertex_{\hat{k}} |\Omega_{\hat{k}}|^2
  + \frac{1}{2}\sum_{\alpha=0}^n {\sum_{\bm{p}}}^\prime
    \frac{{\tilde{\lambda}}^\alpha_{\bm{p}}}{T^\alpha}
    |\Omega_{\bm{p}}^\alpha|^2
  + \frac{1}{2}\sum_{\alpha=0}^n \sum_{\bm{p}}
    \frac{J_\pv }{T^\alpha}
    |Q_{d_1 d_2}^\alpha(\bm{p})|^2
\nonumber\\
&&- \ln\bigg\langle
    \exp\bigg(
    \frac{{\tilde{\eta}}^2}{2 V^n}
    \overline{\sum_{\hat{k}}}
     \vertex_{\hat{k}} \Omega_{\hat{k}}\sum_{s=1,-1} e^{-i\hat{k}\cdot {\hat{c}}_{s}}
\nonumber\\
&&\quad + \frac{i}{2}\sum_{\alpha=0}^n {\sum_{\bm{p}}}^{\prime}
    \frac{{\tilde{\lambda}}^\alpha_{\bm{p}}}{T^\alpha}
     \Omega_{\bm{p}}^\alpha \sum_{s=1,-1} e^{-i\bm{p}\cdot {\bm{c}}_{s}^\alpha}
\nonumber\\
&&\quad + \sum_{\alpha=0}^n \sum_{\bm{p}}
    \frac{J_\pv }{T^\alpha} Q_{d_1 d_2}^\alpha(\bm{p})
    e^{-i\bm{p}\cdot {\bm{c}}^\alpha}
    (n_{d_1}^\alpha n_{d_2}^\alpha - D^{-1}\delta_{d_1 d_2})
    \bigg)
  \bigg\rangle_{1,1+n}.
\ea
By expanding $f_{1+n}$ for small values of the auxiliary fields, one obtains a Landau theory in terms of $\Omega$, $\Omega^\alpha$ and $\nop^\alpha$, which are, respectively, the order-parameter fields for the random solidification transition, the crystallization transition, and the isotropic-nematic transition.
As we assume that IGNEs are incompressible, there will be no fluctuations in the density of dimers (which can be enforced by making $\tilde{\lambda}$ to be extremely large), and thus there will be no corresponding instability in the 1RS.  We shall therefore disregard the contribution from $\Omega^{\alpha}$.
The free energy, Eq.~(\ref{eq:landauwilsonff}), then becomes
\be
\label{eq:landauwilsong}
f_{1+n}(\Omega, \nop)
= \frac{{\tilde{\eta}^2}}{2 V^n}\overline{\sum_{\hat{k}}}
    \vertex_{\hat{k}}
    |\Omega_{\hat{k}}|^2
  + \frac{1}{2}\sum_{\alpha=0}^n \sum_{\bm{p}}
    \frac{J_\pv}{T^\alpha}
    |Q_{d_1 d_2}^\alpha(\bm{p})|^2
  - \ln\langle \exp \big( G_1(\Omega) + G_2(\nopc) \big) \rangle_{1,1+n}\,.
\ee
Note that we have introduced the abbreviations
\ba
G_1(\Omega) &\equiv& \frac{{\tilde{\eta}}^2}{2 V^n} \overline{\sum_{\hat{k}}} \vertex_{\hat{k}} \Omega_{\hat{k}} \sum_{s=1,-1} e^{-i\hat{k}\cdot {\hat{c}}_{s}},
\\
G_2(\nopc) &\equiv& \sum_{\alpha=0}^n \sum_{\bm{p}} \frac{J_\pv}{T^\alpha} Q_{d_1 d_2}^\alpha(\bm{p}) e^{-i\bm{p}\cdot {\bm{c}}^\alpha} \left( n_{d_1}^\alpha n_{d_2}^\alpha - \frac{1}{3}\delta_{d_1 d_2} \right),
\ea
and specialized to three spatial dimensions (i.e., $D=3$).

\section{Proof that $\langle \Omega_{\hat{k}}\rangle_f
=
[\big\langle
\omega_{\hat{k}}
\big\rangle_\chi]$ and
$\langle \nop_\pv^\alpha \rangle_f
=
[\big\langle
\bm{q}_\pv^\alpha
\big\rangle_\chi]$
}
\label{app:c}

We now prove Eq.~(\ref{eq:HS}), by deriving a general expression for $[\langle {\bm{q}}_{\pv_1} \rangle \dots \langle {\bm{q}}_{\pv_P} \rangle]$ ($P\in\{ 1, 2, 3, \dots \}$)
in terms of the conjugate fields $\nop_{\pv}$ and the Landau-Wilson free energy $f_{1+n}$ [cf. Eq.~(\ref{eq:landauwilsonf})]:
\ba
\label{eq:HS_proof}
[\langle {\bm{q}}_{\pv_1} \rangle_\chi \dots \langle {\bm{q}}_{\pv_P} \rangle_\chi ]
&=&
\bigg[
\frac{1}{Z_\chi}
\int \prod_{i=1}^{N}
d{\bm{c}}_{i,1}^{\alpha_1}\,d{\bm{c}}_{i,-1}^{\alpha_1}
{\bm{q}}_{\pv_1}^{\alpha_1}
\exp\left( -\left( H_{\rm{nem}} (\{\bm{n}_i^{\alpha_1}\})
   +H_{\rm{ev}} (\{{\bm{c}}_{i,s}^{\alpha_1}\})
   +H_{\rm{xlink}} (\{{\bm{c}}_{i,s}^{\alpha_1}\}) \right)/T^{\alpha_1} \right)
   \nonumber\\
   &&\quad\times
   \prod_{i=1}^N \delta(|{\bm{c}}_{j,1}^{\alpha_1} - {\bm{c}}_{j,-1}^{\alpha_1}| - \ell)\cdots
\nonumber\\
&&\times
\frac{1}{Z_\chi}
\int \prod_{i=1}^{N}
d{\bm{c}}_{i,1}^{\alpha_P}\,d{\bm{c}}_{i,-1}^{\alpha_P}
{\bm{q}}_{\pv_P}^{\alpha_P}
\exp\left( -\left( H_{\rm{nem}} (\{{\bm{n}_i}^{\alpha_P}\})
   +H_{\rm{ev}} (\{{\bm{c}}_{i,s}^{\alpha_P}\})
   +H_{\rm{xlink}} (\{{\bm{c}}_{i,s}^{\alpha_P}\}) \right)/T^{\alpha_P} \right)
\nonumber\\
&&\quad\times
   \prod_{j=1}^N \delta(|{\bm{c}}_{j,1}^{\alpha_P} - {\bm{c}}_{j,-1}^{\alpha_P}| - \ell)
\bigg]
\nonumber\\
&=& \lim_{n\rightarrow 0} \sum_{\chi}{P_\chi}
\frac{1}{Z_\chi^n}
\prod_{\alpha=0}^{n} \int \prod_{i=1}^{N}
d{\bm{c}}_{i,1}^{\alpha}\,d{\bm{c}}_{i,-1}^{\alpha}
{\bm{q}}_{\pv_1}^{\alpha_1}
\dots
{\bm{q}}_{\pv_P}^{\alpha_P}
   \nonumber\\
&&\quad \times
e^{
-\sum_{\gamma=1}^n
    \left(
    H_{\rm{nem}} (\{\bm{n}_i^\gamma\})
  + H_{\rm{ev}} (\{\bm{c}_{i,s}^\gamma\})
  + H_{\rm{xlink}} (\{\bm{c}_{i,s}^\gamma\}) \right)/{T^{\gamma}}
   }
   \prod_{\gamma=0}^n \prod_{j=1}^N \delta(|{\bm{c}}_{j,1}^\gamma - {\bm{c}}_{j,-1}^\gamma| - \ell),
   \nonumber\\
\ea
where in the second step, we have multiplied numerator and denominator by $n-P$ copies of the factor
\be
Z_\chi
=
\int \prod_{i=1}^{N}
d{\bm{c}}_{i,1} \, d{\bm{c}}_{i,-1}
\exp\left(
-\left( H_{\rm{nem}} (\{\bm{n}_i\})
   + H_{\rm{ev}} (\{\bm{c}_{i,s}\})
   + H_{\rm{xlink}} (\{\bm{c}_{i_e,s_e}\})
   \right)/T^{\gamma}
   \right)
   \prod_{i=1}^N \delta(|{\bm{c}}_{i,1} - {\bm{c}}_{i,-1}| - \ell).
\ee
This produces a factor of $Z_\chi^n$ in the denominator, which goes to unity once the replica limit is taken.
We now perform the average over realizations of quenched disorder:
\ba
[\langle {\bm{q}}_{\pv_1} \rangle_\chi \dots \langle {\bm{q}}_{\pv_P} \rangle_\chi ]
&=&
\lim_{n\rightarrow 0} \sum_\chi \frac{1}{M! Z_{\rm{liq}}^{0}} \left( \frac{V\eta^2}{2 (2\pi b^2)^{D/2} N} \right)^M
e^{H_{\rm{norm}} (\{ \bm{c}_{i,s}^{0} \}) /T^0 }
\prod_{\gamma=0}^n
\prod_{i=1}^N \delta(|{\bm{c}}_{i_1}^\gamma - {\bm{c}}_{i_2}^\gamma| - \ell)
\nonumber\\
&&\times
\int \prod_{i=1}^{N}
d{\bm{c}}_{i,1}^{\alpha_1}\,d{\bm{c}}_{i,-1}^{\alpha_1}
{\bm{q}}_{\pv_1}^{\alpha_1} \dots {\bm{q}}_{\pv_P}^{\alpha_P}
\exp\left( -\sum_{\gamma=0}^n
   \left(
   H_{\rm{nem}} (\{\bm{n}_i^\gamma\})
   + H_{\rm{ev}} (\{\bm{c}_{i,s}^\gamma\})
   + H_{\rm{xlink}} (\{\bm{c}_{i_e,s_e}^\gamma\})
   \right)/T^{\gamma}
   \right)
\nonumber\\
&=&\lim_{n\rightarrow 0} \frac{1}{Z_{\rm{liq}}^{0}}
\prod_{\gamma=0}^n
\int \prod_{i=1}^{N}
d{\bm{c}}_{i,1}^{\alpha_1}\,d{\bm{c}}_{i,-1}^{\alpha_1}
{\bm{q}}_{\pv_1}^{\alpha_1} \dots {\bm{q}}_{\pv_P}^{\alpha_P}
e^{H_{\rm{norm}} (\{ \bm{c}_{i,s}^{0} \}) /T^0 }
\nonumber\\
&&\times
\exp\left(
\sum_{\gamma=0}^n
\left(
\frac{1}{2T^{\gamma}}
\sum_{i,j=1}^{N}
  J_{ij} \big( ({\bm{n}}_i^\gamma\cdot {\bm{n}}_j^\gamma)^2 - \frac{1}{D} \big)
+ \frac{\lambda}{2T^\gamma}
  \sum_{i,j=1}^{N}\sum_{s,t=1,-1}
  \delta({\bm{c}}_{i,s}^\gamma-{\bm{c}}_{j,t}^\gamma)
\right)
\right)
\nonumber\\
&&\times
\exp
\left(
\frac{V\eta^2}{2 (2\pi b^2)^{D/2} N}
\sum_{i,j=1}^{N}
\sum_{s,t=1,-1}
e^{-\frac{1}{2b^2}\sum_{\gamma=0}^{n}|{\bm{c}}_{i,s}^\gamma-{\bm{c}}_{j,t}^\gamma|^2}
\right)
\prod_{i=1}^N \delta(|{\bm{c}}_{i,1}^\gamma - {\bm{c}}_{i,-1}^\gamma| - \ell)
\nonumber\\
&=&
\lim_{n\rightarrow 0}
\frac{B_n \, (4\pi V \ell^2)^{N}}{Z_{\rm{liq}}^{0}}
\Bigg\langle
{\bm{q}}_{\pv_1}^{\alpha_1}
\dots
{\bm{q}}_{\pv_P}^{\alpha_P}
\,
e^{H_{\rm{norm}} (\{ \bm{c}_{i,s}^{0} \}) /T^0}
\nonumber\\
&&\times
\exp\left(
\sum_{\alpha=0}^n {\sum_{\bm{p}}}^{\prime}
\left(
\frac{N J_\pv}{2 T^\alpha}
\{{\bm{q}}_{\pv}^{\alpha}\,{\bm{q}}_{-\pv}^{\alpha}\}
%\right)
%\nonumber\\
%&&\quad\times
- \frac{N {\tilde{\lambda}}_{\bm{p}}}{2 T^\alpha}
|{\omega}_{\bm{p}}^\alpha|^2
\right)
+ \frac{N {\tilde{\eta}}^2}{2} \overline{\sum\limits_{\hat{k}}} \vertex_{\hat{k}} |{\omega}_{\hat{k}}|^2
\right)
\Bigg\rangle_{N,1+n},
\nonumber\\
\ea
where $B_{n}$ is a constant.  Here, $\{ \alpha_{1}, \ldots, \alpha_{P} \} \subseteq \{ 1,\ldots,n \}$, as we are interested in nematic correlators in the measurement ensemble.
In the second step, we have summed over $M$ to obtain an exponential function. To proceed further, we note that the factors of ${\bm{q}}_{\pv}^\alpha$ can be generated by introducing, into the replica partition function $Z_{1+n}$, a source field $\source_{\pv}^\alpha$ that is linearly coupled to ${\bm{q}}_{\pv}^\alpha$; we denote the resulting replica partition function by the symbol $Z_{1+n}[\source]$.  By functionally differentiating $Z_{1+n}[\source]$ with respect to $\source_{\pv}^\alpha$, one can recover the correlators of ${\bm{q}}_{\pv}^\alpha$.  At the end of the computation, one has to take the source limit $\source\rightarrow 0$.  Following this procedure, one obtains
\ba
\label{eq:Qcorrexact}
[ \langle {\bm{q}}_{\pv_1} \rangle_\chi \dots \langle {\bm{q}}_{\pv_P}  \rangle_\chi ]
&=&
\lim_{n\rightarrow 0}\lim_{\source\rightarrow 0}
\frac{(-)^P B_n \, (4\pi V \ell^2)^{N}}{Z_{\rm{liq}}^{0}}
e^{H_{\rm{norm}} (\{ \bm{c}_{i,s}^{0} \}) / T^0 }
\bigg\langle
\frac{\delta}{\delta\source_{\pv_1}^{\alpha_1}}
\dots
\frac{\delta}{\delta\source_{\pv_P}^{\alpha_P}}
e^{\frac{N}{2}\sum_{\alpha=0}^n \sum_{\bm{p}}
    \frac{J_\pv}{T^\alpha}
    \{ {\bm{q}}_{\pv}^{\alpha} \, {\bm{q}}_{-\pv}^{\alpha} \} }
 \nonumber\\
 &&\quad\times
    e^{ -\sum_{\alpha=1}^n \sum_{\bm{p}} \source_{\pv}^\alpha {\bm{q}}_{-\pv}^\alpha
    -\frac{N}{2}\sum_{\alpha=0}^n {\sum_{\bm{p}}}^\prime \frac{{\tilde{\lambda}}_{\bm{p}}}{T^\alpha} |{\widetilde{\Omega}}_{\bm{p}}^{\alpha}|^2 + \frac{N}{2}{\tilde{\eta}}^2 \overline{\sum\limits_{\hat{k}}} \vertex_{\hat{k}} |{\widetilde{\Omega}}_{\hat{k}}|^2}
\bigg\rangle_{N,1+n}
\nonumber\\
&=&
\lim_{n\rightarrow 0}\lim_{\source\rightarrow 0}
\frac{(-)^P B_n \, (4\pi V \ell^2)^{N}}{Z_1}
e^{H_{\rm{norm}} (\{ \bm{c}_{i,s}^{0} \}) / T^0 }
\bigg\langle
\frac{\delta}{\delta\source_{\pv_1}^{\alpha_1}} \dots \frac{\delta}{\delta\source_{\pv_P}^{\alpha_P}}
e^{\frac{N}{2}\sum_{\pv} \frac{J_{\pv}}{T^{0}} \{ {\bm{q}}_{\pv}^\alpha \, {\bm{q}}_{-\pv}^\alpha \}
  }
\nonumber\\
&&\quad\times
e^{\frac{N}{2}\sum_{\alpha=1}^n \sum_{\pv}
\frac{J_{\pv}}{T^\alpha} \big\{ \big( {\bm{q}}_{\pv}^\alpha -\frac{T^\alpha}{N J_{\pv}}\source_{\pv}^{\alpha} \big)
\big( {\bm{q}}_{-\pv}^\alpha -\frac{T^\alpha}{N J_{-\pv}}\source_{-\pv}^{\alpha} \big) \big\}
  }
  \nonumber\\
&&\quad\times
e^{-\sum_{\alpha=1}^n \sum_{\pv} \frac{(T^\alpha)^{2}}{ 2N J_{\pv} \, J_{-\pv} }
  \{ \source_{\pv}^\alpha \, \source_{-\pv}^\alpha \}
  - N \sum_{\alpha=0}^n{\sum_{\bm{p}}}^\prime \frac{{\tilde{\lambda}}_{\bm{p}}}{2T^\alpha} |\omega_{\bm{p}}^{\alpha}|^2 + \frac{N}{2}{\tilde{\eta}}^2 \sum_{\alpha=0}^n \overline{\sum\limits_{\hat{k}}} \vertex_{\hat{k}} |\omega_{\hat{k}}|^2}
\bigg\rangle_{N,1+n}.
\ea
In performing this calculation, we have completed the square in ${\bm{q}}_{\pv}^\alpha$, generating extra quadratic terms $\frac{(T^\alpha)^2}{2N J_{\pv}^{\alpha} \, J_{-\pv}^{\alpha} } \{ \source_{\pv}^{\alpha} \, \source_{-\pv}^{\alpha} \}$.
The Hubbard-Stratonovich transformation can now be performed, with ${\bm{q}}_{\pv}^\alpha -\frac{T^\alpha}{N J_{\pv}}\source_{\pv}^\alpha$, ${\omega}_{\pv}^\alpha$, and ${\omega}_{\hat{k}}$ being the auxiliary fields.
Making use of the following relation (which one can prove)
\ba
&&B_n \bigg\langle
e^{ N \sum_{\pv} \frac{J_{\pv}}{2 T^{0}} \{ {\bm{q}}_{\pv}^\alpha \, {\bm{q}}_{-\pv}^\alpha \}
+ N \sum_{\alpha=1}^n \sum_{\pv} \frac{J_{\pv}}{2 T^\alpha}
    \{ {\bm{q}}_{\pv}^\alpha -\frac{T^\alpha}{N J_{\pv}}\source_{\pv}^\alpha
    \, {\bm{q}}_{-\pv}^\alpha -\frac{T^\alpha}{N J_{-\pv}}\source_{-\pv}^\alpha \}
    }
    \nonumber\\
    &&\quad\times
e^{
- N \sum_{\alpha=0}^n \sum_{\pv}^{\prime} \frac{{\tilde{\lambda}}_{\pv}}{2 T^{\alpha}}
   |{\omega}_{\pv}^{\alpha}|^2
   + \frac{N}{2}{\tilde{\eta}}^2 \overline{\sum\limits_{\hat{k}}} \vertex_{\hat{k}}
   |{\omega}_{\hat{k}}|^{2}}
\bigg\rangle_{N,1+n}
\nonumber\\
&&\qquad=
\int \mathcal{D}\Omega \prod_{\alpha=0}^n \mathcal{D}\Omega^{\alpha} \mathcal{D}\nop^{\alpha}
 e^{-Nf_{1+n} (\Omega_{\hat{k}}, \Omega^\alpha, \nop^\alpha)
    - \sum_{\alpha=1}^n \sum_{\pv} \source_{\pv}^\alpha \nop_{-\pv}^\alpha},
\ea
it follows that
\ba
\label{eq:Qcorreff}
[\langle {\bm{q}}_{\pv_1} \rangle_\chi \cdots \langle {\bm{q}}_{\pv_P} \rangle_\chi ]
&=&
\lim_{n\rightarrow 0}\lim_{\source\rightarrow 0}
\frac{(-)^P (4\pi V \ell^2)^{N}}{Z_{\rm{liq}}^{0}}
\frac{\delta}{\delta\source_{\pv_1}^{\alpha_1}} \cdots \frac{\delta}{\delta\source_{\pv_P}^{\alpha_P}}
\nonumber\\
&&\quad \times
\int \mathcal{D}\Omega \prod_{\alpha=0}^n \mathcal{D}\Omega^\alpha \mathcal{D} \nop^\alpha
e^{-Nf_{1+n} (\Omega, \Omega^\alpha, \nop^\alpha)
    - \sum_{\alpha=1}^n \sum_{\pv} \source_{\pv}^\alpha \nop_{-\pv}^\alpha
    -\sum_{\alpha=1}^n \sum_{\pv} \frac{(T^\alpha)^{2}}{2N J_{\pv} \, J_{-\pv}}
    \{ \source_{\pv}^\alpha \, \source_{-\pv}^\alpha \} }
\nonumber\\
&=&
\lim_{n\rightarrow 0}
\frac{Z_{1+n}}{Z_{\rm{liq}}^{0}}
\frac{
\int \mathcal{D}\Omega \prod_{\alpha=0}^n \mathcal{D}\Omega^\alpha \mathcal{D} \nop^\alpha \,
\big( \nop_{\pv_1}^{\alpha_1} \cdots \nop_{\pv_P}^{\alpha_P} \big)
e^{-Nf_{1+n} (\Omega, \Omega^\alpha, \nop^\alpha) }
}
{
\int \mathcal{D}\Omega \prod_{\alpha=0}^n \mathcal{D} \Omega^\alpha \mathcal{D} \nop^\alpha
e^{-Nf_{1+n} (\Omega, \Omega^\alpha, \nop^\alpha) }
}
\nonumber\\
&\equiv&
\langle \nop_{\pv_1}^{\alpha_1} \cdots \nop_{\pv_P}^{\alpha_P} \rangle_f,
\ea
as we were aiming to establish.
On going from the second to the third line of Eq.~(\ref{eq:Qcorreff}), we have multiplied numerator and denominator by $Z_{1+n}$, and we have also taken the limit $n \rightarrow 0$, which implies that $Z_{1+n} \rightarrow Z_{\rm{liq}}^0$ [as one can see from Eq.~(\ref{eq:pre_HS}) by taking the limit $n\rightarrow 0$].
For the case $P=2$, we recover Eq.~(\ref{eq:HS}).
%On going from Eq.~(\ref{eq:Qcorrexact}) to Eq.~(\ref{eq:Qcorreff}), we have performed the Hubbard-Stratonovich transformation with an extra linear coupling term $\source^\alpha \nop^\alpha$.
%This originates in the extra part $\frac{T^\alpha}{N J_{\pv}}\source_{\pv}^\alpha$ in ${\bm{q}}_{\pv}^\alpha (\bm{p})-\frac{T^\alpha}{N J_{\pv}}\source_{\pv}^\alpha$ which we took to be our collective field. This linear coupling term was originally in the log trace but can be taken out of the log trace straight-forwardly as it does not involve the microscopic statistical variables $\bm{c}_{i,s}$.
%The Hubbard-Stratonovich transformation is implemented after we have taken out the functional derivatives with respect to the source field from the microscopic trace $\langle \ldots \rangle_{N,1+n}$.  After the Hubbard-Stratonovich transformation has been done, we are left with a quantity that the source-field functional derivatives operate on, generating $\nop_{\pv_1}^{\alpha_1} \dots \nop_{\pv_P}^{\alpha_P}$ within the field-theoretic trace.  After this has been done, we take the limit $\source\rightarrow 0$ which gets rid of the quadratic and linear terms in $\source$.

We can also derive Eq.~(\ref{eq:correlrel}) by using Eq.~(\ref{eq:HS_proof}) and taking the replica limit:
\ba
[\langle {\bm{q}}_{\pv_1} \rangle \cdots \langle {\bm{q}}_{\pv_P} \rangle]
&=&
\lim_{n\rightarrow 0}
\frac{
\int \mathcal{D}\Omega \prod_{\alpha=0}^n \mathcal{D}\Omega^\alpha \mathcal{D} \nop^\alpha \,
\big( \nop_{\pv_1}^{\alpha_1} \cdots \nop_{\pv_P}^{\alpha_P} \big)
e^{-Nf_{1+n} (\Omega, \Omega^\alpha, \nop^\alpha) }
}
{
\int \mathcal{D}\Omega \prod_{\alpha=0}^n \mathcal{D} \Omega^\alpha \mathcal{D} \nop^\alpha
e^{-Nf_{1+n} (\Omega, \Omega^\alpha, \nop^\alpha) }
}
\nonumber\\
&\approx&
\lim_{n\rightarrow 0}
\frac{
\int \prod_{\alpha=1}^n \mathcal{D} \nop^\alpha \,
\big( \nop_{\pv_1}^{\alpha_1} \cdots \nop_{\pv_P}^{\alpha_P} \big)
e^{-N H_{\rm{eff}} (\nop^\alpha) }
}
{
\int \prod_{\alpha=1}^n \mathcal{D} \nop^\alpha
e^{-N H_{\rm{eff}} (\nop^\alpha) }
}
\nonumber\\
&\equiv&
\langle\!\langle \nop_{\pv_1}^{\alpha_1} \cdots \nop_{\pv_P}^{\alpha_P} \rangle\!\rangle,
\ea
where in the second step, we have approximated the functional integrals by the values that are stationary with respect to $\Omega$ and exclude density fluctuations $\Omega^\alpha$.

\section{Terms in the Landau-Wilson expansion}
\label{app:d}

In this section we expand the log trace term in Eq.~(\ref{eq:landauwilsong}) for small $\Omega$ and $\nop^\alpha$. This gives us
\ba
\label{eq:log-trace-expansion}
f_{1+n}(\Omega, \nop)
 &=& \frac{{\tilde{\eta}^2}}{2 V^n}
 \overline{\sum\limits_{\hat{k}}}
 \vertex_{\hat{k}} |\Omega_{\hat{k}}|^2
    -\frac{1}{2}\langle G_1(\Omega)^2 \rangle_{1,1+n}
    -\frac{1}{6}\langle G_1(\Omega)^3 \rangle_{1,1+n}
\nonumber\\
 && \quad + \frac{1}{2}\sum_{\alpha=0}^n \sum_{\bm{p}} \frac{J_\pv}{T^\alpha} |Q_{ab}^\alpha(\bm{p})|^2
    -\frac{1}{2}\langle G_2(\nopc)^2 \rangle_{1,1+n}
    -\frac{1}{6}\langle G_2(\nopc)^3 \rangle_{1,1+n}
    -\frac{1}{24}\langle G_2(\nopc)^4 \rangle_{1,1+n}
\nonumber\\
 && \quad
   -\frac{1}{2}\langle G_1(\Omega)^2 G_2(\nopc) \rangle_{1,1+n}
   -\frac{1}{2}\langle G_1(\Omega) G_2(\nopc)^2 \rangle_{1,1+n}
\nonumber\\
 && \quad
   -\frac{1}{4}\big(\langle G_1(\Omega)^2 G_2(\nopc)^2 \rangle_{1,1+n}
   -\langle G_1(\Omega)^2 \rangle_{1,1+n} \langle G_2(\nopc)^2 \rangle_{1,1+n} \big).
\ea

\subsection{Terms proportional to $\Omega\Omega$}

First, we compute the quadratic term for the vulcanization part of the Landau theory:
\ba
\label{eq:G1OmegaSquare}
\langle G_1(\Omega)^2 \rangle_{1,1+n} &=& \frac{{\tilde{\eta}}^4}{4}
{\overline{\sum}}_{\hat{k}_1,\hat{k_2}}
%\sum_{{\hat{k}}_1,{\hat{k}}_2\in HRS}
    \vertex_{{\hat{k}}_1}\vertex_{{\hat{k}}_2} \Omega_{{\hat{k}}_1}\Omega_{{\hat{k}}_2}
    \sum_{s,t}\Big\langle
    e^{-i\big( {\hat{k}}_1\cdot {\hat{c}}_{s}
       + {\hat{k}}_2\cdot {\hat{c}}_{t} \big)} \Big\rangle_{1,1+n}
    \nonumber\\
&=& \frac{{\tilde{\eta}}^4}{2}
%    \overline{\sum\limits_{\hat{k}}}
    {\overline{\sum}}_{\hat{k}}
    \vertex_{\hat{k}}^2 \Omega_{\hat{k}}\Omega_{-\hat{k}}
    \bigg( 1+ \prod_{\alpha=0}^n \frac{\sin(k^\alpha \ell)}{k^\alpha \ell} \bigg).
\ea
At length-scales much larger than $\ell$ or $b$, we can approximate the above result by
\be
\label{eq:G1OmegaSquare_approx}
\langle G_1(\Omega)^2 \rangle_{1,1+n}
\approx
{\tilde{\eta}}^4 {\overline{\sum}}_{\hat{k}}
\Big( 1 - \big( b^2 + \frac{\ell^2}{12} \big) |\hat{k}|^2 \Big) |\Omega_{\hat{k}}|^2.
\ee

\subsection{Terms proportional to $\Omega\Omega\Omega$}

We next compute the cubic term for the vulcanization part of the Landau theory:
\ba
\label{eq:G1OmegaCube}
\langle G_1(\Omega)^3 \rangle_{1,1+n}
&=&
    \frac{{\tilde{\eta}}^6}{8}
    {\overline{\sum}}_{\hat{k}_1,\hat{k}_2,\hat{k}_3}
    %\sum_{{\hat{k}}_1,{\hat{k}}_2,{\hat{k}}_3\in HRS}
    \vertex_{{\hat{k}}_1} \vertex_{{\hat{k}}_2} \vertex_{{\hat{k}}_3}
    \Omega_{{\hat{k}}_1}\Omega_{{\hat{k}}_2}\Omega_{{\hat{k}}_3}
\nonumber\\
&&\times
    \sum_{s_1,s_2,s_3}
    \Big\langle
    e^{-i\big( {\hat{k}}_1\cdot {\hat{c}}_{s_1} + {\hat{k}}_2\cdot {\hat{c}}_{s_2} + {\hat{k}}_3\cdot {\hat{c}}_{s_3} \big)}
    \Big\rangle_{1,1+n}
    \nonumber\\
&=& \frac{{\tilde{\eta}}^6}{8}
    {\overline{\sum}}_{\hat{k}_1,\hat{k}_2,\hat{k}_3}
    %\sum_{{\hat{k}}_1, {\hat{k}}_2 \in HRS}
    \delta_{\hat{k}_1+\hat{k}_2+\hat{k}_3,\hat{0}}
    \vertex_{{\hat{k}}_1} \vertex_{{\hat{k}}_2} \vertex_{{\hat{k}}_3}
    \Omega_{{\hat{k}}_1}\Omega_{{\hat{k}}_2}\Omega_{{\hat{k}}_3}
\nonumber\\
&&\times
    \bigg(
    2+ 3\prod_{\alpha=0}^n \frac{\sin(|-\bm{k}_1^\alpha + \bm{k}_2^\alpha + \bm{k}_3^\alpha| \ell/2)}{|-\bm{k}_1^\alpha + \bm{k}_2^\alpha + \bm{k}_3^\alpha| \ell/2}
    + 3\prod_{\alpha=0}^n \frac{\sin(|-\bm{k}_1^\alpha - \bm{k}_2^\alpha + \bm{k}_3^\alpha| \ell/2)}{|-\bm{k}_1^\alpha - \bm{k}_2^\alpha + \bm{k}_3^\alpha| \ell/2}
    \bigg).
\ea
At length-scales much larger than $\ell$ and $b$, we can approximate this result by
\be
\label{eq:G1OmegaCube_approx}
\langle G_1(\Omega)^3 \rangle_{1,1+n}
\approx
{\tilde{\eta}}^6{\overline{\sum}}_{\hat{k}_1,\hat{k}_2,\hat{k}_3}
\delta_{\hat{k}_1+\hat{k}_2+\hat{k}_3,\hat{0}}
\Omega_{{\hat{k}}_1}\Omega_{{\hat{k}}_2}\Omega_{{\hat{k}}_3}.
\ee

\subsection{Terms proportional to $\nop^\alpha\nop^\alpha$}

Next, we compute the quadratic terms from the nematic part of the Landau theory:
\ba
\langle G_2(\nopc)^2 \rangle_{1,1+n}
&=& \sum_{\alpha,\beta}
    \sum_{{\bm{p}},{\bm{q}}} \sum_{s,t=1,-1}
    \frac{J_\pv J_\qv}{(T^\alpha)^2}
    Q_{d_1 d_2}^\alpha (\bm{p}) Q_{d_3 d_4}^\beta (\bm{q})
    \nonumber\\
&&\quad \times
    \Big\langle e^{-i ( \bm{p}\cdot {\bm{c}}^\alpha + \bm{q}\cdot {\bm{c}}^\beta )}
    \big( n_{d_1}^\alpha n_{d_2}^\alpha - \frac{1}{3}\delta_{d_1 d_2} \big)
    \big( n_{d_3}^\beta n_{d_4}^\beta - \frac{1}{3}\delta_{d_3 d_4} \big)
    \Big\rangle_{1,1+n}.
\ea
Note that this term vanishes for $\alpha\neq\beta$.  For $\alpha=\beta$, we find that
\ba
\label{eq:G2Qsquare}
\langle G_2(\nopc)^2 \rangle_{1,1+n}
&=&  \frac{1}{4\pi} \sum_\alpha \sum_{\bm{p}}
     \frac{|J_\pv|^2}{(T^\alpha)^{2}}
         Q_{d_1 d_2}^\alpha (\bm{p})    Q_{d_3 d_4}^\alpha (-\bm{p})
         \nonumber\\
  &&\quad\times
  \int d{\bm{n}}^\alpha \big( n_{d_1}^\alpha n_{d_2}^\alpha n_{d_3}^\alpha n_{d_4}^\alpha
  -\frac{1}{3} n_{d_1}^\alpha n_{d_2}^\alpha \delta_{d_3 d_4}
  -\frac{1}{3} n_{d_3}^\alpha n_{d_4}^\alpha \delta_{d_1 d_2}
  +\frac{1}{9} \delta_{d_1 d_2} \delta_{d_3 d_4} \big)
\nonumber\\
&=& \frac{1}{15} \sum_\alpha \sum_{\bm{p}}
    \frac{|J^\alpha(\bm{p})|^2}{(T^\alpha)^{2}}
         Q_{d_1 d_2}^\alpha (\bm{p})    Q_{d_3 d_4}^\alpha (-\bm{p})
         \big( \delta_{d_1 d_3}\delta_{d_2 d_4}+\delta_{d_1 d_4}\delta_{d_2 d_3}-\frac{2}{3}\delta_{d_1 d_2}\delta_{d_3 d_4} \big).
\ea
Here, the notation $\int d\bm{n}$ denotes integration over the unit sphere, and we have used the equalities (valid for $D=3$): $\int d\bm{n} \, n_{d_1} n_{d_2} = \frac{4\pi}{3}\delta_{d_1 d_2}$ and $\int d\bm{n} n_{d_1} n_{d_2} n_{d_3} n_{d_4} = \frac{4\pi}{15} (\delta_{d_1 d_2}\delta_{d_3 d_4}+\delta_{d_1 d_3}\delta_{d_2 d_4}+\delta_{d_1 d_4}\delta_{d_2 d_3})$.

\subsection{Coupling terms}

\subsubsection{Terms proportional to $\Omega \nop^\alpha\nop^\beta$ and $\Omega\Omega \nop^\alpha$}

The terms that couple the nematic order parameter to the vulcanization order parameter are the ones that give rise to the physics of nematic elastomers.  At cubic order in $\Omega$ and $\nop$ (i.e., $\Omega \nop^\alpha\nop^\beta$ or $\Omega\Omega \nop^\alpha$) there are two such terms.  They were computed in Ref.~\cite{XPMGZ-NE-VT} and are given by
\begin{subequations}
\ba
\label{eq:OmegaQQlongwavelength}
\langle G_1(\Omega) G_2(\nopc)^2 \rangle_{1,1+n}
&=& \frac{{\tilde{\eta}}^2}{800}\sum_{\alpha\neq\beta}^n \sum_{\bm{p},\bm{q}}
 \vertex_{-{\bm{p}}{\hat{\epsilon}}^\alpha-{\bm{q}}{\hat{\epsilon}}^\beta}
 \frac{J_\pv\, J_\qv}{T^\alpha\, T^\beta}
 \Omega_{-{\bm{p}}{\hat{\epsilon}}^\alpha-{\bm{q}}{\hat{\epsilon}}^\beta}
 Q_{d_1 d_2}^\alpha(\bm{p}) Q_{d_3 d_4}^\beta(\bm{q})
 p_{d_1} p_{d_2} q_{d_3} q_{d_4},
 \nonumber\\
 &&
\\
\label{eq:OmegaOmegaQlongwavelength}
\langle G_1(\Omega)^2 G_2(\nopc) \rangle_{1,1+n}
&=&\frac{{\tilde{\eta}}^4 \ell^2}{5}\sum_{\alpha=0}^n \sum_{\bm{p}}\overline{\sum_{\hat{k}}}
 \vertex_{\hat{k}} \vertex_{-\hat{k}-{\bm{p}}{\hat{\epsilon}}^\alpha}
 \frac{J_\pv}{T^\alpha}
 \Omega_{\hat{k}}\Omega_{-\hat{k}-{\bm{p}}{\hat{\epsilon}}^\alpha}
 Q_{d_1 d_2}^\alpha(\bm{p})
 \nonumber\\
&&\times
 \big( p_{d_1} p_{d_2} + (k^\alpha + \frac{1}{2}p)_{d_1} (k^\alpha + \frac{1}{2}p)_{d_2} \big).
\ea
\end{subequations}

\subsubsection{Terms proportional to $\Omega\Omega\nop^\alpha\nop^\beta$}

%For reasons to be discussed in Appendix~\ref{app:f}, the foregoing coupling terms do not give the leading-order contribution to the novel liquid crystalline behavior that our theory predicts for IGNEs.  This is why we further consider the higher-order terms that are proportional to $\Omega\Omega\nop^\alpha\nop^\beta$.
Terms proportional to $\Omega\Omega\nop^\alpha\nop^\beta$ arise in two forms: connected and disconnected.  We first consider the disconnected type, which is given by $\langle G_1(\Omega)^2 \rangle \langle G_2(\nopc)^2 \rangle$.  By using the expressions obtained for $\langle G_1(\Omega)^2 \rangle_{1,1+n}$ in Eq.~(\ref{eq:G1OmegaSquare}) and $\langle G_2(\nopc)^2 \rangle_{1,1+n}$ in Eq.~(\ref{eq:G2Qsquare}) we obtain
\be
\label{eq:OmegaOmegaQQdisconnected}
\langle G_1(\Omega)^2 \rangle_{1,1+n} \langle G_2(\nopc)^2 \rangle_{1,1+n}
= \frac{8 {\tilde{\eta}}^4}{15}\sum_{\alpha=0}^n
    \overline{\sum\limits_{\hat{k}}}
    %\sum_{\hat{k}\in HRS}
    \sum_{\bm{p}} \vertex_{\hat{k}}^2 \left( \frac{J_\pv}{T^\alpha} \right)^2
    \Omega_{\hat{k}}\Omega_{-\hat{k}}
    Q_{d_1 d_2}^\alpha(\bm{p})Q_{d_1 d_2}^\alpha(-\bm{p}).
\ee
Next, we consider the connected type, $\langle G_1(\Omega)^2 G_2(\nopc)^2 \rangle_{1,1+n}$:
\ba
\label{eq:G1SquareG2Square}
\langle G_1(\Omega)^2 G_2(\nopc)^2 \rangle_{1,1+n}
&=& {\tilde{\eta}}^4
    \sum_{\alpha, \beta=0}^{n} {\overline{\sum}}_{{\hat{k}}_1,{\hat{k}}_2}
    \sum_{\bm{p}, \bm{q}} \sum_{s,t=1,-1}
     \vertex_{{\hat{k}}_1} \vertex_{{\hat{k}}_2}
     \Omega_{{\hat{k}}_1} \Omega_{{\hat{k}}_2}
     \frac{J_\pv J_\qv}{T^\alpha\, T^\beta}
     Q_{d_1 d_2}^\alpha (\bm{p}) Q_{d_3 d_4}^\beta (\bm{q})
     \nonumber\\
   &&\times
   \Big\langle e^{-i\sum_{\gamma=0}^n \big({\bm{k}}_1^\gamma\cdot {\bm{c}}_{s}^{\gamma}
   + {\bm{k}}_2^\gamma\cdot {\bm{c}}_{t}^{\gamma} \big)
   -i\bm{p}\cdot {\bm{c}}^\alpha -i\bm{q}\cdot {\bm{c}}^\beta }
   \nonumber\\
   &&\times
   \big( n_{d_1}^\alpha n_{d_2}^\alpha - \frac{1}{3}\delta_{d_1 d_2} \big)
   \big( n_{d_3}^\beta n_{d_4}^\beta - \frac{1}{3}\delta_{d_3 d_4} \big)
\Big\rangle_{1,1+n}.
\ea
Note that in this expression we have to evaluate terms such as $\langle e^{i\pv\cdot \bm{c}_s} \rangle_{1,1+n}$, $\langle n_{d_1} n_{d_2} \, e^{i\pv\cdot \bm{c}_s} \rangle_{1,1+n}$ and $\langle n_{d_1} n_{d_2} n_{d_3} n_{d_4} \, e^{i\pv\cdot \bm{c}_s} \rangle_{1,1+n}$.  We recall the definitions
\begin{subequations}
\label{eq:definitions}
\be
\bm{c}_s \equiv \bm{c}+(s\ell/2)\bm{n},
\ee
\ba
\langle \cdots \rangle_{1,1+n}
&\equiv&
\prod_{\alpha=0}^n
\int \frac{d{\bm{c}}_{1}^\alpha d{\bm{c}}_{-1}^\alpha}{4\pi V \ell^2} \delta(|\bm{c}_{1}^{\alpha} - \bm{c}_{-1}^{\alpha}| - \ell)
\nonumber\\
&=&\int \prod_{\alpha=0}^{n} \int \frac{d\bm{c}^\alpha\, d\bm{u}^\alpha}{4\pi V\ell^2}\delta(|\bm{u}|-\ell)
\nonumber\\
&=&\int \prod_{\alpha=0}^{n} \int \frac{d\bm{c}^\alpha\, d\bm{n}^\alpha\, d{u}^\alpha\, |\bm{u}^\alpha|^2}{4\pi V\ell^2}\delta(|\bm{u}|-\ell)
\nonumber\\
&=& \int \prod_{\alpha=0}^{n} \int \frac{d\bm{c}^\alpha\, d\bm{n}^\alpha}{4\pi V}.
\ea
\end{subequations}
Here, we have decomposed coordinates into center of mass $\bm{c}=(\bm{c}_{1} + \bm{c}_{-1})/2$ and relative $\bm{u}=\bm{c}_{1} - \bm{c}_{-1}$, and have defined the integral over the unit sphere as
\be
\int \frac{d\bm{n}}{4\pi} \equiv \frac{1}{2}\int_{-1}^{1} d\cos\theta,
\ee
where $\theta$ is an angle that we can define with respect to the direction of the wave-vector.  The following formulas are essential for the next step in our calculation:
\begin{subequations}
\label{eq:operator_formulas}
\be
\int \frac{d\bm{n}}{4\pi} e^{i\pv\cdot \bm{n}} = \frac{\sin |\pv|}{|\pv|},
\ee
\ba
\int \frac{d\bm{n}}{4\pi}
e^{i\bm{p}\cdot \bm{n}}
\big( {\bm{n}}_{d_1} {\bm{n}}_{d_2} - \frac{1}{3}\delta_{d_1 d_2} \big)
&=&
-\int \frac{d\bm{n}}{4\pi}
\left( \frac{\partial}{\partial p_{d_1}}\frac{\partial}{\partial p_{d_2}} + \frac{1}{3}\delta_{d_1 d_2} \right)
    e^{i\bm{p}\cdot \bm{n}}
\nonumber\\
&=&
-\frac{1}{2} \left( \frac{\partial}{\partial p_{d_1}}\frac{\partial}{\partial p_{d_2}} + \frac{1}{3}\delta_{d_1 d_2} \right)
\int_{-1}^{1} d(\cos\theta) e^{i|\pv|\cos\theta}
\nonumber\\
&=&
-\left( \frac{\partial}{\partial p_{d_1}}\frac{\partial}{\partial p_{d_2}} + \frac{1}{3}\delta_{d_1 d_2} \right)
\frac{\sin|\pv|}{|\pv|}
\nonumber\\
&=&
\left( \left( \frac{1}{|\pv|^3}-\frac{3}{|\pv|^5} \right) \sin|\pv| + \frac{3}{|\pv|^4} \cos|\pv| \right)
\left( p_{d_1} p_{d_2} - \frac{|\pv|^2}{3} \delta_{d_1 d_2} \right)
\nonumber\\
&\approx&
-\frac{1}{15}\big( \frac{p_{d_1}}{|\pv|} \frac{p_{d_2}}{|\pv|} - \frac{1}{3}\delta_{d_1 d_2} \big),
\ea
\ba
\int \frac{d\bm{n}}{4\pi} n_{d_1}n_{d_2}n_{d_3}n_{d_4}e^{i\bm{p}\cdot \bm{n}}
&=&
\Big( -\frac{\sin|\pv|}{|\pv|^3} - \frac{3\cos|\pv|}{|\pv|^4} + \frac{3\sin|\pv|}{|\pv|^5} \Big)
\big( \delta_{d_1 d_2}\delta_{d_3 d_4} + \delta_{d_1 d_3}\delta_{d_2 d_4} + \delta_{d_1 d_4}\delta_{d_2 d_3} \big)
\nonumber\\
&&+\Big( -\frac{\cos|\pv|}{|\pv|^4} + \frac{6\sin|\pv|}{|\pv|^5} + \frac{15\cos|\pv|}{|\pv|^6} - \frac{15\sin|\pv|}{|\pv|^7} \Big)
\big( \delta_{d_1 d_2} p_{d_3} p_{d_4} + \delta_{d_1 d_3} p_{d_2} p_{d_4}
\nonumber\\
&&\qquad+ \delta_{d_1 d_4} p_{d_2} p_{d_3}
+ \delta_{d_2 d_3} p_{d_1} p_{d_4} + \delta_{d_2 d_4} p_{d_1} p_{d_3} + \delta_{d_3 d_4} p_{d_1} p_{d_2} \big)
\nonumber\\
&&+
\Big( \frac{\sin|\pv|}{|\pv|^5} + \frac{10\cos|\pv|}{|\pv|^6} - \frac{45\sin|\pv|}{|\pv|^7} - \frac{105\cos|\pv|}{|\pv|^8} + \frac{105\sin|\pv|}{|\pv|^9} \Big)
p_{d_1} p_{d_2} p_{d_3} p_{d_4}
\nonumber\\
&\approx&
\Big( \frac{1}{15} - \frac{|\pv|^2}{210} + \frac{|\pv|^4}{7560} \Big)
\big( \delta_{d_1 d_2}\delta_{d_3 d_4} + \delta_{d_1 d_3}\delta_{d_2 d_4} + \delta_{d_1 d_4}\delta_{d_2 d_3} \big)
\nonumber\\
&&+
\Big( -\frac{1}{105} + \frac{|\pv|^2}{1890} - \frac{|\pv|^4}{83160} \Big)
\big( \delta_{d_1 d_2} p_{d_3} p_{d_4} + \delta_{d_1 d_3} p_{d_2} p_{d_4} + \delta_{d_1 d_4} p_{d_2} p_{d_3}
\nonumber\\
&&\qquad+ \delta_{d_2 d_3} p_{d_1} p_{d_4} + \delta_{d_2 d_4} p_{d_1} p_{d_3} + \delta_{d_3 d_4} p_{d_1} p_{d_2} \big)
\nonumber\\
&&+
\Big( \frac{1}{945} - \frac{|\pv|^2}{20790} + \frac{|\pv|^4}{1081080} \Big)
p_{d_1} p_{d_2} p_{d_3} p_{d_4}.
\ea
\end{subequations}
Using Eqs.~(\ref{eq:definitions}) and (\ref{eq:operator_formulas}), we can now compute $\langle G_1(\Omega)^2 G_2(\nopc)^2 \rangle_{1,1+n}$:
\ba
\label{eq:G1SquareG2Square1}
\langle G_1(\Omega)^2 G_2(\nopc)^2 \rangle_{1,1+n}
&\approx&
\frac{2 {\tilde{\eta}}^4}{15}
\sum_{\alpha=0}^n \sum_{\bm{p},\bm{q}}\overline{\sum_{\hat{k}}}
 \frac{J_\pv\, J_\qv}{(T^\alpha)^2}
 \vertex_{\hat{k}}\vertex_{-\hat{k}-(\bm{p}+\bm{q}){\hat{\epsilon}}^\alpha}
 \Omega_{\hat{k}}\Omega_{-\hat{k}-(\bm{p}+\bm{q}){\hat{\epsilon}}^\alpha}
 Q_{d_1 d_2}^\alpha(\bm{p})Q_{d_3 d_4}^\alpha(\bm{q})
\nonumber\\
&&\times
\Big( \big( 2 - \frac{\ell^2 |\pv + \qv|^2}{56} + \frac{\ell^4 |\pv + \qv|^4}{8064}
- \frac{\ell^2 |\pv + \qv + 4\bm{k}^\alpha|^2}{56} + \frac{\ell^4 |\pv + \qv + 4\bm{k}^\alpha|^4}{8064} \big)
\delta_{d_1 d_3}\delta_{d_2 d_4}
\nonumber\\
&&
+\big(- \frac{1}{14} + \frac{\ell^2 |\pv+\qv|^2}{1008} \big)
\ell^2 (p+q)_{d_1} (p+q)_{d_3} \delta_{d_2 d_4}
\nonumber\\
&&
+\big(- \frac{1}{14} + \frac{\ell^2 |\pv+\qv + 4\bm{k}^\alpha |^2}{1008} \big)
\ell^2 (p+q+4k^\alpha)_{d_1} (p+q+4k^\alpha)_{d_3} \delta_{d_2 d_4}
\nonumber\\
&&+\frac{\ell^4}{2016}(p+q)_{d_1} (p+q)_{d_2}(p+q)_{d_3} (p+q)_{d_4}
\nonumber\\
&&
+\frac{\ell^4}{2016}(p+q+4k^\alpha)_{d_1} (p+q+4k^\alpha)_{d_2}(p+q+4k^\alpha)_{d_3} (p+q+4k^\alpha)_{d_4}
\Big)
\nonumber\\
&&+
\frac{2 {\tilde{\eta}}^4 \ell^4}{225}
\sum_{\alpha,\beta=0\atop{(\alpha\neq\beta)}}^{n} \sum_{\bm{p},\bm{q}}\overline{\sum_{\hat{k}}}
 \frac{J_\pv\, J_\qv}{T^\alpha\, T^\beta}
 \vertex_{\hat{k}}\vertex_{-\hat{k}-{\bm{p}}{\hat{\epsilon}}^\alpha-{\bm{q}}{\hat{\epsilon}}^\beta} \Omega_{\hat{k}}\Omega_{-\hat{k}-{\bm{p}}{\hat{\epsilon}}^\alpha-{\bm{q}}{\hat{\epsilon}}^\beta}
 Q_{d_1 d_2}^\alpha(\bm{p})Q_{d_3 d_4}^\beta(\bm{q})
\nonumber\\
&&\times
 \Big( k_{d_1}^\alpha k_{d_2}^\alpha k_{d_3}^\beta k_{d_4}^\beta + \frac{1}{4}(p_{d_1} k_{d_2}^\alpha + k_{d_1}^\alpha p_{d_2})(q_{d_3} k_{d_4}^\beta + k_{d_3}^\beta q_{d_4}) + \frac{1}{8} p_{d_1} p_{d_2} q_{d_3} q_{d_4} \Big),
\ea
in which the $\nop^\alpha$ matrices are constrained to be symmetric and traceless.

\section{Deriving the effective Hamiltonian for liquid crystallinity}
\label{app:eff_theory}
In this Appendix we derive the effective Hamiltonian for liquid crystallinity, Eq.~(\ref{eq:H_eff0}).  The first step in the derivation is to set $\Omega$ to its saddle-point value $\bar\Omega$ in $f_{1+n}$ of Eq.~(\ref{eq:log-trace-expansion}) whilst retaining fluctuations of $\nop$, and evaluate the terms $\langle G_1(\Omega)^{2} G_2(\nopc)\rangle_{1,1+n}$, $\langle G_1(\Omega)G_2(\nopc)^{2}\rangle_{1,1+n}$, $\langle G_1(\Omega)^{2} G_2(\nopc)^{2}\rangle_{1,1+n}$ and $\langle G_1(\Omega)^{2}\rangle_{1,1+n} \langle G_2(\nopc)^{2}\rangle_{1,1+n}$ with $\Omega = \bar\Omega$.  As we are considering length-scales large compared with $b$ or $\ell$,
we shall make the approximations $\vertex_{\hat{k}}\approx 1$ and $J_\pv \approx J_{\bm{0}}$.
(The wave-vector dependent parts of $\vertex_{\hat{k}}$ and $J_\pv$ would generate corrections of higher order in the wave-vector.)
By further arguing that fluctuation modes with $p > \xi_L^{-1}$ can be neglected in $f_{C}(\bar{\Omega},\nop)$ (see App.~\ref{app:eff_Hamiltonian}), we shall arrive at the effective Hamiltonian (\ref{eq:H_eff0}).

\subsection{$\langle G_1(\bar\Omega)^{2} \, G_2(\nopc)\rangle_{1,1+n}$}
\label{app:eff_theory_1}
This term is given by
\ba
\label{eq:OmegaOmegaQ}
\langle G_1(\bar{\Omega})^2 \, G_2(\nopc) \rangle_{1,1+n}
&\approx&\frac{{\tilde{\eta}}^4 \ell^2}{5}\sum_{\alpha=0}^n \sum_{\bm{p}}\overline{\sum_{\hat{k}}}
 \frac{J_{\bm{0}}}{T^\alpha}
 {\bar\Omega}_{\hat{k}}{\bar\Omega}_{-\hat{k}-{\bm{p}}{\hat{\epsilon}}^\alpha}
 Q_{d_1 d_2}^\alpha(\bm{p})
% \nonumber\\
%&&\times
 \big( p_{d_1} p_{d_2} + (k^\alpha + \frac{1}{2}p)_{d_1} (k^\alpha + \frac{1}{2}p)_{d_2} \big).
\ea
The product
${\bar\Omega}_{\hat{k}}{\bar\Omega}_{-\hat{k}-{\bm{p}}{\hat{\epsilon}}^\alpha}$
involves a product of two Kronecker deltas, $\delta_{\sum_{\alpha=0}^{n}\bm{k}^\alpha,\bm{0}}\delta_{-\pv-\sum_{\alpha=0}^{n}\bm{k}^\alpha,\bm{0}}$, that enforces macroscopic translational invariance, and implies that $\pv=\bm{0}$.  Equation~(\ref{eq:OmegaOmegaQ}) then becomes
\be
\label{eq:OmegaOmegaQ1}
\langle G_1(\bar{\Omega})^2 G_2(\nopc) \rangle_{1,1+n}
\approx
\sum_{\alpha=0}^{n}
\sum_{\hat{k}}
\frac{{\tilde{\eta}}^4 \, \ell^2 \, J_{\bm{0}}}{5T^\alpha}
{\bar\Omega}_{\hat{k}}
\,
{\bar\Omega}_{-\hat{k}}
k_{d_1}^\alpha\,
k_{d_2}^\alpha.
\ee
By decomposing the full wave-vector sum $\sum_{\hat{k}}$ into three separate parts corresponding to contributions from the HRS, 1RS and 0RS, the right hand side of Eq.~(\ref{eq:OmegaOmegaQ1}) becomes proportional to
\be
\sum_{\alpha=0}^{n}
\left( \sum_{\hat{k}} - \sum_{\hat{k}\in 1RS} - \sum_{\hat{k}\in 0RS} \right)
%\frac{{\tilde{\eta}}^4 \, \ell^2 \, J_{\bm{0}}}{5T^\alpha}
{\bar{\Omega}}_{\hat{k}}
\,
{\bar{\Omega}}_{-\hat{k}}
k_{d_1}^\alpha k_{d_2}^\alpha.
\ee
Here, the 0RS part vanishes, and the 1RS part also vanishes, because ${\bar\Omega}_{\hat{k}}$ is proportional to $\delta_{\sum_{\gamma=0}^{n} \bm{k}^\gamma, \bm{0}}$, which is zero if $\hat{k}$ belongs to the 1RS.  We are thus left with the full wave-vector sum; it is proportional to
\ba
\sum_{\hat{k}}
\int\frac{d\bm{z}_1\,d\bm{z}_2}{V^2}
k_{d_1}^\alpha\, k_{d_2}^\alpha\, e^{-\xi_L^2|\hat{k}|^2
+ i\sum_{\gamma=0}^n \bm{k}^\gamma\cdot (\bm{z}_1-\bm{z}_2)}
&=&
\sum_{\hat{k}}
\int\frac{d\bm{c}\,d\bm{u}}{V^2}
k_{d_1}^\alpha\, k_{d_2}^\alpha\, e^{-\xi_L^2\sum_{\gamma}|\bm{k}^\gamma
- i\frac{\bm{u}}{2\xi_L^2}|^2
- \frac{(n+1)|\bm{u}|^2}{4\xi_L^2}}
\nonumber\\
&\propto&
\delta_{d_1 d_2} \frac{1}{2\xi_L^2}
\left( \frac{\pi}{\xi_L^2} \right)^{(n+1)D/2}
\left( \frac{4\pi\xi_L^2}{n+1} \right)^{D/2}
\big( 1 - \frac{1}{n+1} \big)
\nonumber\\
&=&0,
\ea
where in the last step we have taken the replica limit.  On going from the first to the second step, we have changed to center-of-mass coordinate $\bm{c}$ and relative coordinate $\bm{u}$, and on going from the second to the third step we have shifted the wave-vector $\bm{k}^\gamma\rightarrow \bm{k}^\gamma - i\bm{u}/(2\xi_L^2)$ and integrated over $\bm{k}^\gamma$ ($\gamma=0,1,\ldots,n$) and $\bm{r}$.
Thus, the term
$\langle G_1(\bar\Omega)^{2} \, G_2(\nopc)\rangle_{1,1+n}$
vanishes.

\subsection{$\langle G_1(\bar\Omega) \, G_2(\nopc)^{2}\rangle_{1,1+n}$}

To lowest order in wave-vector, this term is given by
\ba
\langle G_1(\Omega) \, G_2(\nopc)^2 \rangle_{1,1+n}
&\approx&
\frac{{\tilde{\eta}}^2 \ell^4}{800}\sum_{\alpha\neq \beta} \sum_{\pv,\qv} \frac{J_0^2}{T^\alpha\, T^\beta}
p_{d_1} p_{d_2} q_{d_3} q_{d_4}
{\bar\Omega}_{-\pv\epsilon^\alpha-\qv\epsilon^\beta} \nopc_{d_1 d_2}^\alpha(\pv) \nopc_{d_3 d_4}^\beta(\qv)
\nonumber\\
&=&
\frac{{\tilde{\eta}}^2 \ell^4}{800}\sum_{\alpha\neq \beta}
\sum_{\pv,\qv} \int \frac{d\bm{z}}{V}
\frac{J_0^2\, G}{T^\alpha\, T^\beta}
p_{d_1} p_{d_2} q_{d_3} q_{d_4}
e^{-i(\pv+\qv)\cdot \bm{z} - \frac{1}{2}\xi_L^2(p^2+q^2)}
\nopc_{d_1 d_2}^\alpha(\pv) \nopc_{d_3 d_4}^\beta(\qv)
\nonumber\\
&=&
\frac{G {\tilde{\eta}}^2 \, \ell^4}{800}\sum_{\alpha\neq \beta}
\sum_{\pv} \frac{J_0^2}{T^\alpha\, T^\beta}
p_{d_1} p_{d_2} p_{d_3} p_{d_4} e^{-p^2\xi_L^2}
\nopc_{d_1 d_2}^\alpha(\pv) \nopc_{d_3 d_4}^\beta(-\pv)
%\nonumber\\
%&\equiv&
%\frac{G {\tilde{\eta}}^2 J_0^2\, \ell^4}{800}\sum_{\alpha\neq \beta}
%\sum_{\pv,\qv} \frac{p^4 e^{-p^2\xi_L^2}}{T^\alpha\, T^\beta}
%\phi^\alpha(\pv) \phi^\beta(-\pv)
\ea
%In the last line, we have defined for simplicity the field $\phi(\pv)\equiv p_{d_1} p_{d_2} \nopc_{d_1 d_2}(\pv)$.  This can be regarded as the \lq\lq longitudinal\rq\rq\ component of $\nop$.

\subsection{$\langle G_1(\bar\Omega)^{2}\rangle_{1,1+n} \, \langle G_2(\nopc)^{2}\rangle_{1,1+n}$}
\label{sec:appG_OOQQ}
This term vanishes.  To show this, it is sufficient to consider the value of $\overline{\sum\limits_{\hat{k}}}\,{\bar{\Omega}}_{\hat{k}}\,{\bar{\Omega}}_{-\hat{k}}$.
Using the formula (\ref{eq:barredOPmom}) for $\bar{\Omega}_{\hat{k}}$ and the decomposition $\overline{\sum\limits_{\hat{k}}}\equiv \sum_{\hat{k}}-\sum_{\hat{k}\in 1RS} - \sum_{\hat{k}\in 0RS}$, we have that
\ba
\label{eq:diagonalvanish}
&&\overline{\sum\limits_{\hat{k}}}
{\bar{\Omega}}_{\hat{k}}
{\bar{\Omega}}_{-\hat{k}}
=
\left( \sum_{\hat{k}}-\sum_{\hat{k}\in 1RS} - \sum_{\hat{k}\in 0RS} \right)
  G^2 \int\frac{d{\bm{z}_1} d{\bm{z}}_2}{V^2}
  e^{-\xi_L^2|\hat{k}|^2 -i\sum_{\gamma=0}^n {\bm{k}}^\gamma\cdot ({\bm{z}}_1-{\bm{z}}_2)}
  \nonumber\\
&&\qquad
= G^2 \left( -1 + V^{1+n}\int \prod_{\gamma=0}^n \frac{d\bm{k}^\gamma}{(2\pi)^D}
  \int\frac{d{\bm{c}} \, d{\bm{r}}}{V^2}
  e^{-\sum_{\gamma=0}^n \Big( \xi_L^2|{\bm{k}}^\gamma + \frac{i\rv}{2\xi_L^2}|^2 \Big)
  -\frac{(1+n)r^2}{4\xi_L^2}}
  \right)
  \nonumber\\
&&\qquad
= G^2 \left( -1 + \frac{V^{1+n}\cdot V}{V^2} \frac{1}{(2\pi)^{(1+n)D}}
    \Big(\frac{4\pi\xi_L^2}{1+n}\Big)^{D/2}
    \Big( \frac{\pi}{\xi_L^2} \Big)^{(1+n)D/2}
    \right)
    \nonumber\\
&&\qquad
=n\ln \left( \frac{V}{(2\pi)^D \xi_L^D} \right) G^2 + O(n^2).
\ea
After taking the replica limit $n\rightarrow 0$, the above result vanishes.
Here, we have used the definition $\sum_{\hat{k}}\equiv \frac{V^{1+n}}{(2\pi)^{(1+n)D}}\int d\hat{k}$ when performing the integral over $\hat{k}$.  In the last step, we have integrated over $\hat{k}$, $\bm{r}$, and $\bm{c}$.
The term $\langle G_1(\bar{\Omega})^2 \rangle_{1,1+n} \langle G_2(\nopc)^2 \rangle_{1,1+n}$ is therefore zero in the replica limit.

\subsection{$\langle G_1(\bar\Omega)^{2} \, G_2(\nopc)^{2}\rangle_{1,1+n}$}

We now consider the contribution $\langle G_1(\bar{\Omega})^2 \, G_2(\nopc)^2 \rangle$.  Despite its being of higher order in $\bar{\Omega}$, this term has long wavelength properties that are sufficiently hard, and it is therefore important for the purpose of studying the polydomain state of IGNEs.  It contains a part that is diagonal in replica space (i.e., proportional to $\nop^\alpha\, \nop^\alpha$) and a part that is off-diagonal in replica space (i.e., proportional to $\nop^\alpha\, \nop^\beta$ for $\alpha \neq \beta$).

First, let us consider the replica-diagonal part.  From Eq.~(\ref{eq:G1SquareG2Square1}), with $\Omega$ set to its saddle-point value $\bar{\Omega}$, and by making use of the fact that ${\bar{\Omega}}_{\hat{k}}{\bar{\Omega}}_{-\hat{k}-(\pv+\qv)\epsilon^\alpha}\propto \delta_{\sum_{\gamma}\bm{k}^\gamma,\bm{0}}\delta_{-\sum_{\gamma}\bm{k}^\gamma-\pv-\qv,\bm{0}}$, which implies that $\pv+\qv=\bm{0}$, we see that replica-diagonal terms with prefactors in powers of $|\pv+\qv|$ vanish.
Thus the replica-diagonal contribution to $\langle G_1(\Omega)^2 \, G_2(\nopc)^2 \rangle$ (with $\Omega$ set to ${\bar{\Omega}}$) becomes proportional to the quantity,
\ba
&&\sum_{\alpha=0}^{n}\sum_\pv \overline{\sum\limits_{\hat{k}}}
{\bar{\Omega}}_{\hat{k}}
{\bar{\Omega}}_{-\hat{k}}
Q_{d_1 d_2}^\alpha(\pv)Q_{d_3 d_4}^\alpha(-\pv)
\Big( \big( 2 - \frac{2}{7}\ell^2|\bm{k}^\alpha|^2 + \frac{2}{63}\ell^4 |\bm{k}^\alpha|^4 \big) \delta_{d_1 d_3}\delta_{d_2 d_4}
\nonumber
\\
&&\quad
+16\big( -\frac{1}{14} + \frac{\ell^2}{63} |\bm{k}^\alpha|^2 \big) \ell^2 k_{d_1}^\alpha k_{d_3}^\alpha \delta_{d_2 d_4}
+\frac{8}{63}\ell^4 k_{d_1}^\alpha k_{d_2}^\alpha k_{d_3}^\alpha k_{d_4}^\alpha
\Big).
\ea
The term proportional to $\delta_{d_1 d_3}\delta_{d_2 d_4}$ (without the wave-vector prefactors) vanishes, as we showed in Eq.~(\ref{eq:diagonalvanish}).
On the other hand, the remaining terms should also vanish, as ${\bar{\Omega}}_{\hat{k}}{\bar{\Omega}}_{-\hat{k}}$ is isotropic in wave-vector space.
The vanishing of the term in
$\sum_{\alpha=0}^{n}
\overline{\sum_{\hat{k}}}
k_{d_1}^\alpha k_{d_2}^\alpha
{\bar{\Omega}}_{\hat{k}}
{\bar{\Omega}}_{-\hat{k}}
$
comes about by the same argument as was presented in App.~\ref{app:eff_theory_1}.
By means of a calculation similar to that performed in App.~\ref{app:eff_theory_1}, one can also show that the following result holds in the replica limit:
\be
\overline{\sum_{\hat{k}}}
k_{d_1}^\alpha k_{d_2}^\alpha k_{d_3}^\alpha k_{d_4}^\alpha
{\bar{\Omega}}_{\hat{k}}
{\bar{\Omega}}_{-\hat{k}}
\equiv
\left( \sum_{\hat{k}} - \sum_{\hat{k}\in 1RS} - \sum_{\hat{k}\in 0RS} \right)
k_{d_1}^\alpha k_{d_2}^\alpha k_{d_3}^\alpha k_{d_4}^\alpha
{\bar{\Omega}}_{\hat{k}}
{\bar{\Omega}}_{-\hat{k}}
=0.
\ee
Next, we compute the replica off-diagonal contribution to $\langle G_1(\bar{\Omega})^2 \, G_2(\nopc)^2 \rangle$, whose terms we divide into the following three classes: (i)~a term with prefactor proportional to $kkkk$, (ii)~terms with prefactors proportional to $kk$ but not $kkkk$, and (iii)~terms with prefactors that do not depend on $k$ [see Eq.~(\ref{eq:G1SquareG2Square1})].

First, we compute the replica-off-diagonal term proportional to $kkkk$:
\ba
\label{eq:couplingSIdots}
&&\sum_{\alpha\neq\beta}^n \sum_{\bm{p},\bm{q}}\overline{\sum_{\hat{k}}}
 {\bar{\Omega}}_{\hat{k}}
 {\bar{\Omega}}_{-\hat{k}-{\bm{p}}{\hat{\epsilon}}^\alpha-{\bm{q}}{\hat{\epsilon}}^\beta}
 Q_{d_1 d_2}^\alpha(\bm{p})Q_{d_3 d_4}^\beta(\bm{q})\,
 k_{d_1}^\alpha k_{d_2}^\alpha k_{d_3}^\beta k_{d_4}^\beta
\nonumber\\
&=& G^2 V^{n+3} \int \prod_{\gamma=0}^{n} \frac{d^3 k^\gamma}{(2\pi)^3}\frac{d^3 p}{(2\pi)^3}\frac{d^3 q}{(2\pi)^3}
\int \frac{d\bm{z}_1}{V}\frac{d\bm{z}_2}{V}
\,\sum_{\alpha,\beta=0\atop{(\alpha\neq\beta)}}^{n}
\nopc_{d_1 d_2}^\alpha(\pv)\nopc_{d_3 d_4}^\beta(\qv)
\,
k_{d_1}^\alpha k_{d_2}^\alpha k_{d_3}^\beta k_{d_4}^\beta
\nonumber\\
&&
\times
\exp
\bigg(
-\sum_{\gamma\neq\alpha,\beta}\xi_L^2\left| \bm{k}^\gamma + i\frac{\bm{z}_1-\bm{z}_2}{2\xi_L^2} \right|^2
-\frac{(n-1)|\bm{z}_1-\bm{z}_2|^2}{4\xi_L^2}
-\frac{1}{2}\xi_L^2(p^2+q^2) - i(\pv+\qv)\cdot \bm{z}_2
\nonumber\\
&&\quad
-\xi_L^2\left| \bm{k}^\alpha + i\frac{\bm{z}_1-\bm{z}_2}{2\xi_L^2} + \frac{1}{2}\pv \right|^2
+\xi_L^2\left| i\frac{\bm{z}_1-\bm{z}_2}{2\xi_L^2} + \frac{1}{2}\pv\right|^2
-\xi_L^2\left| \bm{k}^\beta + i\frac{\bm{z}_1-\bm{z}_2}{2\xi_L^2} + \frac{1}{2}\qv \right|^2
\nonumber\\
&&\quad
+\xi_L^2\left| i\frac{\bm{z}_1-\bm{z}_2}{2\xi_L^2} + \frac{1}{2}\qv \right|^2
\bigg).
\ea
We replace the coordinates $\bm{z}_1$ and $\bm{z}_2$ by the relative coordinate $\bm{R}$ and center-of-mass coordinate $\bm{C}$, defined respectively by $\bm{R}=\bm{z}_1-\bm{z}_2$ and $\bm{C}=(\bm{z}_1+\bm{z}_2)/2$, and integrate over $\bm{C}$, which results in a factor of $(2\pi)^{d}V\delta_{\pv+\qv,\bm{0}}$.
Next, we integrate over $\qv$ to enforce the equality $\qv=-\pv$.
We then integrate out $\bm{k}^{\gamma}$ (for $\gamma\neq\alpha,\beta$).
The right-hand side of Eq.~(\ref{eq:couplingSIdots}) then becomes
\ba
\label{eq:term}
&&\frac{G^2 V^{n+2}}{(2\pi)^{(n-1)D}}
\left( \frac{\pi}{\xi_L^2} \right)^{(n-1)D/2}
\sum_{\alpha,\beta=0\atop{(\alpha\neq\beta)}}^{n}
\int \frac{d\bm{k}^\alpha}{(2\pi)^{D}}
\frac{d\bm{k}^\beta}{(2\pi)^{D}}
\frac{d\pv}{(2\pi)^{D}}
\int \frac{d\bm{R}}{V}
\nopc_{d_1 d_2}^\alpha(\pv)\nopc_{d_3 d_4}^\beta(-\pv)
\nonumber\\
&&\quad\times
\exp\Big( -\frac{1}{2}p^2\xi_L^2 - \frac{(n+1)|\bm{R}|^2}{4\xi_L^2}-\big(|\bm{k}^{\alpha}|^2 + |\bm{k}^{\beta}|^2\big)\xi_L^2 \Big)
\nonumber\\
&&\quad\times
\big(k^\alpha - \frac{i\bm{R}}{2\xi_L^2} -\frac{p}{2}\big)_{d_1}
\big(k^\alpha - \frac{i\bm{R}}{2\xi_L^2} -\frac{p}{2}\big)_{d_2}
\big(k^\beta - \frac{i\bm{R}}{2\xi_L^2} +\frac{p}{2}\big)_{d_3}
\big(k^\beta - \frac{i\bm{R}}{2\xi_L^2} +\frac{p}{2}\big)_{d_4}.
\ea
Because the exponent is invariant under rotations of $\bm{R}$, $\bm{k}^\alpha$, $\bm{k}^\beta$ and $\pv$, any terms having prefactors proportional to $R_{d_1}R_{d_2}$, $k_{d_1}^\alpha k_{d_2}^\alpha$, $k_{d_1}^\beta k_{d_2}^\beta$ or $p_{d_1}p_{d_2}$ will, respectively, become proportional to $\delta_{d_1 d_2}$ on integrating over $\bm{R}$, $\bm{k}^\alpha$, $\bm{k}^\beta$ and $\bm{p}$.  However, on contracting with $\nop_{d_1 d_2}$, these terms will yield no contributions, owing to the tracelessness of $\nop_{d_1 d_2}$.  Thus, the term (\ref{eq:term}) effectively becomes
\ba
&&\frac{G^2 V^{n+2}}{(2\pi)^{(n-1)D}}
\left( \frac{\pi}{\xi_L^2} \right)^{(n-1)D/2}
\sum_{\alpha,\beta=0\atop{(\alpha\neq\beta)}}^{n}
\int \frac{d\bm{k}^\alpha}{(2\pi)^{D}}
\frac{d\bm{k}^\beta}{(2\pi)^{D}}
\frac{d\pv}{(2\pi)^{D}}
\int \frac{d\bm{R}}{V}
\nopc_{d_1 d_2}^\alpha(\pv)\nopc_{d_3 d_4}^\beta(-\pv)
\nonumber\\
&&\quad\times
\exp\Big( -\frac{1}{2}p^2\xi_L^2 - \frac{(n+1)|\bm{R}|^2}{4\xi_L^2}-\big(|\bm{k}^{\alpha}|^2 + |\bm{k}^{\beta}|^2\big)\xi_L^2 \Big)
\nonumber\\
&&\quad\times
\Big(
\frac{1}{16\xi_L^4}
\big(
R_{d_1} R_{d_3} p_{d_2} p_{d_4} + R_{d_1} R_{d_4} p_{d_2} p_{d_3}
+ R_{d_2} R_{d_3} p_{d_1} p_{d_4} + R_{d_2} R_{d_4} p_{d_1} p_{d_3}
\big)
\nonumber\\
&&\qquad+
\frac{1}{16\xi_L^8} R_{d_1} R_{d_2} R_{d_3} R_{d_4}
+ \frac{1}{16} p_{d_1} p_{d_2} p_{d_3} p_{d_4}
\Big)
\ea
Then, by integrating over $\bm{R}$, $\bm{k}^{\alpha}$ and $\bm{k}^{\beta}$, we obtain the following result:
\ba
&&\sum_{\alpha,\beta=0\atop{\alpha\neq\beta}}^{n} \sum_{\bm{p},\bm{q}}\overline{\sum_{\hat{k}}}
 {\bar{\Omega}}_{\hat{k}}
 {\bar{\Omega}}_{-\hat{k}-{\bm{p}}{\hat{\epsilon}}^\alpha-{\bm{q}}{\hat{\epsilon}}^\beta}
 Q_{d_1 d_2}^\alpha(\bm{p})Q_{d_3 d_4}^\beta(\bm{q})\,
 k_{d_1}^\alpha k_{d_2}^\alpha k_{d_3}^\beta k_{d_4}^\beta
\nonumber\\
&&\quad=
\frac{G^2}{2\xi_L^4}
\sum_{\alpha,\beta=0\atop{\alpha\neq\beta}}^{n}
\sum_{\pv}
\big(
\delta_{d_1 d_3}\delta_{d_2 d_4} + \delta_{d_1 d_3}p_{d_2} p_{d_4}\xi_L^2 + \frac{1}{8}p_{d_1}p_{d_2}p_{d_3}p_{d_4}\xi_L^4
\big)
e^{-\frac{1}{2}p^2\xi_L^2}
\nopc_{d_1 d_2}^\alpha(\pv)\, \nopc_{d_3 d_4}^\beta(-\pv).
\ea
%From this, one can deduce that the coupling term $-\frac{1}{4}\langle G_{1}(\Omega)^{2} \, G_{2}(\nop)^{2} \rangle$ is given by
%\be
%-\frac{1}{4}\langle G_{1}(\Omega)^{2} \, G_{2}(\nop)^{2} \rangle =
%-\frac{{\tilde{\eta}}^2 \ell^4 J_0^2 G^2 }{ 900 \xi_L^4\, T^\alpha\, T^\beta }
%\sum_{\pv} \exp(-p^2\xi_L^2/2) \{ \nop_{\pv}^{\alpha} \, \nop_{-\pv}^{\beta} \}.
%\ee

By means of similar calculations, we obtain the following results:
\ba
&&\sum_{\alpha,\beta=0\atop{\alpha\neq\beta}}^{n} \sum_{\bm{p},\bm{q}}\overline{\sum_{\hat{k}}}
 {\bar{\Omega}}_{\hat{k}}
 {\bar{\Omega}}_{-\hat{k}-{\bm{p}}{\hat{\epsilon}}^\alpha-{\bm{q}}{\hat{\epsilon}}^\beta}
 Q_{d_1 d_2}^\alpha(\bm{p})Q_{d_3 d_4}^\beta(\bm{q})\,
 \big( p_{d_1} k_{d_2}^\alpha + k_{d_1}^\alpha p_{d_2} \big)
 \big( q_{d_3} k_{d_4}^\beta + k_{d_3}^\beta q_{d_4} \big)
\nonumber\\
&&\quad=
\frac{G^2}{\xi_L^4}
\sum_{\alpha,\beta=0\atop{\alpha\neq\beta}}^{n}
\sum_{\pv}
\big(
2 \delta_{d_1 d_3} p_{d_2} p_{d_4} \xi_L^2 + p_{d_1} p_{d_2} p_{d_3} p_{d_4} \xi_L^4
\big)
e^{-\frac{1}{2}p^2\xi_L^2}
\nopc_{d_1 d_2}^\alpha(\pv)\, \nopc_{d_3 d_4}^\beta(-\pv),
\ea
\ba
&&\sum_{\alpha,\beta=0\atop{\alpha\neq\beta}}^{n} \sum_{\bm{p},\bm{q}}\overline{\sum_{\hat{k}}}
 {\bar{\Omega}}_{\hat{k}}
 {\bar{\Omega}}_{-\hat{k}-{\bm{p}}{\hat{\epsilon}}^\alpha-{\bm{q}}{\hat{\epsilon}}^\beta}
 Q_{d_1 d_2}^\alpha(\bm{p})Q_{d_3 d_4}^\beta(\bm{q})\,
 p_{d_1} p_{d_2} q_{d_3} q_{d_4}
\nonumber\\
&&\quad=
G^2
\sum_{\alpha,\beta=0\atop{\alpha\neq\beta}}^{n}
\sum_{\pv}
 p_{d_1} p_{d_2} p_{d_3} p_{d_4}
 e^{-\frac{1}{2}p^2\xi_L^2}
\nopc_{d_1 d_2}^\alpha(\pv)\, \nopc_{d_3 d_4}^\beta(-\pv).
\ea
Putting the pieces together, we then obtain
\ba
\langle G_1(\Omega)^2 \, G_2(\nopc)^2 \rangle_{1,1+n}
&\approx&
\frac{G^2 {\tilde{\eta}}^4 \, \ell^4}{225\xi_L^4}
\sum_{\alpha,\beta=0\atop{\alpha\neq \beta}}^{n}
\sum_{\pv}
\frac{J_0^2}{T^\alpha\, T^\beta}
e^{-p^2\xi_L^2}
\nonumber\\
&&\times
\big(
\delta_{d_1 d_3}\delta_{d_2 d_4} + 5\delta_{d_1 d_3}p_{d_2}p_{d_4}\xi_L^2 + \frac{33}{8}p_{d_1}p_{d_2}p_{d_3}p_{d_4}
\big)
\nopc_{d_1 d_2}^\alpha(\pv) \nopc_{d_3 d_4}^\beta(-\pv).
\ea

\subsection{Effective Hamiltonian for liquid crystallinity}
\label{app:eff_Hamiltonian}
The effective Hamiltonian for liquid crystallinity, $H_{1+n}[\{ \nop^\alpha \}_{\alpha=0}^{n}]$ in Eq.~(\ref{eq:H_eff0}), can now be derived using the coefficients obtained in App.~\ref{app:d}.
Defining $H_{1+n}[\{ \nop^\alpha \}_{\alpha=0}^{n}]$ by
\ba
H_{1+n}[\{ \nop^\alpha \}_{\alpha=0}^{n}]
&\equiv&
f_{Q}(\nop) + f_{C}(\bar{\Omega},\nop),
\ea
we have
\ba
H_{1+n}[\{ \nop^\alpha \}_{\alpha=0}^{n}]
&\approx&
\frac{1}{2}\sum_{\alpha=0}^n \sum_{\bm{p}} \frac{J_\pv}{T^\alpha} |Q_{ab}^\alpha(\bm{p})|^2
    -\frac{1}{2}\langle G_2(\nopc)^2 \rangle_{1,1+n}
\nonumber\\
&&- \frac{1}{2}\langle G_1(\bar{\Omega})^2 G_2(\nopc) \rangle_{1,1+n}
   -\frac{1}{2}\langle G_1(\bar{\Omega}) G_2(\nopc)^2 \rangle_{1,1+n}
\nonumber\\
&&- \frac{1}{4}\big(\langle G_1(\bar{\Omega})^2 G_2(\nopc)^2 \rangle_{1,1+n}
   -\langle G_1(\bar{\Omega})^2 \rangle_{1,1+n} \langle G_2(\nopc)^2 \rangle_{1,1+n} \big)
\nonumber\\
&\approx&
 \sum_{\bm{p}}
 \left(
 \frac{1}{2T^0}(\LDLa^0 t^0 + \stiff^0 p^2)
 \{ \nop_{\pv}^{0} \, \nop_{-\pv}^{0} \}
%\nonumber\\
%&&\!\!\!\!\!\!\quad
+\sum_{\alpha=1}^n
%\sum_{\bm{p}}
 \frac{1}{2T}(\LDLa t + \stiff p^2)
 \{ \nop_{\pv}^{\alpha} \, \nop_{-\pv}^{\alpha} \}
 \right)
\nonumber\\
&&-\sum_\pv
\left(
- \frac{1}{T^0}\sum_{\alpha=1}^{n}\sum_{\bm{p}}
  \smear_{\pv} \{ \nop_{\pv}^{0} \, \nop_{-\pv}^{\alpha} \}
%\nonumber\\
%&&\!\!\!\!\!\!\quad
+ \frac{1}{2T}\sum_{\alpha,\beta=1\atop{(\alpha\neq\beta)}}^{n}
%\sum_{\bm{p}}
  \smear_{\pv} \{ \nop_{\pv}^{\alpha} \, \nop_{-\pv}^{\beta} \}
  \right),
  \label{eq:E17}
\ea
where $\smear_\pv$ is defined as in Eq.~(\ref{eq:smear_definition}).  Eq.~(\ref{eq:E17}) is Eq.~(\ref{eq:H_eff0}).
In performing this derivation we have made two approximations.
Our first approximation consists in neglecting terms of cubic or quartic order in $\nop$, as we are considering an IGNE that is both prepared and measured at high temperatures.
Our second approximation consists in neglecting those terms in Eq.~(\ref{eq:f_C}) that vanish with $p\xi_L$ whilst retaining those terms that do not vanish with $p\xi_L$. It may appear that terms with factors of larger powers in $p\xi_L$ would dominate over those with smaller powers of $p\xi_L$ for large $p\xi_L$; however, the same terms also come equipped with exponential-damping factors of $\exp(-p^2\xi_L^2/2)$, which means that the large $p\xi_L$ contributions can effectively be neglected. We are thus left with the small $p\xi_L$ modes, and for these, the terms in $f_{C}(\bar{\Omega},\nop)$ that vanish with $p\xi_L$ are obviously smaller than those that do not vanish with $p\xi_L$.

The effective Hamiltonian thus obtained allows us to predict a novel regime in which correlators $\mycorrel^T$ and $\mycorrel^G$ undergo spatial decay and oscillation~(see Sec.~\ref{sec:oscillatory_decay}).
As we shall see in the following subsection, the terms neglected by our second approximation would not result in a qualitative change to this prediction.

\subsection{Corrections arising from neglected terms}
\label{app:h}

In the previous subsection the effective Hamiltonian $H_{1+n}[\{ \nop^\alpha \}_{\alpha=0}^{n}]$ was derived under the approximation that we neglect terms originating in $f_{C}(\bar{\Omega},\nop)$ that contain factors of $p\xi_L$.  In this subsection we consider the modifications to the thermal and glassy correlators $\mycorrel^T$ and $\mycorrel^G$ arising from these neglected terms.  For simplicity, we consider the case of a system prepared at a high temperature, so that the nematic order parameter in the zeroth replica, $\nop^0$, may be set to zero.
Taking the neglected terms into account, one has the effective Hamiltonian
\ba
\label{eq:modified_Heff}
H_{{\rm{eff}}}\left[ \{ \nop^\alpha \}_{\alpha=1}^{n} \right]
&=&
\frac{1}{2T}\sum_\alpha\sum_\pv \big( \LDLa t + \stiff p^2 \big) \{ \bm{W}_\pv^\alpha\, \bm{W}_\pv^\alpha \}
-\frac{\nu_1}{2T^2}\sum_{\alpha\neq \beta}\sum_\pv e^{-\frac{1}{2}p^2\xi_L^2} \{ \bm{W}_\pv^\alpha\, \bm{W}_\pv^\beta \}
\nonumber\\
&&+
\frac{1}{2T}\sum_\alpha\sum_\pv \big( \LDLa t + \stiff p^2 \big) \bm{X}_\pv^\alpha \cdot \bm{X}_\pv^\alpha
-\frac{\nu_1}{2T^2}\sum_{\alpha\neq \beta}\sum_\pv \big( 1 + 10p^2\xi_L^2 \big) e^{-\frac{1}{2}p^2\xi_L^2} \bm{X}_\pv^\alpha \cdot \bm{X}_\pv^\beta
\nonumber\\
&&+
\frac{1}{2T}\sum_\alpha\sum_\pv \big( \LDLa t + \stiff p^2 \big) \phi_\pv^\alpha \, \phi_\pv^\alpha
\nonumber\\
&&
-\frac{1}{2T^2}\sum_{\alpha\neq \beta}\sum_\pv
\Big( \nu_1 \big( 1 + \frac{17}{4} p^4\xi_L^4 \big) e^{-\frac{1}{2}p^2\xi_L^2}
+ \nu_2 p^4\xi_L^4 e^{-p^2\xi_L^2} \Big) \phi_\pv^\alpha \cdot \phi_\pv^\beta,
\ea
where we have introduced $\nu_1\equiv (G^2 {\tilde{\eta}}^4 J_0^2/225)(\ell/\xi_L)^{4}$ and $\nu_2\equiv (G {\tilde{\eta}}^2 J_0^2/800)(\ell/\xi_L)^{4}$.  We have also made use of the fact that any symmetric, traceless tensor $\nop_\pv$ can be decomposed (in a wave-vector-dependent manner) into its five independent modes $\phi_\pv$, $\bm{X}_\pv$ and $\bm{W}_\pv$: viz.,
\be
\nopc_{d_1 d_2}(\pv) = \frac{p_{d_1}}{|\pv|}\, \frac{p_{d_2}}{|\pv|} \phi_\pv
+ P_{d_1 d_3}^T\, \frac{p_{d_2}}{|\pv|} \, X_{d_3}(\pv)
+ P_{d_2 d_3}^T\, \frac{p_{d_1}}{|\pv|} \, X_{d_3}(\pv)
+ W_{d_1 d_2}(\pv),
\ee
where the modes $\phi_\pv$, $\bm{X}_\pv$ and $\bm{W}_\pv$ are given by
\begin{subequations}
\ba
\phi_\pv &\equiv& \frac{p_{d_1}}{|\pv|}\, \frac{p_{d_2}}{|\pv|}\, \nopc_{d_1 d_2}(\pv),
\\
W_{d_1 d_2}(\pv) &\equiv& P_{d_1 d_3} \, P_{d_2 d_4} \, \nopc_{d_3 d_4}(\pv)
- \frac{1}{2} \big( P_{d_5 d_3} \, P_{d_5 d_4} \, \nopc_{d_3 d_4} + \phi_\pv \big)
P_{d_1 d_2},
\\
\bm{X}_{d_1}(\pv) &\equiv& P_{d_1 d_2} \, \frac{p_{d_3}}{|\pv|} \, \nopc_{d_2 d_3}(\pv),
\ea
\end{subequations}
and the projection operator $P_{d_1 d_2}^T$ is defined via
\be
P_{d_1 d_2}^T\equiv \delta_{d_1 d_2} - \frac{p_{d_1}}{|\pv|}\, \frac{p_{d_2}}{|\pv|}.
\ee
By using Eq.~(\ref{eq:modified_Heff}) we can compute the thermal and glassy correlators for the five independent modes in Fourier space.
The thermal fluctuation correlators are given by
\begin{subequations}
\label{eq:mode_thermal_correlators}
\ba
\left[ \{ \langle \bm{W}_\pv\, \bm{W}_{-\pv} \rangle - \langle \bm{W}_\pv\rangle \langle \bm{W}_{-\pv} \rangle \} \right]
&=& \frac{2T}{\LDLa t + \stiff p^2
+
\nu_1 T^{-1} e^{-\frac{1}{2}p^2\xi_L^2}},
\\
\left[ \langle \bm{X}_\pv \cdot \bm{X}_{-\pv} \rangle - \langle \bm{X}_\pv\rangle \cdot \langle \bm{X}_{-\pv} \rangle \right]
&=& \frac{2T}{\LDLa t + \stiff p^2 + \nu_1 T^{-1} \big( 1 + 10 p^2\xi_L^2 \big) e^{-\frac{1}{2}p^2\xi_L^2} },
\\
\left[
\langle \phi_\pv\, \phi_{-\pv} \rangle - \langle \phi_\pv \rangle \langle \phi_{-\pv} \rangle
\right]
&=& \frac{T}{\LDLa t + \stiff p^2
+ T^{-1}
\Big(
\nu_1\big( 1 + \frac{17}{4}p^4\xi_L^4 \big) e^{-\frac{1}{2}p^2\xi_L^2}
+ \nu_2 p^4\xi_L^4 e^{-p^2\xi_L^2}
\Big) }.
\ea
\end{subequations}
We note that a given mode's correlator undergoes oscillatory decay in real space if the least stable mode has a nonzero wave-vector.  Equivalently, if we Taylor expand the denominator of that correlator in powers of the wave-vector about zero and find that the effective stiffness (i.e., the coefficient of the quadratic term) is negative then the correlator undergoes oscillatory decay in real space.  We can check that the effective stiffness does indeed become negative at sufficiently large values of the disorder strength (which is proportional to $\nu_1$) for the modes $\bm{W}_\pv$ and $\phi_\pv$, although it does not for $\bm{X}_\pv$.  The full thermal fluctuation correlator $\mycorrel^T(\rv)$ is the sum of the inverse Fourier transforms of the correlators for the modes in Eq.~(\ref{eq:mode_thermal_correlators}).  Thus, at sufficiently large disorder strength, $\mycorrel^T(\rv)$ will undergo oscillatory decay.

To determine whether the glassy correlator $\mycorrel^G(\rv)$ undergoes oscillatory decay at sufficiently large values of the disorder strength we apply a similar analysis, computing the glassy correlator for each mode:
\begin{subequations}
\label{eq:mode_glassy_correlators}
\ba
\left[
\{ \langle \bm{W}_\pv\rangle \langle \bm{W}_{-\pv} \rangle \}
\right]
&=& \frac{2\nu_1 e^{-\frac{1}{2}p^2\xi_L^2}}{\big( \LDLa t + \stiff p^2
+
\nu_1 T^{-1} e^{-\frac{1}{2}p^2\xi_L^2} \big)^2};
\\
\left[
\langle \bm{X}_\pv\rangle \cdot \langle \bm{X}_{-\pv} \rangle
\right]
&=& \frac{2\nu_1 (1+10p^2\xi_L^2)e^{-\frac{1}{2}p^2\xi_L^2}}{\Big( \LDLa t + \stiff p^2 + \nu_1 T^{-1} \big( 1 + 10 p^2\xi_L^2 \big) e^{-\frac{1}{2}p^2\xi_L^2} \Big)^2};
\\
\left[
\langle \phi_\pv \rangle \langle \phi_{-\pv} \rangle
\right]
&=& \frac{\nu_1\big( 1 + \frac{17}{4}p^4\xi_L^4 \big) e^{-\frac{1}{2}p^2\xi_L^2}
+ \nu_2 p^4\xi_L^4 e^{-p^2\xi_L^2}
}{\Big( \LDLa t + \stiff p^2
+ T^{-1}
\big(
\nu_1\big( 1 + \frac{17}{4}p^4\xi_L^4 \big) e^{-\frac{1}{2}p^2\xi_L^2}
+ \nu_2 p^4\xi_L^4 e^{-p^2\xi_L^2} \big)
\Big)^2 }.
\ea
\end{subequations}
The full thermal fluctuation correlator $\mycorrel^G(\rv)$ is the sum of the inverse Fourier transforms of the correlators for the modes in Eq.~(\ref{eq:mode_glassy_correlators}).  As for the thermal correlator, the denominators of the glassy correlators for the modes $\bm{W}_\pv$ and $\phi_\pv$ (but not $\bm{X}_\pv$) can acquire minima at non-zero wave-vectors for sufficiently large disorder strengths. In real space, such nonzero-wave-vector minima imply that the correlator $\mycorrel^G(\rv)$ undergoes oscillation as it decays.

\section{A conventional random field approach}
\label{app:rf_approach}
%**How is this approach "conventional"?  Does the conventionality refer to the use of only one random field?  Should we call it "naive random field approach"? Should we write something about using two random fields?

In this Appendix we provide the steps leading to the thermal and glassy correlators that one would obtain from a conventional random field approach.  In this approach, an IGNE subject to a given realization of quenched disorder $H$ can be described by the Landau free energy
\be
H_{\rm{rf}} = \frac{1}{2}\sum_\pv \big( \LDLa t + \stiff p^2 \big)\{ \nop_\pv\, \nop_{-\pv} \} - \sum_\pv \{ \bm{H}_\pv\, \nop_{-\pv} \}.
\ee
Here, $\bm{H}$ is a quenched random field having a Gaussian distribution defined by mean and variance
\begin{subequations}
\ba
\left[ H_{d_1 d_2}(\pv) \right] &=& 0,
\\
\left[ \{ H_{d_1 d_2}(\pv)\, H_{d_3 d_4}(-\pv) \} \right]
&=& \Delta (\delta_{d_1 d_2}\delta_{d_3 d_4} + \delta_{d_1 d_2}\delta_{d_3 d_4} + \delta_{d_1 d_2}\delta_{d_3 d_4}).
\ea
\end{subequations}
By using the replica technique to eliminate the quenched disorder, one obtains the effective Hamiltonian
\be
H_{\rm{ef}} = \frac{1}{2}\sum_{\alpha=1}^{n} \sum_\pv \big( \LDLa t + \stiff p^2 \big) \{ \nop_\pv^\alpha\, \nop_{-\pv}^\alpha \}
+ \frac{1}{2}\sum_{\alpha, \beta=1}^{n} \sum_\pv \Delta \{ \nop_\pv^\alpha\, \nop_{-\pv}^\beta \}.
\ee
The inverse of the Hessian matrix corresponding to $H_{\rm{ef}}$ is given by (for $\alpha,\beta=1,\ldots,n$)
\be
\label{eq:hessian_inverse}
\left( \frac{\delta^{2}H_{\rm{ef}}}{\delta Q_{d_1 d_2}^\alpha(\pv) \delta Q_{d_1 d_2}^\beta(-\pv)} \right)^{-1}
= \frac{1}{\LDLa t + \stiff p^2}\delta^{\alpha\beta} + \frac{\Delta}{\big( \LDLa t + \stiff p^2 \big)^2}\mathbf{1}^{\alpha\beta},
\ee
where repeated Cartesian indices $d_1$, $d_2$ mean that they are to be summed over, and $\mathbf{1}^{\alpha\beta}$ is an $n\times n$ matrix with each entry having the value of unity.
Equation~(\ref{eq:hessian_inverse}) yields for the thermal and glassy correlators the results
\begin{subequations}
\ba
\mycorrel_\pv^T &=& \frac{5T}{\LDLa t + \stiff p^2}\,,
\\
\mycorrel_\pv^G &=& \frac{5T \, \Delta}{\big(\LDLa t + \stiff p^2\big)^2}\,.
\ea
\end{subequations}
Note that the denominators of $\mycorrel_\pv^T$ and $\mycorrel_\pv^G$, calculated via a conventional random field approach developed in this appendix, do not feature the length-scale-dependent function $\smear_\pv$ that appears in the denominators of (and plays an essential role in determining the behavior of) the correlators (\ref{eq:correl}) derived from the microscopic dimer-and-springs model.

%\newpage

\end{widetext}


\begin{thebibliography}{99}
\bibitem{KF}
The Kupfer-Finkelmann procedure is as follows.  The nematic polymer melt is weakly cross-linked to form a weak gel, afterwhich it is uniaxially stretched in the nematic state until it becomes a macroscopic nematically-ordered state. Keeping the sample stretched, a second cross-linking reaction is initiate, which fixes the enforced uniaxial alignment.
\bibitem{ev-footnote}
One could allow the entire rod volume to be excluded, not just the ends. However, this is technically cumbersome and does not add add anything essential to the key effect of the rod-end excluded-volume interaction, which is to penalize density fluctuations.
\bibitem{ref:massfunction}
From the above expression for $\mass(T^{\alpha},\pv)$, it may appear that for temperatures $T^{\alpha}>T^\ast$, fluctuation modes with very short wavelengths (i.e., $p \gg a^{-1}$) are always marginally stable whereas the zero wave-vector mode can become more stable than those modes only at $T$ close to $T^\ast$.  However, the marginally stable, short-wavelength modes are really an artefact of the approximation we have made, in Eq.~(\ref{eq:Hnem}), to the true interaction potential, $\mathcal{J}(\bm{c})$, of nematogens.
In the general spirit of Landau theory, which considers length-scales larger than the shortest length-scales (for example, the range of the nematogen--nematogen interaction $a$), we shall exclude consideration of those fluctuation modes that have wave-vector magnitudes larger than $1/a$.
\bibitem{instability-footnote}
This critical value ${\tilde{\eta}}^2$ implies that the critical average linking number $[M]/N$ is $1/2$, which is consistent with that found for the randomly-linked particle model~\cite{xiaoming}.  This is not surprising, as the connectivity of the network is a topological characteristic that should not vary with the shape of the connected constituents (e.g., if one were to replace point particles with extended rods).
\end{thebibliography}
\end{document}